\documentclass[aps,twocolumn,superscriptaddress,longbibliography]{revtex4-1}

\usepackage{mathrsfs}
\usepackage{slashed}
\usepackage{amsmath}
\usepackage{amsfonts}
\usepackage{graphicx,color}
\usepackage{times}
\usepackage[caption=false]{subfig}
\usepackage[colorlinks,linkcolor=red,citecolor=blue]{hyperref}
\usepackage{amssymb}
\usepackage{color}
\usepackage{graphicx}
\usepackage{dcolumn}
\usepackage{bm}
\usepackage[header,title,page,titletoc]{appendix}

\begin{document}

\preprint{...}

\title{Fractionally charged anyon generated by topological path fusion in magnetic flux lattice}

\author{Tieyan Si}

\affiliation{ School of Physics, Harbin Institute of Technology, Harbin, 150080, China}

\date{\today}

\begin{abstract}

Anyon usually exists as collective excitation of two dimensional electron gas subjected to strong magnetic field, carrying fractional charges and exotic statistical character beyond fermion and boson. Fractional quantum Hall effect (FQHE) is the only experimental system showing solid evidence of anyon and a serial of fractional charges so far. Searching for new serial of fractional charges in FQHE or other physical system is still a challenge for both theoretical and experimental study. Here a topological fusion theory of propagating paths winding around a pair of fluxes is proposed to explore the physical origin of fractional charges. This topological path fusion theory not only generated all of the existed serial of fractional charges in FQHE and found the exact correspondence between FQHE and integral quantum Hall effect (IQHE), but also predicted new serial of fractional charges in FQHE. Further more, serial irrational charges like $2/(3+\sqrt{2})$ in one dimensional lattice of magnetic fluxes as well as that in two dimensional lattice of magnetic fluxes, such as $(1+\sqrt{2})$, are predicted. Even in three dimensional network of magnetic fluxes, a serial of fractionally charged anyon is predicted by this topological path fusion theory, which has exactly correspondence with the knot lattice model of anyon. In fact, in a multi-connected space time without magnetic field, this topological path fusion theory still holds, revealing an universal existence of fractional charge and mass in quantum material with strong confinement of particles (such as photonic crystal with porous nano-structures) and paving a new way for topological quantum computation.

\end{abstract}

\pacs{73.43.Cd, 03.75.Ss, 05.30.Fk, 71.10.Fd}

\maketitle


\section{Introduction}

The collective excitations of two dimensional electron gas in strong magnetic field carry a serial of fractional fractional charges, which are measured by the fractional Hall conductance \cite{Stormer1999} and explained by Laughlin wavefunction \cite{Laughlin1983} as well as composite fermion theory (i.e., one electron binding together with a pair of magnetic flux) \cite{Jain1989}. Topological order inspired by FQHE have attracted longstanding research interest on fractionally charged quasiparticles (which sometimes behave like anyons) in condensed matter physics \cite{Murthy2003}\cite{Kapfer2019}\cite{Nayak2008}\cite{Blok1990}. However, unlike the serial filling fractions in FQHE, only a few fractionally charged states are found in other many body physics theory, such as 0-charged spinon in resonance valence bond state \cite{Rice2007}, kinks with e/2 in Polyacetylene chain \cite{Su1979}, 1/3 filling states in interacting boson system on Kagome lattice \cite{Zhang2013}, irrational charge in quantum dimer model on hypercubic lattices \cite{Moessner2010}, and fractional quasi-excitation states in one-dimensional optical superlattice\cite{Lang2012}. Fractional filling states with non-trivial topological order has promising application in topological quantum computation \cite{Nayak2008} and exploring new physical phases in topological matters \cite{Kou2016}. In fact, many fractionally charged states in FQHE or other quantum lattice model are still not fully understood from an unified root of physical principal.

Here we proposed a topological path fusion mechanism of propagating electrons in magnetic flux lattice to generate a serial of fractionally charged states. These fractional charges cannot be explained by the Aharonov-Bohm effect (AB-effect) caused by the interference of the wave functions of two possible paths for an electron passing around one magnetic flux \cite{Aharonov1959}. When an electron meets the magnetic flux lattice, besides the two paths keeping the fluxes to its one side, there still exist many other paths that penetrate through the domain between the two fluxes. The scattering amplitude of an electron passing through this flux lattice, according to Feynman's path integral theory \cite{Feynman1959}, must take into account of all possible paths. The conventional quantum interference of different paths only considered the paths that are well separated and propagates monotonically in one direction, it always leads to Aharonov-Bohm effect. However, in some extremely confined cases, some path may winds back or gets too close to avoiding its neighboring path, it would inevitably interference with itself or fuse into other paths. This path fusion process is the quantum origin of fractional charges in this topological path fusion theory.

There are two ways to count all topologically non-equivalent paths which can not map into one other under continuous topological transformation. The first way is viewing each flux as a forbidden hole and the whole space as a multi-connected domain, then different paths are characterized by their local winding number around a flux within the flux lattice. Another equivalent way is continuously braiding a flux attached by an unbroken initial path with other selected fluxes. These two approaches are equivalent, because the winding motion of an electron around a flux is the relative motion of a flux carrying an unbroken electric current around to exchange its position with other fluxes. Mapping the unbroken electric current into a simple closed curve and the flux into a genus under the mathematical constraint that the curve avoid crossing itself everywhere, the braiding operations of fluxes enclosed by a loop current can be well quantified by Thurston's train track theory \cite{Thurston1988}\cite{Mosher2003}, which is applied to design the optimal mixing strategy of two fluids with low Reynolds number \cite{Gouillart2006}\cite{Allshouse2012}, study the topological fluid mechanics of point vortex \cite{Boyland2003} and topological chaos in dynamics systems \cite{Thiffeault2005}\cite{Stremler2011}. Here this topological path fusion model can be implemented by topological mixing of two quantum fluids, one is charged superfluid which is experimentally realizable by charged superfluid helium \cite{Laimer2019}, the other is normal viscous fluid helium. The charged superfluid helium acts as conducting channel in which an electron can move around freely but keeps the total probability conserved simultaneously.

The paper is organized as follows: in section II, the topological path fusion is first introduced by quantum interference of three paths around a flux pair followed by a quantum field description of winding tracks by Abelian Chern-Simons field theory. Then proposed the exact correspondence between winding train track and curves on torus as well as the knot lattice. Different serials of fractional charges are derived from this topological path fusion theory and irrational charges are predicted around triple flux cluster. In section III, the topological path fusion model are expanded into one dimensional lattice of fluxes. In section IV, the winding train track pattern in two dimensional lattice of flux pairs are generated by translation operation as well as topological transformation of two dimensional knot lattice. In section V, the fractionally charged anyon are well-constructed in three dimensional lattice of magnetic fluxes, disclosing a new phenomena beyond the widespread belief that anyon does not exist in three dimensional space. The last section is a brief summary and outlook.

\section{Fractional charges from the topological fusion of paths around magnetic flux cluster}

\subsection{Fractional charges generated by topological fusion of paths around magnetic flux pair}

\subsubsection{Topological path fusion of an electron passing through magnetic flux pair}

$\textbf{The gauge symmetry of braiding a flux pair}$

\begin{figure}
\begin{center}
\includegraphics[width=0.40\textwidth]{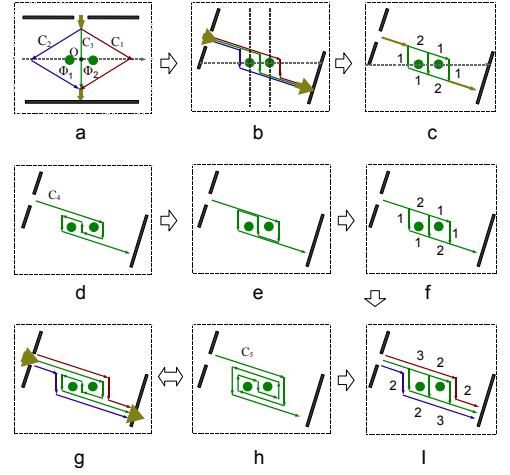}
\caption{\label{ABphase} (a) The interference of three electron beams out of the same source is induced by the magnetic flux pair that is placed in between the source screen and detector screen. The Aharonov-Bohm phase in this case is similar to triple-slits interference effect. (b) The three electron beams fuses at a small injecting angle that between the input middle beam and the axis passing through the center of two fluxes. (c) The fused electron beam loops around two magnetic fluxes with anisotropic weight on each edge. (d) The electron beam passing through the middle gap between two fluxes experienced one clockwise braiding on the two fluxes. (e - f) The braided middle path fuses at a small injecting angle. (g) The three paths with a braided middle path are generated out of the same source. (h) The middle path experienced two clockwise braiding operations on the two fluxes. (I) The three paths with a fused middle path that experienced only one clockwise braiding operation.}
\end{center}
\vspace{-0.5cm}
\end{figure}

An electron beam passing through a magnetic flux shows Aharonov-Bohm effect\cite{Aharonov1959}. Here we consider the three electron beams emitted by the same source, passing a pair of magnetic fluxes (represented by the green disc labeled by $\Phi_{1}$ and $\Phi_{2}$ in Fig. \ref{ABphase} (a)) to interfere with one another on the detector screen. When the three topologically inequivalent paths, labeled by  $C_{1}$ (the red path), $C_{2}$ (the blue path) and $C_{3}$ (the green path) in Fig. \ref{ABphase} (a), travel through the electromagnetic potential field in the surrounding region around the two magnetic fluxes, an electron wave function gains three different phases along the three paths,
\begin{eqnarray}\label{psi1psi2}
|\psi_{_{1}}\rangle&=& \psi_{_{10}}e^{\frac{ie}{\hbar{c}}\int_{C_{1}}{\vec{A}{(\vec{x})}}d\vec{x}},\;\;|\psi_{_{2}}\rangle= \psi_{_{20}}e^{\frac{ie}{\hbar{c}}\int_{C_{2}}{\vec{A}{(\vec{x})}}d\vec{x}}.\nonumber\\
|\psi_{_{3}}\rangle&=& \psi_{_{30}}e^{\frac{ie}{\hbar{c}}\int_{C_{3}}{\vec{A}{(\vec{x})}}d\vec{x}}.
\end{eqnarray}
The initial state is the superposition of the three wave functions, $|\Psi_{c}\rangle=|\psi_{_{1}}\rangle+|\psi_{_{2}}\rangle+|\psi_{_{3}}\rangle$ with respect to the three independent paths. The probability density distribution on the detector screen is determined by the inner product of the incoming state $|\Psi_{c}\rangle$ and the final outgoing state $\langle\Psi_{c}|$,
\begin{eqnarray}\label{psi1psi2psi3}
P_{c}&=&\langle\Psi_{c}|\Psi_{c}\rangle=|\psi_{_{10}}|^2+|\psi_{_{20}}|^2+|\psi_{_{30}}|^2 \nonumber\\
&+&2Re[\psi^{\ast}_{_{10}}\psi_{_{30}}\exp[{\frac{ie}{\hbar{c}}(\Phi_{_{2}})}]]\nonumber\\
&+&2Re[\psi^{\ast}_{_{20}}\psi_{_{30}}\exp[{\frac{ie}{\hbar{c}}(-{\Phi_{_{1}}})}]]\nonumber\\
&+&2Re[\psi^{\ast}_{_{10}}\psi_{_{20}}\exp[{\frac{ie}{\hbar{c}}({\Phi_{_{1}}}+{\Phi_{_{2}}})}]].
\end{eqnarray}
The interference pattern is governed by two independent phase differences, which varies with two independent magnetic field strength, even though it is still a technological challenge to tune the strength of two nearest neighboring magnetic fluxes.

The probability distribution Eq. (\ref{psi1psi2psi3}) is solid within a wide range of incoming angle between electron beam and the plane expanded by the two parallel flux tubes. However for a very small incoming angle as showed in Fig. \ref{ABphase} (b), the three independent paths may overlap one another when they pass the the same route with a spatial scale of the distance between two magnetic fluxes. Then the weight of wave functions along the four edges of a rectangular loop path around one flux becomes anisotropic as Fig. \ref{ABphase} (b) showed. The upper edge around the first flux is composed of $C_{1}$ and $C_{3}$, with its left edge and bottom edge $C_{2}$ and the right edge $C_{3}$. The weight distribution around the second flux is similar to that of the first flux but rotated by $\pi$ around their middle point. We define the topological path fusion as the extreme case that the two path segments above the first flux are confined in a small space and get too close to distinguish from each other until they inevitable fuse into one. The number of fused paths are labeled on the edges around the flux pair in Fig. \ref{ABphase} (c). This path fusion process is not detectable by phase difference in Aharonov-Bohm effect, because the detector screen only received the resultant interference pattern of all paths instead of the branch process before the resultant interference.

In the conventional combination of three independent paths monotonically passing through the flux pair,
each path can be replaced by many other possible configurations with knot. Any unknotted curve in the outer region far away from the flux pair is topologically equivalent either to path $C_{1}$ or $C_{2}$. If we confine paths exactly in two dimensional space, the knotted paths are forbidden to exist unless it intersects with itself. In the outer region, any continuous path that obeys the self-avoid rule is topologically equivalent to the monotonic path $C_{1}$ or  $C_{2}$. However, the path $C_{3}$ going through the middle gap between the two fluxes still has many possible configurations with non-trivial topology, generated by braiding operation on the two fluxes and keeping the path continuous. For example, the path $C_{4}$ in Fig. \ref{ABphase} (d) as a topological transformation of $C_{3}$ first wind around the second flux and turn back to the first flux, implemented by exchanging the location of two fluxes, which does not affect the topology of the monotonic path $C_{1}$ and $C_{2}$ (Fig. \ref{ABphase} (g)). The interference pattern of the three paths ( $C_{1}$, $C_{2}$ and $C_{4}$) turns out to be the same as that of ( $C_{1}$, $C_{2}$ and $C_{3}$), because the newly added segments of the vectorial integration of electromagnetic field along $C_{4}$ cancelled each other. Thus Aharonov-Bohm phase does not contain the information that distinguish $C_{3}$ from $C_{4}$ in Fig. \ref{ABphase}. However this lost information revealed an internal gauge symmetry of the superposition of wave function, $|\Psi_{c}\rangle=|\psi_{_{1}}\rangle+|\psi_{_{2}}\rangle+|\psi_{_{3}}\rangle$. A continuous rotation around the middle axis that lies in between two flux tubes (labeled as 'O' in Fig. \ref{ABphase} (a)) without breaking the middle path $C_{3}$ keeps the wave function $|\Psi_{c}\rangle$ invariant,
\begin{eqnarray}\label{Uptheta=u}
\Psi_{c}[\vec{\Phi}] = \Psi_{c}[U({\theta})\vec{\Phi}],
\end{eqnarray}
where $\vec{\Phi}=(\Phi_{1},\Phi_{2})^{T}$, $U({\theta})$ is a group element of the proper rotation group SO(2). Because SO(2) group is isomorphic to one dimensional unitary transformation group $U(1)$. The rotation of two fluxes is equivalent to introducing a phase factor into the resultant wave function $\Psi_{c}$ under the action of U(1) group $U({\theta})=\exp[{i\theta}]$. The generator of this U(1) group is the $z$-component of angular momentum operator,
\begin{eqnarray}\label{Lz}
\hat{L}_{z} = -i\frac{\partial}{\partial\theta}, \;\;\;\hat{L}_{z} \Psi_{m}=m\Psi_{m}, \;\;\;\Psi_{m} = e^{im\theta}.
\end{eqnarray}
The resultant wave function after transformation can be simplified as, $\Psi'_{c}=U({\theta})\Psi_{c}=\Psi_{m}\Psi_{c}$. The eigenvalue of angular momentum $\hat{L}_{z}$ is in fact an integer, ($m=\pm1,\pm2,\pm3,\cdots$), that counts many periods the two fluxes are exchanged either in clockwise or in counter-clockwise direction, which is also the winding number in topology theory. This number is exactly the eigenvalue of braiding operator that results in fractional charges in knot lattice model\cite{{SiT2019}} as well as following sections.

The U(1) gauge symmetry of the source wave is broken when the incoming wave bombards the plane confining two flux tubes, but is restored in the resultant wave function on detector screen. The path fusion breaks the U(1) symmetry during the collision process between electron wave and two flux tubes. The resulted path configuration can distinguish different braiding operations over the two fluxes, and generates a hierarchy of fractional charges. For example, the path fusion of $C_{3}$ is still $C_{3}$ itself. However the path fusion of $C_{4}$ alone (as showed in Fig. \ref{ABphase} (e) (f)) leads to the same track distribution as the fusion of the three paths, $C_{1}$, $C_{2}$ and $C_{3}$ (Fig. \ref{ABphase} (c)), generating the fractional charge of $1/3$. A further path fusion of $C_{1}$, $C_{2}$ and $C_{4}$ generates fractional charges $2/5$ (Fig. \ref{ABphase} (I)), which also is the same track distribution generated by the fusion of $C_{5}$ alone (Fig. \ref{ABphase} (h)). This is because one clockwise braiding on flux pair upon the curve $C_{3}$ leads to $C_{4}$, and two clockwise braiding leads to $C_{5}$, and so on. Therefore the unfused path through the middle gap between the two fluxes is characterized by the number of braiding operations, which equals to the topological winding number of the vortex path. The exemplar fusion strategy above generates suggests two different ways of constructing a hierarchy of fractional charges, the first way is fusing paths under continuous braiding operations, the second way is continuously adding $C_{1}$, $C_{2}$ upon the fused paths of ($C_{1}$, $C_{2}$ and $C_{3}$) and then fuse all paths. These two approaches outcome the same track distribution and fractional charges. From the point view of quantum mechanic, every path carries one unit of probability weight, a topological braiding operation does not change the ultimate probability distribution on the detector screen. However the probability distribution on the propagating path may oscillates between different fractions before they reach the detector screen. This probability redistribution not only splits one elementary charge into fractional charges, but also splits the mass carried by the propagating beam into fractional mass.

\begin{figure}
\begin{center}
\includegraphics[width=0.4\textwidth]{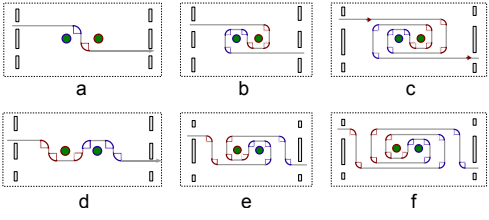}
\caption{\label{ChernSimon} (a) The initial path that penetrate through the middle gap region between two fluxes. (b) The path after one clockwise braiding on the two fluxes and (c) after two clockwise braiding. (d) The path under a counterclockwise braiding on the initial path. (e) The path under two counterclockwise braiding on the initial path and (f) under three counterclockwise braiding.}
\end{center}
\vspace{-0.5cm}
\end{figure}

$\textbf{Quantum field theory of topological path fusion}$

The winding propagation path around the flux pair can be effectively described by Abelian Chern-Simons field theory. In a classical physics theory, an electron propagating in an external electromagnetic field under the propulsion of Lorentz force, $\vec{f} = q (\vec{E}+\vec{v}\times{\vec{B}})$, with the magnetic field oriented in z - direction parallel to magnetic flux tube in Fig. \ref{ChernSimon}. The magnetic field bends the current from x-axis to y-axis or vice versa, inducing a Hall current, $J^{\mu}=\sigma_{xy}\epsilon^{\mu\lambda}E_{\lambda}$, which is characterized by the turning arcs in Fig. \ref{ChernSimon}. Here the Hall conductance coefficient $\sigma_{xy} = {\nu{e^2}}/{2\pi}$ is proportional to the filling factor $\nu$. Since both the electric field $\vec{E}$ and magnetic field $\vec{B}$ can be expressed by the external electromagnetic tensor $F_{\mu\nu} =\partial_{\mu}A_{\nu}- \partial_{\nu}A_{\mu}$, where $A$ is electromagnetic potential. The linear response of electric current to the external electromagnetic field potential in quantum field theory reads,
\begin{eqnarray}\label{Ju=dA}
{\delta}J^{\mu}=\sigma_{xy}\epsilon^{\mu\nu\lambda}\partial_{\nu}\delta{A}_{\lambda}.
\end{eqnarray}
Besides the external electromagnetic potential $A$, a gauge potential $a_{\mu}$ induced by the U(1) symmetry of electron wavefunction under the braiding operation of two fluxes, also introduced a gauge field current,
\begin{eqnarray}\label{Ju=daa}
J_{a}^{\mu} = \frac{1}{2\pi}\epsilon^{\mu\nu\lambda}\partial_{\nu}{a}_{\lambda} = \frac{1}{2\pi}b_{\mu}.
\end{eqnarray}
These two currents above both contribute to the Ginsburg-Landau Lagrangian for Laughlin state, which is composed of two parts \cite{Blok1990}, $\mathcal{L} = \mathcal{L}_{0}+ \mathcal{L}_{cs}$,
\begin{eqnarray}\label{lagrangian=L0+Lcs}
\mathcal{L}_{0} = \psi^{\dag}i(\partial_{0}+ia_{0}-ieA)\psi +\frac{\psi^{\dag}(\partial_{\mu}+ia_{\mu}-ieA_{\mu})^2\psi}{2m_{e}} ,\nonumber
\end{eqnarray}
The fractional filling factor $\nu$ is governed by the Chern-Simons terms in the second part of Lagrangian, $\mathcal{L}_{cs}$. The path of a moving electron in Fig. \ref{ChernSimon} is composed of two coupling terms,
\begin{eqnarray}\label{lagrangianCS}
\mathcal{L}_{cs} &=& -\frac{m}{4\pi}\epsilon^{\mu\nu\lambda}{a}_{\mu}\partial_{\nu}{a}_{\lambda}+ \frac{e}{2\pi}\epsilon^{\mu\nu\lambda}{A}_{\mu}\partial_{\nu}{a}_{\lambda}, \nonumber\\
&=& -\frac{m}{4\pi}{a}_{\mu}{b}_{\mu}+ \frac{e}{2\pi}{A}_{\mu}{b}_{\mu},
\end{eqnarray}
where $m$ defines the filling factor $\nu=1/m$. The first term on the right hand side of Eq. (\ref{lagrangianCS}) is the coupling between gauge potential and gauge field tensor, the second term couples the external electromagnetic potential to gauge field. This Chern-Simons Lagrangian is the sum of helicity action---a topological invariant of knot. The path of Fig. \ref{ChernSimon} (a) corresponds to the current of an integral charge with its dynamics governed by the Lagrangian equation $\mathcal{L}_{m=1}$. The Lagrangian with $m=3$ governs the dynamic motion of fractional charge $1/3$ with respect to the trajectory showed in Fig. \ref{ChernSimon} (b). The integer $m$ is directly read out by counting number of the turning arcs around one of the two fluxes in Fig. \ref{ChernSimon}. For instance, there is one turning arc around $\Phi_{1}$ in Fig. \ref{ChernSimon} (a),  three arcs in Fig. \ref{ChernSimon} (b) and five arcs in Fig. \ref{ChernSimon} (c).
It is also computable by choosing a hybrid symmetric gauge, i.e.,  the symmetric gauge potential vector field around the left flux is oriented into the opposite direction as that around the right flux,
\begin{eqnarray}\label{symmetrica1a2}
\vec{a}_{l} &=& -\frac{y}{2}b \mathbf{e}_{x} + \frac{x}{2}b \mathbf{e}_{y},\;\;x<0,\;\;\;\nonumber\\
\vec{a}_{r} &=& \frac{y}{2}b \mathbf{e}_{x} - \frac{x}{2}b \mathbf{e}_{y}, \;\;x>0,\nonumber\\
\vec{a}_{l} &=& \vec{a}_{r} =\frac{|y|}{2}b \mathbf{e}_{y}, \;\;x=0.
\end{eqnarray}
A monodirectional gauge vector is introduced on the interface border between two domains at $x = 0$ to ensure the continuity of gauge vector field. These gauge vectors together form convective vector flows that eject out of the north pole ($y>0$) and sinks into the south pole ($y<0$). The complete electromagnetic potential vector is composed of two domains, $\vec{a} = \vec{a}_{l} + \vec{a}_{r}$. $ \vec{a}_{l} $ generates magnetic field $b_{l}$ in the left half-plane, $x<0$. $ \vec{a}_{r} $ generates $b_{r}$ in the right half-plane, $x>0$,
\begin{eqnarray}\label{a1a2}
\vec{b}_{l} H_{s}(-x) = \nabla\times\vec{a}_{l} ,\;\;\;-\vec{b}_{r} H_{s}(x) = \nabla\times\vec{a}_{r},
\end{eqnarray}
where $H_{s}(x)$ is the Heaviside function. $[H_{s}(x) =1, x > 0; \;\;H_{s}(x) = 0, x < 0].$ The integer $m$ in Lagrangian Equation is counted by the winding number of turning arcs in the braided paths,
\begin{eqnarray}\label{m= xa1a2}
m = \frac{1}{\pi}\nabla\times\vec{a}.
\end{eqnarray}
The paths under counterclockwise braiding in Fig. \ref{ChernSimon} (d-f) yield a negative $m$ but equal absolute value as that of clockwise braiding after the same periods of braiding operations. This is because the turning arcs that bends in opposite direction cancelled each other during path fusion process.

The whole serial of filling fractions is spontaneously generated by sequent braiding operation on flux pair and path fusion process thereafter. Take the $1/3$ charge state as an example, its corresponding winding track is Fig. \ref{ChernSimon} (b), one more clockwise breading generated two more turning arcs in the same direction as before (Fig. \ref{ChernSimon} (c)). The two new arcs carry new gauge potential field $\bar{a}_{\mu}$ which generate new gauge field $\bar{b}_{\mu} = \epsilon^{\mu\nu\lambda}\partial_{\nu}\bar{a}_{\lambda}$. In the meantime, the new gauge potential $\bar{a}_{\mu}$ also couples to the old gauge field ${b}_{\mu}$. The complete Lagrangian for the fused new path is
\begin{eqnarray}\label{lagrangianCSmm}
\mathcal{L}_{m,{m}_{1}} &=& \frac{e}{2\pi}\epsilon^{\mu\nu\lambda}{A}_{\mu}\partial_{\nu}{a}_{\lambda}-\frac{m}{4\pi}\epsilon^{\mu\nu\lambda}{a}_{\mu}\partial_{\nu}{a}_{\lambda} \nonumber\\
&-& \frac{{m}_{1}}{4\pi}\epsilon^{\mu\nu\lambda}\bar{a}_{\mu}\partial_{\nu}\bar{a}_{\lambda}+ \frac{1}{2\pi}\epsilon^{\mu\nu\lambda}{a}_{\mu}\partial_{\nu}\bar{a}_{\lambda}\nonumber\\
&=& \frac{e}{2\pi}{A}_{\mu}{b}_{\mu}-\frac{m}{4\pi}{a}_{\mu}{b}_{\mu}
 -\frac{{m}_{1}}{4\pi}\bar{a}_{\mu}\bar{b}_{\mu}+ \frac{1}{2\pi}{a}_{\mu}\bar{b}_{\mu}.
\end{eqnarray}
This Lagrangian governs the fusion of the two new track segments into that of  $1/3$ quantum Hall state, resetting the weight distribution of current layers from that of $1/3$ to $2/5$, as showed in Fig. \ref{ABphase} (f)(I). The corresponding filling fraction derived from this Lagrangian is
\begin{eqnarray}\label{m= mb}
\nu = \frac{1}{m-\frac{1}{{m}_{1}}}.
\end{eqnarray}
For the special case of Fig. \ref{ABphase} (f)(I),
$m=3$, $\bar{m}=2$, it yileds $\nu = 2/5$. A serial of filling fractions is constructed by repeating the Lagrangian construction above and match it with the corresponding winding path. This topological path fusion method agrees with hierarchy construction of fractional quantum Hall effect based on Abelian Chern-Simons field theory \cite{Blok1990},
\begin{eqnarray}\label{lagrangianCSkIJ}
\mathcal{L}_{K} = -\frac{1}{4\pi}K_{IJ}\epsilon^{\mu\nu\lambda}{a}_{I\mu}\partial_{\nu}{a}_{J\lambda} + \frac{e}{2\pi}q_{I}\epsilon^{\mu\nu\lambda}{A}_{\mu}\partial_{\nu}{a}_{I\lambda},
\end{eqnarray}
 Where $K_{IJ}$ is a matrix witn its diagonal terms asigned with the integer of filling factors, i.e., $K_{11} = m = 3$, $K_{ii} = {m_{i-1}} = 2={m_{i-1}} = 2,K_{i,i-1} =K_{i-1,i}=-1;  i = 2,3,\cdots.$ This lagrangian yields a general filling fraction,
\begin{eqnarray}\label{m=1many}
\nu = \frac{1}{m-\frac{1}{{m}_{1}-\frac{1}{m_{2}-\cdots}}}.
\end{eqnarray}
A similar but different fractional hierarchy from above also exist in the splitting sequence in Thurston train track theory \cite{Thurston1988}\cite{Mosher2003}. Here we showed the first example of fractional hierarchy of train track that matches physical reality, in which every fraction carries an odd denominator.

$\textbf{Mapping the knot on torus into train track}$

\begin{figure}
\begin{center}
\includegraphics[width=0.45\textwidth]{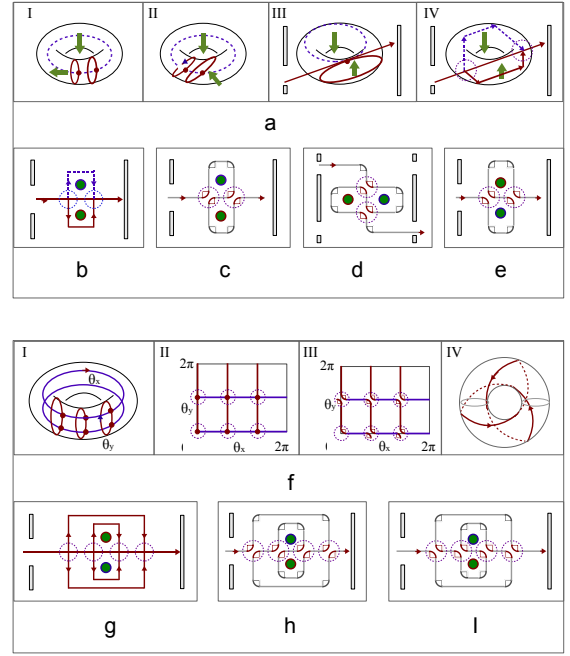}
\caption{\label{rotor} (a) I-IV, The scheme to show the procedure of mapping three electron path loops on torus into path loops around two magnetic fluxes. One loop is cut to create an open path. (b) The final pattern of the path on torus (a) IV is projected into plane and reformed into two square loops surrounding two fluxes respectively with an open path in between. (c) The crossings split into a pair of self-avoiding vacuum arcs, $|O_{\pi/4}\rangle$=$({^{\lrcorner}}{_\ulcorner})$. (d) Rotation of two fluxes by an angle $\pi/2$ in clockwise direction. (e)  The splitting of crossing current is realized by the left vacuum state,  $|O_{3\pi/4}\rangle$=$({_\urcorner}{^{\llcorner}})$. (f) I-IV, The flow chart of mapping the initial five loops on torus into trefoil knot on torus by track splitting using vacuum state.  (g) The double layered square loops, as reformed representation of two x-loop and three y-loop on torus, have four crossings. (h) The splitting action of right vacuum state $|O_{\pi/4}\rangle$=$({^{\lrcorner}}{_\ulcorner})$ map out a winding train track around two fluxes for the $2/5$ state. (I) The left splitting by left vacuum state $|O_{3\pi/4}\rangle$=$({_\urcorner}{^{\llcorner}})$ generates the fractional charge state of $3/5$.}
\end{center}
\vspace{-0.5cm}
\end{figure}

In this section, it will be shown that, Abelian Chern-Simons field theory is a topological invariant not only for knot, but also for winding train tracks around flux pair. In fact, the knot on torus can be mapped into braided electron path around two magnetic fluxes by the topological vacuum states in knot lattice model \cite{{SiT2019}}, which matches the splitting algebra in train track theory \cite{Mosher2003}. Take the $1/3$ filling state as an example, it originates from three path loops of electron on torus in Fig. \ref{rotor} (a), i.e., two vertical loops (red loop) wrap the horizontal hole and one horizontal loop (yellow loop) wraps the vertical hole. The two independent holes are equivalent to two magnetic fluxes (as showed by the bold green arrow in Fig. \ref{rotor} (a)-I). The two vertical loop are first continuously tilted into the same plane of the horizontal loop by keeping the topology of the curves invariant (Fig. \ref{rotor} (a)-II), then one of the two vertical loop is cut to create two open ending points that are connected to the input source and the output detector (Fig. \ref{rotor} (a)-III). The vector of magnetic flux tube is kept perpendicular to the plane of tilted vertical loop. then the three loops are transformed into an open channel that is sandwiched in between the two closed loops (Fig. \ref{rotor} (a)-IV).
The irregular loop scheme in Fig. \ref{rotor} (a)-IV is reshaped into rectangular loops around two fluxes oriented into a vertical ordering (Fig. \ref{rotor} (b)). Even though the train track theory\cite{Thurston1988}\cite{Mosher2003} only describes closed curves, here we fixed the two ending points to infinity, which is equivalent to a closed curve. The loop path in Fig. \ref{rotor} (b) can be mapped into train track curve by replacing every crossing point (enclosed by the dashed circle in Fig. \ref{rotor} (b)) with a right or a left vacuum state, $|O_{\pi/4}\rangle$=$({^{\lrcorner}}{_\ulcorner})$ and $|O_{3\pi/4}\rangle$=$({_\urcorner}{^{\llcorner}})$, which matches exactly the vacuum states in the knot lattice model \cite{{SiT2019}}. The train track curve generated by two right vacuum states in Fig. \ref{rotor} (c) is essentially equivalent to a train track curve generated by exchanging the position of two fluxes in clockwise direction. A further rotation of the flux pair by an angle of $\pi/2$ in clockwise direction produces a train track curve (Fig. \ref{rotor} (d)) that is exactly the same as the winding path of electron beam in Fig. \ref{ABphase} (d) and Fig. \ref{ChernSimon} (b). The fused train track of Fig. \ref{rotor} represents the fractional charge state $1/3$. Its dual fractional charge state $2/3$ is represented by the train track generated out of two left vacuum states, $|O_{3\pi/4}\rangle$=$({_\urcorner}{^{\llcorner}})$, in (Fig. \ref{rotor} (e)), with respect to a counterclockwise rotation upon the flux pair.

The topological transformation theory above offers a new construction method for a serial of fractional charges. For example, by adding one more vertical loop and one more horizontal loop upon the initial loop pattern on torus for $1/3$ state, it leads to the initial loop pattern for $2/5$ state (Fig. \ref{rotor} (f)-I). Every loop is represented by a straight line in the angle-coordinate system $\theta_{x}-\theta_{y}$ with periodical boundary condition (the three red vertical lines Fig. \ref{rotor} (f)-II correspond to the three vertical loop in Fig. \ref{rotor} (f)-I, the two horizontal yellow lines indicate the two horizontal loops in Fig. \ref{rotor} (f)-I). Replacing the six crossing point (enclosed by the dashed circle) by the left vacuum state $|O_{3\pi/4}\rangle$=$({_\urcorner}{^{\llcorner}})$ fuses the five initial loops into one knot (Fig. \ref{rotor} (f)-III)), which is exactly a trefoil knot on torus (Fig. \ref{rotor} (f)-IV)). Thus vacuum state induced the fusion of path loops, driving the free loops into a connected knot state. On the other side, performing the same topological transformation procedure of Fig. \ref{rotor} (a) on the five loops on torus of Fig. \ref{rotor} (f)-I) maps equivalently the the initial loop pattern into a double layer loop track around the two fluxes, Fig. \ref{rotor} (f)-g). The $2/5$ charge state are generated by splitting the six crossings with four right vacuum states $|O_{\pi/4}\rangle$=$({^{\lrcorner}}{_\ulcorner})$ (Fig. \ref{rotor} (h)). While the $3/5$ charge state are result of track splitting by four right vacuum states $|O_{3\pi/4}\rangle$=$({_\urcorner}{^{\llcorner}})$ as showed in Fig. \ref{rotor} (I). These train tracks coincide exactly with the winded paths in Fig. \ref{ABphase} and
Fig. \ref{ChernSimon} after one more $\pi/2$ rotation upon the flux pair.

\begin{figure}
\begin{center}
\includegraphics[width=0.45\textwidth]{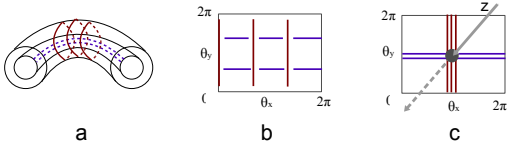}
\caption{\label{torus} (a) The x- and y-loop are placed on two separated torus surface respectively. (b) The knot lattice of the crossing loop segments in $\theta_{x}-\theta_{y}$ coordinate system. (c) The fused track loops with an input and an output end.}
\end{center}
\vspace{-0.5cm}
\end{figure}

The topological transformation from knot on torus to train track around a flux pair also sets up a route for a quantum mechanical theory of fractional charges. Based on the path loops in $\theta_{x}-\theta_{y}$ coordinate of Fig. \ref{torus} (a), we represent each horizontal path by a wave function $|\psi_{x}\rangle = \psi_{x}|{x}\rangle$ and each vertical path by $|\psi_{y}\rangle = \psi_{y}|{y}\rangle$. The wave function of an electron in this path grid is
 \begin{eqnarray}\label{psixy}
|\psi\rangle & =&  \psi_{x}|{x}\rangle + \psi_{y}|{y}\rangle, \;\; \nonumber\\
 \psi_{\alpha}(t)& = & e^{-i\theta_{\alpha}(t)}\sqrt{n_{\alpha}(t)},\;\;\; \alpha= x, y.
\end{eqnarray}
with a normalized probability density, $|\psi_{x}|^2+|\psi_{y}|^2 = 1$. This wave function evolves under the action of a hopping Hamiltonian, $\hat{H}=H_{\alpha\beta}|\alpha\rangle\langle\beta|$,
\begin{eqnarray}
H=\begin{pmatrix}
q_{x}V_{x} & t_{xy}  \\
t_{yx} & q_{y}V_{y}  \\
\end{pmatrix}.
\end{eqnarray}
where $V_{\alpha}$ are the respective external potential due to the oriented path loops, with the respective electric charges $q_{\alpha}$ running in the path, $t_{yx}=-t_{xy}$ indicating the anti-symmetric character of fermion wave function. The Schr$\ddot{o}$dinger equation of motion is
\begin{eqnarray}
\frac{d}{dt}\begin{pmatrix}
\psi_{x} \\
\psi_{y}  \\
\end{pmatrix}=-i\hbar\begin{pmatrix}
-eV_{x} & t_{xy}  \\
t_{yx} & -eV_{y}  \\
\end{pmatrix}
\begin{pmatrix}
\psi_{x} \\
\psi_{y}  \\
\end{pmatrix}.
\end{eqnarray}
Substituting the complex wave function Eq.\ref{psixy} into the Schr$\ddot{o}$dinger equation above yields the dynamic equation of two phases,
 \begin{eqnarray}\label{dtheta}
\frac{d\theta_{\alpha}}{dt} = -\frac{e}{\hbar}V_{\alpha}(t), \;\;\;\alpha=x,y.
\end{eqnarray}
Because the total number of particles is conserved, $n_{x}+n_{y}=1$, the particles lost in $y$ loop join in x-loop,  the tunneling current is
 \begin{eqnarray}\label{dny}
\frac{dn_{y}}{dt} &=& \frac{it_{xy}}{\hbar} (\psi_{x}^{\ast}\psi_{y}-\psi_{y}^{\ast}\psi_{x})\nonumber\\
&=&\frac{2t_{xy}}{\hbar} \sqrt{(1-n_{y})n_{y}}\sin(\theta_{x}-\theta_{y}).
\end{eqnarray}
The dynamic Eq. (\ref{dtheta}) of the two phases depicts periodical or quasi-periodical trajectories on torus, which relies on the ratio of the two voltage components. For a constant voltage generated by fractional charges,
 \begin{eqnarray}\label{vxvy}
V_{x}= \hbar(\frac{-e}{q_{x}})^{-1},   \;\;\; V_{y} = \hbar(\frac{-e}{q_{y}})^{-1},
\end{eqnarray}
the solution of the dynamic equation of the two phases,
 \begin{eqnarray}\label{dqalpha}
\frac{d\theta_{\alpha}}{dt} = q_{\alpha},\;\;\;, \alpha=x,y;
\end{eqnarray}
is a knot on torus. For example, $(q_{x} = 2, q_{y} = 3)$ leads to a trefoil knot on torus as showed in Fig. \ref{rotor} (f). If the ratio of  $q_{x}$ to $q_{y}$ is a rational number, the trajectory of the electron is always a knotted curve on torus. The tunneling current Eq. (\ref{dny}) admits a solution,
 \begin{eqnarray}\label{ny}
n_{y}= \cos^{2}[-\frac{t_{xy}}{\hbar}\sin(q_{x}-q_{y})t],\;\; n_{x}=1-n_{y}.
\end{eqnarray}
The tunneling current Eq. (\ref{dny}) describes the tunneling current along the borderline between two magnetic fluxes, represented by single track in the winding path as along as $q_{x}-q_{y}=1$. After the topological transformation and track fusion operations, the topological route grid (Fig.\ref{torus}. (c)) splits an electron into fractional charges,
 \begin{eqnarray}\label{QxQy}
Q_{x}= \frac{q_{x}}{q_{x}+q_{y}}, \;\;\;\; Q_{y}= \frac{q_{y}}{q_{x}+q_{y}}.
\end{eqnarray}
These knotted paths result in the quantized Hall resistance. For an irrational number of ratio $q_{x}$ to $q_{y}$, the trajectory on torus is not a closed curve, instead it draws an endless open curve that never intersect with itself. These open path lead to the classical behavior of Hall resistance,
 \begin{eqnarray}\label{qedge}
R_{_{H}}= {V_{y}}/{vQ_{x}}.
\end{eqnarray}
The tunneling current Eq. (\ref{dny}) describes the edge current from X-channel into Y-channel, which are located in separated torus surface (Fig.\ref{torus} (a)). The tunneling edge current along the borderline carries a fractional charge, $Q_{t}= \frac{q_{x}-q_{y}}{q_{x}+q_{y}}$. This tunneling edge current only exists for the case that the two effective fluxes in Fig.\ref{rotor} (a) are oriented in opposite directions. In this fermionic case, the current on the interface between two fluxes is the sum of two current segments flowing in the same direction, one comes from left loop and the other one is from the right loop. In the bosonic case, the two fluxes are oriented in the same direction, the interface current from the the left loop runs exactly in the opposite direction of that from the right loop. As a result, the two track segments along the interface cancelled each other. Combining the action of left and right vacuum states generates one arbitrary knot lattice, revealing a deep connection between train track and knot lattice model of anyons \cite{{SiT2019}}.

$\textbf{Projecting a knot lattice into a train track}$

\begin{figure}
\begin{center}
\includegraphics[width=0.36\textwidth]{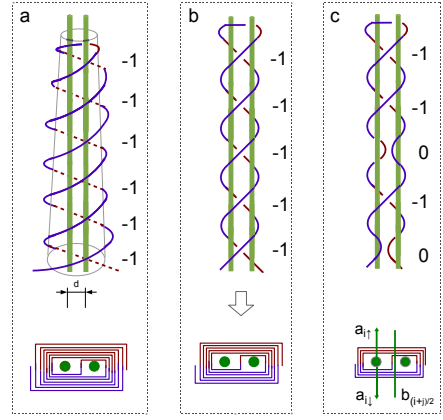}
\caption{\label{1+1Dknot1} (a) The double helix electron path around an asymmetric cylinder surface projects a winding train track around two fluxes. (b) The asymmetric cylinder surface is transformed into a normal cylinder to map the helical path into a normal lattice of many knots. (c) An exemplar state of one dimensional knot lattice around two fluxes with rotated vacuum states, $|O_{\pi/2}\rangle =\;\;\rangle \langle $.}
\end{center}
\vspace{-0.5cm}
\end{figure}

A more straightforward way of visualizing the relation between knot and train track is topological projection. For the simplest case of Fig. \ref{1+1Dknot1}, the projection of a spiral track around two magnetic flux tubes into two dimensional plane maps exactly the corresponding train track, winding around the outer region of two fluxes with a borderline current sandwiched in between two fluxes (Fig. \ref{1+1Dknot1}). The double helix track in Fig. \ref{1+1Dknot1} is essentially a one dimensional knot lattice\cite{{SiT2019}}. The two ending points of the double helix are fixed to a source point and a drain point respectively, which merge into one point at infinity to fulfil the conservation law of mass and close curve in train track theory. A decreasing magnetic field is applied from top to down to continuously expand the circle orbital (Fig. \ref{1+1Dknot1}), demonstrating a double helical track on an asymmetric cylinder (Fig. \ref{1+1Dknot1} (a)). The projected train track to the bottom plane depicts a vortex path around a double core of two fluxes. This projection smeared out the spatial distribution of magnetic field strength along the longitudinal axis of the flux tube. Without changing the topology of the helical curve, the asymmetric magnetic field can be replaced by an uniform magnetic field distribution as Fig. \ref{1+1Dknot1} (b) shows. Each vortex pattern of train track in the bottom plane can also be characterized by the same topological linking number of the knot lattice, which is defined by the total number of positive crossings $N_{+}$ that minus the total number of negative crossings $N_{-}$, $L_{link} = ({N_{+}-N_{-}})$. Because each braiding generates one crossing in the knot lattice, this linking number equals to the total number of braiding operations, it is also exactly the winding number of vortex flow of the train track. The vacuum state $|O_{\pi/2}\rangle =\;\;\rangle\langle $ eliminates the crossings and reduces the number of track layers around the the two flux cores as well (Fig. \ref{1+1Dknot1} (c)).

Projecting a knot lattice of double helix current into a vortex of train track around two fluxes is not only a mathematical projection, but also provide a physical mechanism of track fusion. When there exist two flux tubes oriented in opposite direction, the chirality of the circling orbital of electron around one flux is exactly in the opposite direction of that around the other one. The two circling electric currents in the border region between two fluxes naturally fused into one, since the current segments along the border flows in the same direction. If the initial velocity of electron is tilted out of the bottom plane, the electron moves along a helical trajectory like that in Fig. \ref{1+1Dknot1} (b). When the spiral orbital (Fig. \ref{1+1Dknot1}) is pressed into a two dimensional plane strong confinement, the strong magnetic field binds many layers of circular tracks into one circular bundle. If the thickness of the circular bundle is smaller than the matter wavelength of electron, two or more layers of track are covered simultaneously by one electron wave. This lead to the fusion of many stacked current tracks. This fusion also reduced the electromagnetic energy of the circular bundle, because current segments in opposite orientation repel each other, while those in the same direction attract each other. In winding tracks at the bottom plane of Fig. \ref{1+1Dknot1}, the electric current segment on one side of the flux shows an alternating orientation with respect to the odd or even stages of helical stair winding around $z$-axis. The electric current in the nearest neighboring tracks are always in opposite direction before the track fusion, but switched to the same direction after track fusion. Every continuous current must first spirals into the edge and then turns back to form a double helix current. This topological constraint reveals the special role of edges and the double core of vortex paths.

The track fusion mechanism generates fractional charges and fractional Hall conductance. The number of track layers is still preserved after the track fusion, indicating an incompressible fluid of electron gas in quantum Hall effect. The number of layers of track segments at the two sides of one flux is not homogeneously distributed. As showed in Fig. \ref{1+1Dknot1} (c), there are three layers above flux No. 1 with four layers below, three layers to the left and one layer to the right. An input elementary charge from the left bottom corner sent $3e/7$ upward into the left channel and $4e/7$ to the bottom channel. For a general case, we draw a vertical line passing through the center of the $i$th flux in the flux pair of Fig. \ref{1+1Dknot1} (c) to count the number of tracks that intersect the track segments above, $a_{{i}\uparrow}$, and that below $a_{{i}\downarrow}$. Another vertical line is place in between the two fluxes to count the total number of horizontal tracks, $b_{{(i+j)/2}}$. These weight factors obey the equation,
 \begin{eqnarray}\label{b=a+a}
b_{{{(i + j)}/{2}}} = {a_{{i}\uparrow} + a_{{i}\downarrow}},
\end{eqnarray}
which is in fact the conservation equation of charge. The fractional charge above and below the flux are quantified by equation,
 \begin{eqnarray}\label{a/b=a+a}
Q_{_{{i}\uparrow}} = \frac{a_{{i}\uparrow}}{b_{{{(i + j)}/{2}}}},\;\;\;\;\;Q_{_{{i}\downarrow}} = \frac{a_{{i}\downarrow}}{b_{{{(i + j)}/{2}}}}.
\end{eqnarray}
By tracing a helical path back upward from the entrance at the bottom, Fig. \ref{1+1Dknot1} clearly shows that a current in the front side of the flux (represented by the solid blue lines) switches its direction on the edge and flows into the perpendicular tracks in the back (represented by the dashed red lines). This defines Hall resistance in quantum Hall effect,
\begin{eqnarray}\label{Rxy=VI}
R_{yx} = V_{y}/{I_{x}} =V_{y}/v_{e} Q_{_{{i}\uparrow}},
\end{eqnarray}
where $V$ is the applied voltage, $v_{e}$ is the velocity of the charge. The Hall current only exist when the edge on top or the current that is in between the double core exists (as shown in Fig. \ref{1+1Dknot1}). This topological conclusion is coincide with the observation in FQHE experiment that the Hall resistance is not detectable unless the edge is connected \cite{Stormer1999}.

It costs a very strong magnetic field to observe the fractional quantum Hall effect \cite{Stormer1999}. This topological path fusion model also provides a geometric quantification of magnetic field strength based on classical orbital of electron in magnetic field, because the track fusion here only occurs in strong magnetic field. The effective magnetic field strength $B$ is proportional to the number of magnetic fluxes in unit area, i.e.,  $B = N(\Phi)/S$, $S$ is the area of cross section surface. The distance between two nearest neighboring fluxes is denoted as $d$, then the unit area in one dimensional lattice of fluxes is $S = d$, and $ S = d^{2} $ in two dimensional square lattice. For the two magnetic fluxes in Fig. \ref{1+1Dknot1}, the corresponding magnetic field strength reads,
\begin{eqnarray}\label{B=phid}
B = \Phi_{0} /d,
\end{eqnarray}
where $ \Phi_{0} = h/2e = 2.07 \times 10^{-15}$ Weber is the flux quanta, $h$ is Planck's constant and e is the electron charge. A large separation distance $d$ indicates a weak magnetic field strength. A moving electron at an initial velocity $v$ draws an isolated circular track around one flux. The radius of circle is proportional to inverse of magnetic field strength, $r_{e} = (m_{e}v)/(eB)$, here $m_{e}$ is the mass of electron. Strong(weak) magnetic field confines an electron to a small(large) circle. For a given magnetic field strength, every flux is surrounded by a fixed number of concentric circles. The flux tube in weak magnetic field is surround by more concentric orbital circles than that in strong magnetic field. When two identical fluxes meet each other, if the distance between two fluxes is larger than the maximal diameter of the outmost circle, $d>2r_{e}$, the untouched concentric circles represent free electrons that does not collide each other. Increasing the magnetic field strength shortens the distance between two fluxes, driving the orbital circles to meet and fuse into a single track. The stronger magnetic field there is, the more layers of orbital circle are fused. Along the fused spiral track (like that in Fig. \ref{1+1Dknot1}), we assume the election moves at the same speed everywhere. The distance between the outmost current segment and the flux center is utterly determined by magnetic field strength. Specifically speaking, the train track formed by the fusion of seven concentric circles is showed in Fig. \ref{1+1Dknot1} (a), and six circles fused in Fig. \ref{1+1Dknot1} (b), and four circles fused in Fig. \ref{1+1Dknot1} (c). When the distance between two fluxes continuous to decrease with respect to an increasing strength of magnetic field, the circular tracks around the two fluxes are strongly bind together to form single bundle that winds around the flux pair with a small radius. This track fusion continuous until there is only one detectable bundle that collects all of the other tracks. Then it reaches the ultimate track pattern with respect to fractional charge $1/3$ as showed in Fig. \ref{ABphase} (f). This geometrically quantified magnetic field strength explained why fractional charge $1/3$ is only observed in the strongest magnetic field region \cite{Stormer1999}.

Every winding current track contributes an additional magnetic filed upon the external magnetic field, because each circular electric current is effectively a magnetic dipole. The total magnetic field is strengthened if the magnetic dipole points in the same direction as external magnetic field, otherwise, the total magnetic field strength is reduced. The orientation of the magnetic dipole is determined by the winding number $m$, which counts the winding period of the spiral track as well as the number of braiding operations over the flux pair. $m$ is also the integer index in Abelian Chern-Simons field theory. It also counts the number of concentric loops before the track fusion and is proportional to the radius of the outmost orbital circle around the one flux  before track fusion (Fig. \ref{1+1Dknot1}). Therefore the effective magnetic field strength $B^{\ast}$ in this train track model is defined as,
\begin{eqnarray}\label{B=b-b0m}
B^{\ast} = \frac{B}{1-2p\;m},
\end{eqnarray}
where $B$ is the external magnetic field. $2p$ counts the number of flux quanta absorbed by the electron path. This effective magnetic field strength is exactly coincide with the effective magnetic field in the composite fermion theory of FQHE \cite{Jain1989}. Specifically, for the winding paths showed in Fig. \ref{1+1Dknot1} (a), there are six braiding periods in Fig. \ref{1+1Dknot1} (a), five in Fig. \ref{1+1Dknot1} (b) and three in Fig. \ref{1+1Dknot1} (c). The corresponding effective magnetic field strength for the train tracks in Fig. \ref{1+1Dknot1} (a) is quantified by $B^{\ast} = B /(1- 2p6)$,  $B^{\ast} = B/(1 - 2p5)$ in Fig. \ref{1+1Dknot1} (b) and$B^{\ast} = B/(1 - 2p3)$ in Fig. \ref{1+1Dknot1} (c). The winding number for the cases above is positive because the two fluxes are braided in counterclockwise direction. If the two fluxes are braided in clockwise direction, the winding number $m$ is a negative integer, leading to an increasing term upon the external magnetic field. This quantification rule of effective magnetic field holds for arbitrary number of braiding periods. Thus the train track model offers a topological explanation on the effective magnetic field in the composite fermion theory of FQHE \cite{Jain1989}.

\subsubsection{The train tracks for the fractional conductance with odd denominator}

\begin{figure}
\begin{center}
\includegraphics[width=0.48\textwidth]{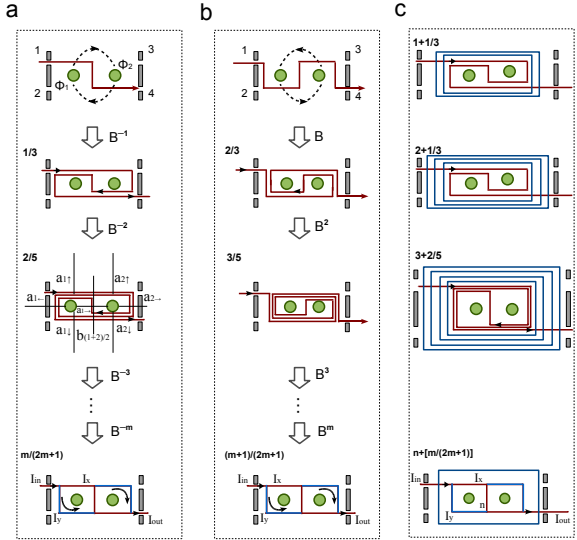}
\caption{\label{fracond} (a) The flux pair, represented by the two green discs, is in between two screens with double slits represented by the gray bars. (b) A serial braiding operations in counter-clockwise direction on the flux pair generates the dual stacks as Fig (a). (c) The single channel representation of the fused stacked electric currents around each flux.}
\end{center}
\vspace{-0.5cm}
\end{figure}

The fractional conductance with odd denominator is effectively constructed by Abelian Chern-Simons field theory \cite{Blok1990}. Here we provide a different approach by winding paths around flux pair, which visualizes the real space track pattern for fractional Hall conductance with odd denominator and extends into a much more general scope beyond quantum field theory. The flux pair is placed in between two screens with double slits in Fig. \ref{fracond}. The slits No. 1 and No. 3 are at the back of the flux, and the slits No. 2 and No. 4 are in the front of the flux. Because the currents in front of flux run in the perpendicular direction of that in the back (Fig. \ref{1+1Dknot1}), whenever a current runs from  No. 1 to No. 4 (or from  No. 2 to No. 3), it contributes to Hall resistance (or termed as off-diagonal resistance) in Eq. (\ref{Rxy=VI}). The diagonal resistance is measured by the current from No. 1 to No. 3( or from  No. 2 to No. 4) in Fig. \ref{fracond}, as defined following,
\begin{eqnarray}\label{Rxx=VI}
R_{xx} = V_{x}/{I_{x}},\;\;\;  R_{yy} = V_{y}/{I_{y}}.
\end{eqnarray}
According to electromagnetic field theory, the electric current $I_{\alpha}$ is proportional to the amount of charges $Q_{\alpha}$ that passes the tracks at speed of $v$, $I_{\alpha}= v Q_{\alpha}$.

Folding a simple track by braiding flux pair generates a stack of laminar tracks around each flux. When these laminar tracks get closer enough to one another, a track fusion is inevitable due to quantum tunneling effect. resulting in fractional charges running in the fused track bundle. We start from the simplest case of single track that runs from the slit No. 1 to No. 4 by passing the border region between the two fluxes (Fig. \ref{fracond} (a)). By performing a clockwise braiding on the flux pair $\hat{B}^{-1}$, the initial single track folds into asymmetric laminar tracks around the two fluxes (Fig. \ref{fracond} (a) 1/3), where the number of track layers above and below the two fluxes are listed as following:
\begin{eqnarray}\label{a12updown}
a_{1,\uparrow}&=& 2,\;\;\; a_{1,\downarrow} = 1,\;\; a_{2,\uparrow}= 1, \;\;\;a_{2,\downarrow} = 2.  \nonumber\\
a_{1,\leftarrow}&=& 1,\;\; a_{1,\rightarrow} = 1,\;\; a_{2,\leftarrow}= 1, \;\;\;a_{2,\rightarrow} = 1.
\end{eqnarray}
Each track represents one unit of passing probability of electron. When the two track segments above $\Phi_{1}$ and below $\Phi_{2}$ gets too close to distinguish from each other, they fuse into one resultant track but still keeps two units of pass probability. Thus the total number of stack layers on the cross section of each side of flux represents the passing probability of one electron. When an electron with an elementary charge $e$ is injected into the track from slits No. 1, a fractional charge $2e/3$ goes into the track $a_{1,\uparrow}$ and a fractional charge $e/3$ runs into $a_{1,\leftarrow}$ to fulfill the conservation law of charges. The fractional charge $2e/3$ splits into two tracks across the border region of two flues, one $e/3$ winds around the flux $\Phi_{2}$ from above, the other $e/3$ runs along the border line. In mind of the quantized Hall conductance in fractional quantum Hall system, here the Hall resistance can be formulated in the same way,
\begin{eqnarray}\label{Rxy=VIQ}
R_{xy} = R_{0}\frac{1}{Q_{y}},  \;\;\;  Q_{y} = \frac{a_{1,\leftarrow}}{a_{1,\leftarrow}+a_{1,\uparrow}}= \frac{a_{1,\downarrow}}{a_{1,\downarrow}+a_{1,\uparrow}},
\end{eqnarray}
here $R_{0}=h/e^2$ is the quantum of resistivity in quantum Hall systems. Here the train tracks in Fig. \ref{fracond} (a) (1/3) offers a train track explanation to the fractional quantum Hall effect with a filling factor of $1/3$.

Other serial fractional charges are generated by systematic braiding on the flux pair. For instance, one more braiding in clockwise direction on the train track pattern in Fig. \ref{fracond} (a) (1/3) leads to the fractional charges of $2e/5$ and $3e/5$ (Fig. \ref{fracond} (a) (2/5)). Note here the number of track layers on the left side $a_{1,\leftarrow}$ of flux $\Phi_{1}$ is always equal to that below $a_{1,\downarrow}$. While its dual flux $\Phi_{2}$ shows the opposite case, $a_{2,\rightarrow} =a_{2,\uparrow}$. For simplicity, we use one thick track to summarize the stacked tracks around each flux, as showed in Fig. \ref{fracond} (a) (m). The total number of original tracks labels its weight. After $m$ round of braiding, the weight of tracks around the flux pair counts as follows:
\begin{eqnarray}\label{ma12updown}
a_{1,\uparrow}&=& m+1,\;\; a_{1,\downarrow} = m, \;\; a_{2,\uparrow}= m, \;\;\;a_{2,\downarrow} = m+1.  \nonumber\\
a_{1,\leftarrow}&=& m,\;\;\;\;\;\;\; a_{1,\rightarrow} = 1,\;\; a_{2,\leftarrow}= 1, \;\;\;a_{2,\rightarrow} = m,
\end{eqnarray}
which still admits a conservation equation, $a_{i,\uparrow} + a_{j,\uparrow} = b_{(i+j)/2} = 2m+1.$ These train track patterns correspond to the fractional quantum Hall states with fractional charges,
\begin{eqnarray}\label{Qma12updown}
Q_{1,\leftarrow}&=& Q_{2,\rightarrow} = \frac{m}{2m+1},\;\; Q_{1,\rightarrow} = Q_{2,\leftarrow}= \frac{1}{2m+1}, \nonumber\\
Q_{1,\uparrow}&=&Q_{2,\downarrow} =\frac{m+1}{2m+1}, \;\; Q_{1,\downarrow} = Q_{2,\uparrow}= \frac{m}{2m+1},
\end{eqnarray}
The fractional charge serial above converges to a half charge $e/2$ under infinite times of braiding. The fractional charges above obeys a special linear group transformation, $SL(2,m)$,
\begin{eqnarray}\label{QudUk}
 Q_{2,\uparrow}(m)& =&\frac{m}{2m+1}=
\left(
\begin{array}{cc}
1 & 0 \\
2 & 1 \\
\end{array}
\right)(m)=U_{2,\uparrow}(m),\nonumber\\
 Q_{1,\downarrow}(m)& =&\frac{m+1}{2m+1}=
\left(
\begin{array}{cc}
 1 & 1 \\
2 & 1 \\
\end{array}
\right)(m)=U_{1,\downarrow}(m).
\end{eqnarray}
The group elements of $SL(2,m)$ is denoted by matrix $U_{i,\uparrow}$ and $U_{i,\downarrow}$, which maps an integer $m$ into a fractional number. The limit of this fractional serial is 1/2 when $m$ approximates to infinity. Note the fractional charge ${1}/({2m+1})$ always runs along the border line between the two fluxes. The clockwise braiding only generates the fractional Hall resistance serial,
\begin{eqnarray}\label{Rxy=m}
R_{xy} = R_{0}\frac{1}{Q_{y}} = R_{0} \frac{2m+1}{m}.
\end{eqnarray}
In order to reach the other fractional Hall resistance serial,
\begin{eqnarray}\label{Rxy=m+1}
R_{xy} = R_{0}\frac{1}{Q_{y}} = R_{0} \frac{2m+1}{m+1}.
\end{eqnarray}
a counterclockwise braiding must be performed on the initial simple track, as showed in Fig. \ref{fracond} (b). In that case, the fractional charge  $Q =\frac{m+1}{2m+1}$ turns into its perpendicular direction without losing charge. The fractional Hall resistance is proportional to the effective magnetic field strength $B^{\ast}$ as defined in Eq. \ref{B=b-b0m}. This theoretical conclusion is exactly coincide with the experimental measurement of Hall resistance \cite{Stormer1999}.

Serial of fractional charges near other integral filling state is also observed in FQHE experiment \cite{Murthy2003}. Here the integral filling state is represented by concentric circles around the flux pair outside the fractional train tracks. The layer number of concentric circles equals to the number of electrons. Then the fractional filling serial around integer $n$ is generated by the train tracks of braiding two fluxes, enveloped by $n+1$ layers of concentric circles. Fig. \ref{fracond} (c) shows the typical train tracks for filling states of $1+(1/3)$, $2+(1/3)$, and $3+(2/5)$. The trains tracks of a general fractional charges,
\begin{eqnarray}\label{n+m2m+1}
Q_{1,\leftarrow}&=& Q_{2,\rightarrow} = Q_{1,\downarrow} = Q_{2,\uparrow}=n+\frac{m}{2m+1},     \nonumber\\
Q_{1,\uparrow}&=&Q_{2,\downarrow} =n+\frac{m+1}{2m+1}, \nonumber\\
Q_{1,\rightarrow} &=& Q_{2,\leftarrow}= \frac{1}{2m+1},
\end{eqnarray}
is represented by fused track rectangles around each flux, with its four edges assigned with different weight. The brown edges represent a current that switch its direction into its perpendicular direction without losing or gaining any charges (Fig. \ref{fracond} (c)) The blue circle outside around the flux pair in Fig. \ref{fracond} (c) is assigned with a number $n+1$ that tells how many electrons filled in the flux pair. The Hall resistance of these fractional charged states obeys similar equations as Eq. (\ref{Rxy=m})
 \begin{eqnarray}\label{Rxy=n+m}
R_{xy} &=& R_{0}\frac{1}{Q_{y}} = R_{0} \frac{1}{n+\frac{m}{2m+1}},\;\; n=0,1,2,3\cdots \;\;\;
\end{eqnarray}
and $m=1,2,3,\cdots$. The other serial of fractional resistance,
 \begin{eqnarray}\label{Rxy=n+m2}
R_{xy} &=& R_{0}\frac{1}{Q_{y}} = R_{0} \frac{1}{n+\frac{m+1}{2m+1}},  \;\; n=0,1,2,3\cdots \;\;\;
\end{eqnarray}
corresponds to the braiding operations in counterclockwise direction.

\begin{figure}
\begin{center}
\includegraphics[width=0.40\textwidth]{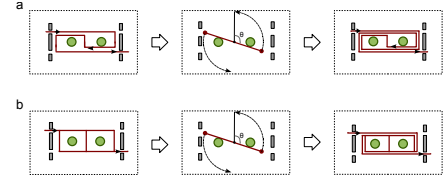}
\caption{\label{trackfusion} (a) A trains track segment in a decreasing magnetic field splits into more layers, i.e., from $e/3$ to $2e/5$. (b) The train track splitting for a general train track pattern around the two fluxes. }
\end{center}
\vspace{-0.5cm}
\end{figure}

The Hall resistance as we derived from braiding train tracks and knot lattice model has an exact one-to-one correspondence with the measured Hall resistance for two dimensional electron gas in strong magnetic field, where the Hall resistivity increases for an increasing magnetic field \cite{Murthy2003}. Here the effective magnetic field strength for a general fractional charge is quantified by
\begin{eqnarray}\label{b=bm-n}
B^{\ast} = \frac{B-\frac{1}{n}}{( 1- 2pm)}, \;\;\;n=1,2,3\cdots \;\;\;
\end{eqnarray}
here $m$ is winding number of train tracks, it is also the total number of braiding operations. $n$ is total number of layers of concentric circles around the outer region of the flux pair. The maximal magnetic field strength generates fractional charge of $e/3$ with $m = -1$ and $n=0$, here all layers of train track stack fused into an ultimate single track around the flux pair, there is no more simpler tracks for fractional charges than $e/3$ (as shown in Fig. \ref{fracond} (a)).

A reduction of magnetic field strength result in track splitting, the electron is confined by less Lorentz force to wind around the flux pair over longer distance (Fig. \ref{trackfusion}). For instance, if the electron winds around the center of flux pair over one more period, it drives $e/3$ state into $2e/5$ state (Fig. \ref{trackfusion} (a)). For a more general case in which the bonds around each flux are assigned with general weights (Fig. \ref{trackfusion} (b)), one period of train track splitting obeys the rule as follows,
\begin{eqnarray}\label{deltaa12}
\delta a_{1,\uparrow}&=& 1,\;\;\delta a_{1,\downarrow} = 1, \;\; \delta a_{2,\uparrow}= 1, \;\; \delta  a_{2,\downarrow} = 1; \nonumber\\
\delta  a_{1,\leftarrow}&=& 1,\;\; \delta a_{1,\rightarrow} = 0,\;\; \delta a_{2,\leftarrow}= 0, \;\; \delta a_{2,\rightarrow} = 1. \;\;\;\;\;
\end{eqnarray}
The effective magnetic field reduces to zero when the external magnetic field reaches $1/n$. The strongly confined track bundle would be completely released from the core region of the flux pair.  Since there exist both a large number of layers of train track above and below the flux pair, the elementary charge splits evenly into half charge $e/2$ when it passes the flux pair. Thus the half filling states is the limit case of fractional charge serial. However there always exist an infinitely small fractional charge, $e/(2m+1)$, along the border line of the flux pair. When the magnetic field continuous to decrease, a number of isolated circles enclose the flux pairs along the outer border without penetrating through the border region between the two fluxes. In that case, the fractional charges running on the top or bottom boundary is $e(n+(1/2))$. But the fractional charge sandwiched in between two fluxes is still $e(\frac{1}{2m+1})$.

\begin{figure}
\begin{center}
\includegraphics[width = 0.35\textwidth]{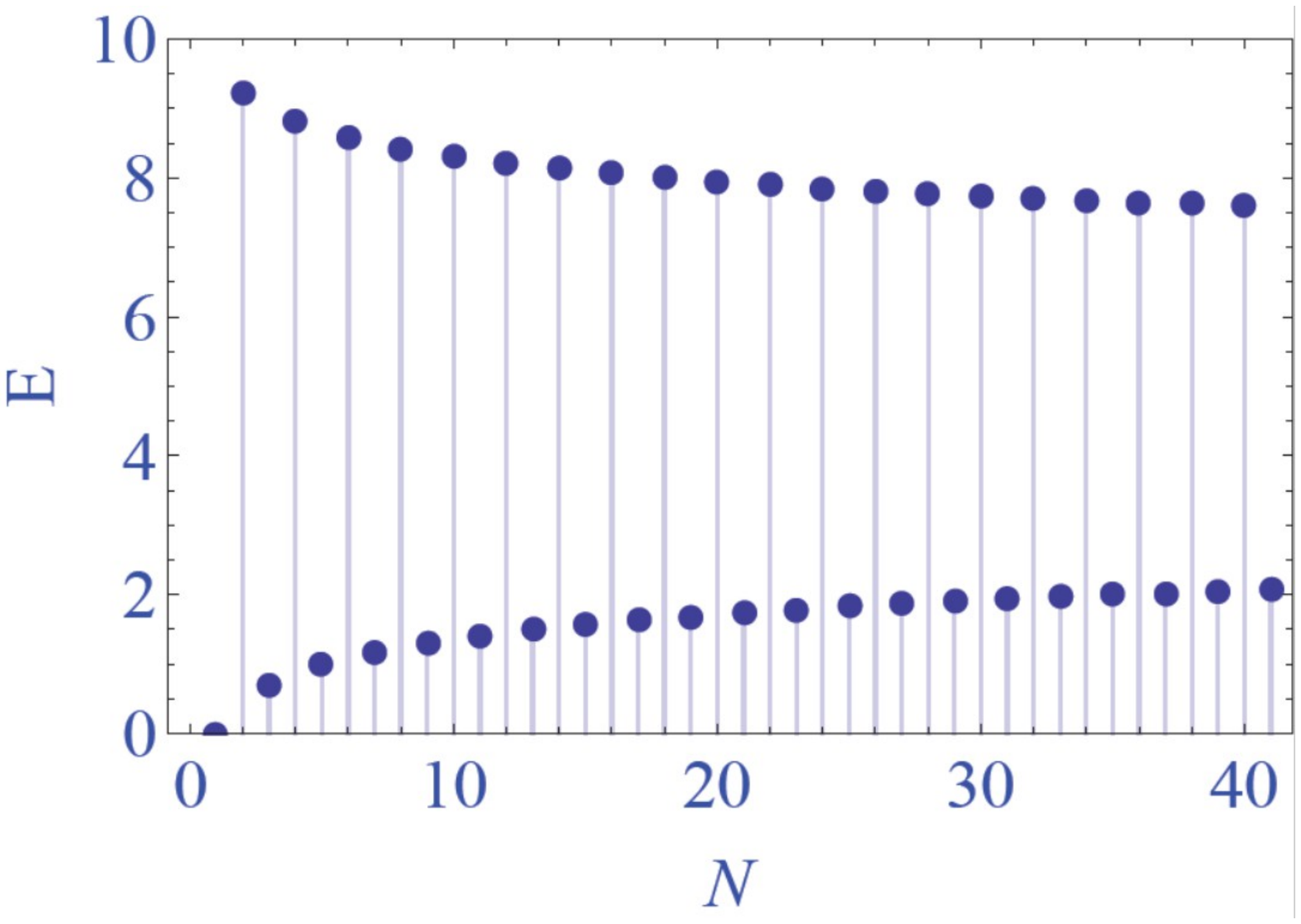}
\caption{ \label{energy} The total electromagnetic energy for 41 layers of ant- parallel currents, $E_{i}$, as the eigenvalue of Hamiltonian Equation $H_{i}$. Here the distance d =  0.0001. }
\end{center}
\vspace{-0.5cm}
\end{figure}

In the train track patterns around the flux pair, the current in the nearest neighboring train track segments always run in antiparallel direction. Fractional charges prefer to stay on the fused track composed of odd number of current tracks due to its lower electromagnetic energy. Braiding operation generates many stacked anti-parallel currents. Even number of stacked current always has higher energy than odd number of stacks. For N layers of stacked antiparallel currents, their total electromagnetic energy is summarized into the Hamiltonian equation,
 \begin{eqnarray}\label{HamiE}
H_{i}= - \frac{\mu_{0}l_{i}}{2\pi}\sum_{r>l=1}^{N}(-1)^{r-l}\ln[(r-l)d](I_{i,y+l}I_{i,y+r}).
\end{eqnarray}
Here $\mu_{0} = 1.26*10^{-6} (T m/A)$ is dielectric coefficient. $l_{i}$ is the horizontal length of the current track at the $i$th site. $d$ is the perpendicular distance between the nearest neighboring electric current segment. Since the distance between two moving electrons in one dimensional electronic system usually falls in micro/nanometer scale, it is reasonable to set $d<<1$ to make sure the repulsive energy increases when two antiparallel currents get closer. We set the value of current operators as $I_{i}=1$ for simplicity.

For a stack of $N$ layers of antiparallel currents,  the numerical computation of energy showed in Fig. \ref{energy}, the energy curve of odd number of tracks is always below that of even number of tracts. Two antiparallel currents bear the maximal initial energy. The energy of even number of current stacks decreases as the total number of stack layers grows. While on the opposite side, odd number of current stacks shows an increasing energy curve (Fig. \ref{energy}). The energy curve of odd stack and even stack finally converges to a fixed point when the total number of stacked currents approaches to infinity under infinite number of braiding. That fixed point energy is the eigenenergy of the half-charged states in the limit of $m\rightarrow\infty$, which is 4.95174 for the numerical setting above. The electromagnetic energy for finite fractional charges are listed as the discrete points in Fig. \ref{energy}. The stack of $(n+1)$ layers of concentric circles corresponds to the case of integral filling states $\nu = n$, in which electrons circling around the flux pair in the same chirality. The electric current in these orbital circles run in the same direction by attracting each other to reduce the electromagnetic energy.

At low temperature around 1K, the mean free path of electron reaches 1/5 mm \cite{Stormer1999}. A rising temperature reduces the electron mobility as well as the maximal length of the winding path, preventing the generation of fractional charges near the half charge state which only exist for the maximal mean free path. When the temperature grows above critical value, the mean free path of electron is not long enough to complete the minimal winding operation around a flux pair. As a result, the electron falls into the range of ballistic transportation, demonstrating a classical transportation behavior. Fractional Hall conductance is no long observable above critical temperature. The longer winding tracks has higher probability to survive at lower electromagnetic energy. As showed before, longer winding tracks generated by more braiding operations indeed have lower energy. Thus the existence probability of topologically braided state with energy $H(c)$ obeys the Boltzmann-Maxwell distribution,
\begin{eqnarray}\label{pro}
P=\frac{1}{Z}\exp[{-H(c)/k_{b}T}],
\end{eqnarray}
where $k_{b}$ is Boltzmann constant, $T$ is temperature and H(c) is the Hamiltonian for the topological current track pattern $c$. $Z$ is partition function,
\begin{eqnarray}\label{partition}
 Z = \sum_{c} \exp[{-H(c)/k_{b}T}].
\end{eqnarray}
For the simplest case of single state with N anti-parallel currents, the partition function term is directly computed by substituting the eigen-energy equation (\ref{HamiE}) into the partition function Eq. (\ref{partition}) above,
\begin{eqnarray}\label{partition0}
 Z_{0} = \exp[\frac{\mu_{0}a}{2\pi{k_{b}T}}]\prod_{r>l=1}^{N}[(r-l)d]^{(-1)^{r-l}}.
\end{eqnarray}
The general partition function of a more complex winding current pattern is derived following the same principals above, which also holds in two and three dimensional lattice of magnetic fluxes.

\subsubsection{The train tracks for the fractional conductance with even denominator}

\begin{figure}
\begin{center}
\includegraphics[width=0.40\textwidth]{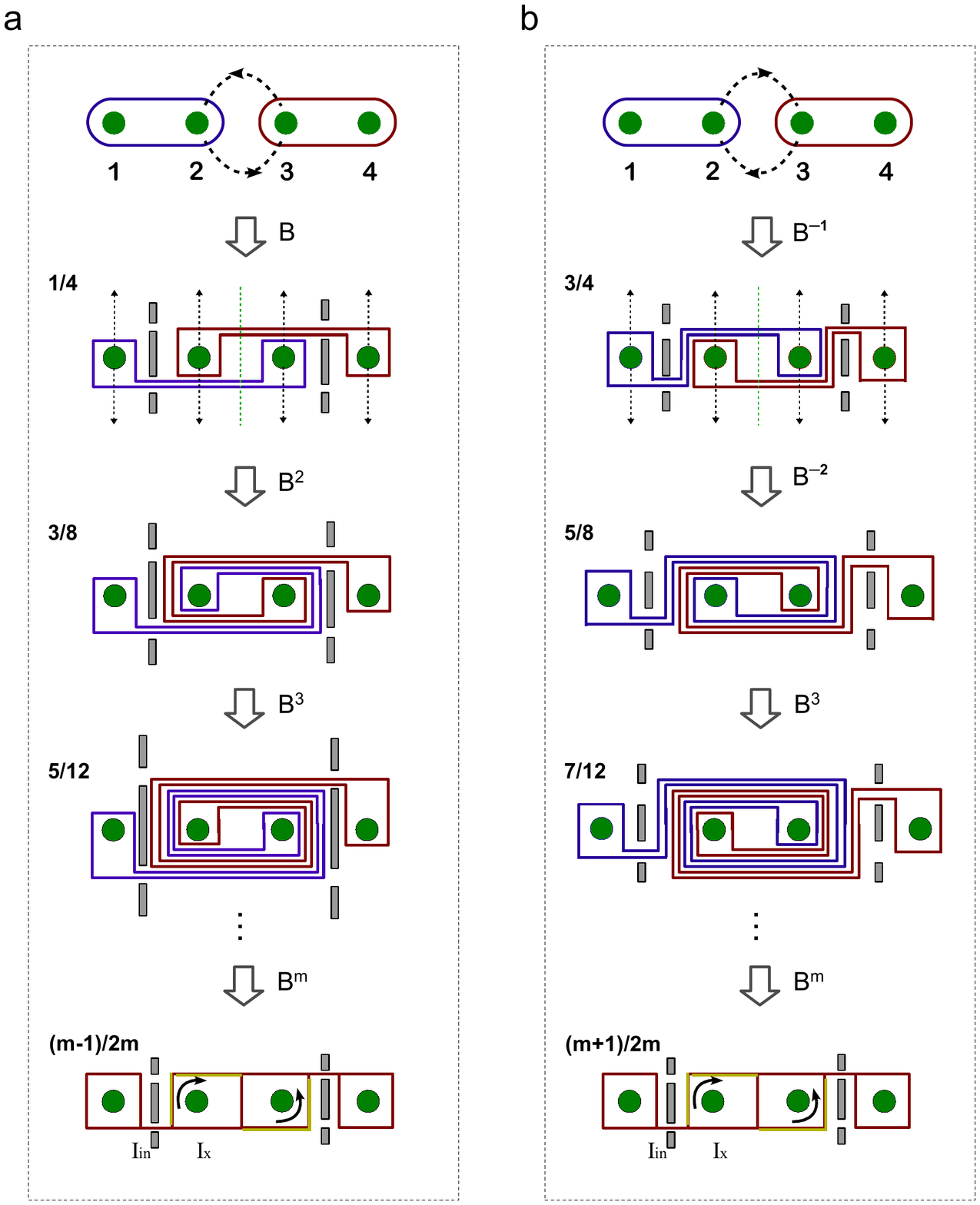}
\caption{\label{dimercon} (a) Counterclockwise braiding over the nearest neighboring fluxes that belong to different loop dimer generates the fractional charges with even denominator, $\frac{m-1}{2m}$. (b). Clockwise braiding over the nearest neighboring fluxes leads to the fractional charge serial, $\frac{m+1}{2m}$.}
\end{center}
\vspace{-0.5cm}
\end{figure}

Besides the fractional charges with odd denominator, fractional Hall resistance with certain even denominators is also observed in quantum Hall effect, such as $5e/2$ \cite{Stormer1999} \cite{Murthy2003}. The Abelian Chern-Simons field theory cannot effectively describe the fractional Hall resistance with even denominators. It is generally believed $5e/2$ state is non-Abelian state but is still not confirmed so far. Here we provide a systematic construction of fractional charges with a serial of even denominators along the same route of topological path fusion theory.

Unlike the train tracks around one flux pair for the fractional charges with odd denominators, it takes at least two flux pairs to construct fractional charges with even denominators, as shown in Fig. (\ref{dimercon}). In the beginning, the tracks winds around the outer region of the flux pair without penetrating through the border region within the flux pair (or termed as flux dimer). The first dimer is composed of flux No. 1 and No. 2, which is separated from the second flux dimer of No. 3 and No. 4.
The two tracks that bridge the fluxes No. 1 and No. 2 are squeezed into the front slit of the left double slits, while the bridge track between flux No. 3 and No. 4 is squeezed into the back slit of the right double slits screen. Then we fix the flux No. 1 and No. 4, because they falls outside of the two screens. One counterclockwise braiding on flux No. 2 and No. 3 results in the track stack distribution as follows,
\begin{eqnarray}\label{ma23updown}
a_{2,\uparrow}&=& 1,\;\; a_{2,\downarrow} = 3, \;\; a_{3,\uparrow}= 3, \;\;\;a_{3,\downarrow} = 1.  \nonumber\\
a_{2,\leftarrow}&=& 1,\;\; a_{2,\rightarrow} = 2,\;\; a_{3,\leftarrow}= 2, \;\;\;a_{3,\rightarrow} = 1.
\end{eqnarray}
After track fusion, the elementary charge splits into fractional charges along the edges around flux No. 2 and No. 3,
\begin{eqnarray}\label{Qma23updown1i4}
Q_{2,\leftarrow}&=& Q_{3,\rightarrow} = \frac{1}{4},\;\; Q_{2,\rightarrow} = Q_{3,\leftarrow}= \frac{2}{4}, \nonumber\\
Q_{1,\uparrow}&=&Q_{2,\downarrow} =\frac{1}{4}, \;\; Q_{1,\downarrow} = Q_{2,\uparrow}= \frac{3}{4}.
\end{eqnarray}
Two counterclockwise braiding results in the fractional charge serial around the flux pair,
\begin{eqnarray}\label{Qma23updown3i8}
Q_{2,\leftarrow}&=& Q_{3,\rightarrow} = \frac{3}{8},\;\; Q_{2,\rightarrow} = Q_{3,\leftarrow}= \frac{2}{8}, \nonumber\\
Q_{2,\uparrow}&=&Q_{3,\downarrow} =\frac{5}{8}, \;\; Q_{1,\downarrow} = Q_{2,\uparrow}= \frac{5}{8}.
\end{eqnarray}
Fractional charge $5e/12$ is generated by three counterclockwise braiding on flux No. 2 and No. 3 ( As shown in Fig. \ref{dimercon} (a)). The fractional charge generated by $m$ rounds of braiding are list as follows,
\begin{eqnarray}\label{Qm23updownm-12m}
Q_{2,\leftarrow}&=& Q_{3,\rightarrow} = \frac{m-1}{2m},\;\; Q_{2,\rightarrow} = Q_{3,\leftarrow}= \frac{2}{2m}, \nonumber\\
Q_{2,\uparrow}&=&Q_{3,\downarrow} =\frac{m+1}{2m}, \;\;\; Q_{2,\downarrow} = Q_{3,\uparrow}= \frac{m+1}{2m}.\;\;\;\;
\end{eqnarray}
Note there is a small fraction of charge $1/m$ running along the border line between flux No. 2 and No. 3. Similar to the fractional charges with odd denominator, a clockwise braiding on the fluxes No. 2 and No. 3 results in the dual charges distribution,
\begin{eqnarray}\label{Qm23updownm+12m}
Q_{2,\leftarrow}&=& Q_{3,\rightarrow} = \frac{m+1}{2m},\;\; Q_{2,\rightarrow} = Q_{3,\leftarrow}= \frac{2}{2m}, \nonumber\\
Q_{2,\uparrow}&=&Q_{3,\downarrow} =\frac{m+1}{2m}, \;\;\; Q_{2,\downarrow} = Q_{3,\uparrow}= \frac{m-1}{2m}.\;\;\;\;
\end{eqnarray}
This fractional serial is not the only serial that can be generated by braiding the four fluxes. Braiding the fluxes No. 1 and No. 2, or No. 3 and No. 4, does not generate fractional charges because the fluxes are enveloped into the same domain. However, fixing No. 1 and No. 3 and braiding No. 2 and No. 4 generates another serial of fractional charges with even denominator. Therefore it is straightforward to introduce two or three independent braiding operations on the four fluxes to generate other serial of fractional charges. The non-commutable character of braiding matrices is directly read out from distribution of fractional charges on the bonds around the four fluxes.

\subsubsection{The correspondence between integral Hall conductance and fractional Hall conductance}

The composite fermion theory of FQHE suggested an accurate correspondence between FQHE and IQHE \cite{Jain1989}. This correspondence is supported by experimental observation of fractional Hall resistance with odd denominator. However there still lacks a rigours understanding so far on why such a correspondence exists. Here we proposed a topological surgery method to map a train track pattern of integral Hall resistance into fractional Hall resistance with both odd denominator and even denominator.

\begin{figure}
\begin{center}
\includegraphics[width=0.40\textwidth]{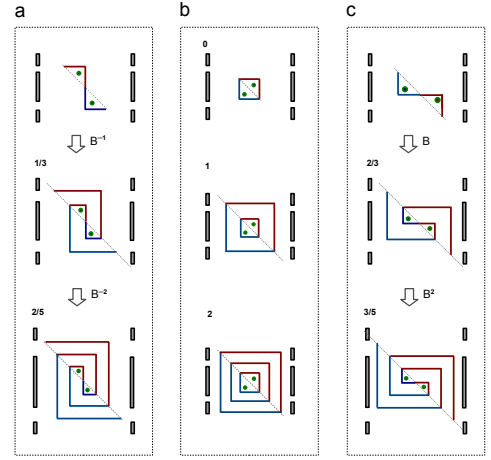}
\caption{\label{fluxtrans} (a) The train tracks generated by translating the upper arcs to the left and docking them together. (b) The train tracks represented the integral filling states. (c) The train tracks generated by translating the upper red arcs to the right and docking them together.}
\end{center}
\vspace{-0.5cm}
\end{figure}

The integral filling states are represented by a pile of concentric loop currents that envelopes a flux pair without penetrating through the border region between them ( as shown in Fig. \ref{fluxtrans} (b)). The zero filling states $\nu = 0$ is represented by one loop current around the flux pair ( Fig. \ref{fluxtrans} (b) - 0). The filling state $\nu = 1$ corresponds to two loops around the flux pair ( Fig. \ref{fluxtrans} (b) - 1), and three loops for $\nu = 2$ ( Fig. \ref{fluxtrans} (b) - 2) , and so on. There are $n+1$ loops enveloping the flux pair for $\nu = n$ filling state. We cut the concentric loop currents around the flux pair along the border line between two fluxes into two set of current arcs simultaneously (as shown in  Fig. \ref{fluxtrans} (b) ). The upper (lower) current arcs are represented by red (blue) lines.

The fractional charges are generated by translating the current arcs together with the flux they enveloped along the cutting line, and dock each red arc with the corresponding blue arc in the new locations along the cutting line. Translating the upper arcs to the left hand side over one step and and docking them together generates the fractional filling states with $\nu = m/(2m+1)$ (as shown in  Fig. \ref{fluxtrans} (a)). While translating the upper arcs to the right hand side generates the fractional fillings $\nu = (m+1)/(2m+1)$ (as shown in  Fig. \ref{fluxtrans} (c)). The shifting direction determines the chirality of braiding operations. Translating to the left (right) induces a counterclockwise (clockwise) braiding over the flux pair. The number of braiding operations is exactly equal to the integral filling factor $\nu = n$. Therefore, the correspondence between integral Hall effect and fractional Hall effect has rigorous geometric foundation.

\begin{figure}
\begin{center}
\includegraphics[width=0.40\textwidth]{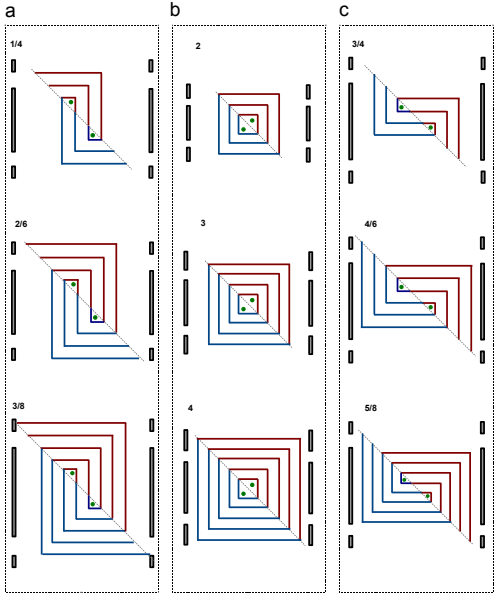}
\caption{\label{evenfluxtrans} (a) The upper semi-circles are translated by two steps to the left and glued together with the bottom semi-circles at the new contacting point. (b) The electric current track around a pair of fluxes, represented by the two small green discs. (c) The upper semi-circles arcs translated to the right by two steps and reunite with bottom semi-circles.}
\end{center}
\vspace{-0.5cm}
\end{figure}

This topological surgery method not only generates fractional charges with odd denominator, but also predicted fractional charges with even denominators, and so on. If the Translation distance covers an odd number of steps, it leads to fractional charges with odd denominator. While an even number of translation operations results in fractional charges with even denominators.  For instance, two-step translation to the left generated the fractional serial $(m-1)/2m$ (as shown in  Fig. \ref{evenfluxtrans} (a) ), while two-step translation to the right leads to the fractions $(m+1)/2m$ (as shown in  Fig. \ref{evenfluxtrans} (c) ). Note here the $e/3$ state also exist as $2e/6$ in the even denominator serial but is different from the $e/3$ in odd denominator serial which corresponds to the integral filling $n=1$  (as shown in  Fig. \ref{fluxtrans} (b) ). While the $2e/6$ here corresponds to the integral filling factor $n=3$  (as shown in  Fig. \ref{evenfluxtrans} (b) ). Thus $e/3$ state is highly degenerated state, which can also be generated by p (p$>$2) steps of translation operations.

\begin{figure}
\begin{center}
\includegraphics[width=0.40\textwidth]{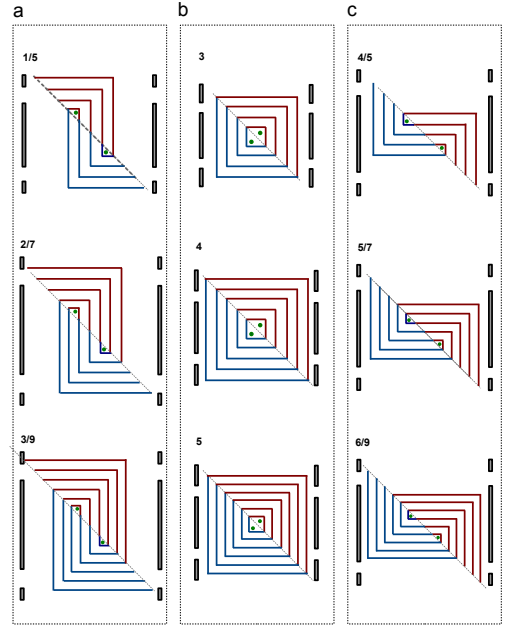}
\caption{\label{3fluxtrans} (a) The upper semi-circles are translated to the left by three steps and reunite with the bottom ones. (b) The electric current track around a pair of fluxes, represented by the two small green discs. (c) The upper semi-circles are translated to the right by three steps and reunite with bottom ones. }
\end{center}
\vspace{-0.5cm}
\end{figure}

Different odd number of translation operations lead to different serial of fractional charges. Fig. \ref{3fluxtrans} shows the train tracks generated by three-step translation operations. The minimal integral filling states for three-step translation is $n=3$. Three-step translation to the left results in the fractional serial (Fig. \ref{3fluxtrans} (a))
\begin{eqnarray}\label{n=3left}
Q_{1,\leftarrow}&=& Q_{2,\rightarrow} = Q_{1,\downarrow} = Q_{2,\uparrow} =\frac{m-2}{2m-1},   \nonumber\\
Q_{1,\uparrow}&=&Q_{2,\downarrow} =\frac{m+2}{2m-1}, \nonumber\\
Q_{1,\rightarrow} &=& Q_{2,\leftarrow}= \frac{3}{2m-1},
\end{eqnarray}
While three-step translation to the right hand side leads to the dual serial fractional charges (Fig. \ref{3fluxtrans} (c)),
\begin{eqnarray}\label{n=3right}
Q_{1,\leftarrow}&=& Q_{2,\rightarrow} = Q_{1,\downarrow} = Q_{2,\uparrow} =\frac{m+2}{2m-1},   \nonumber\\
Q_{1,\uparrow}&=&Q_{2,\downarrow} =\frac{m-2}{2m-1}, \nonumber\\
Q_{1,\rightarrow} &=& Q_{2,\leftarrow}= \frac{3}{2m-1}.
\end{eqnarray}
Note here there exist an open track sandwiched in between the dimer tracks. This fractional serial is hybrid combination of the one-step translation with the two-step translation cases. The $e/3$ state now corresponds to $3e/9$ for the integral filling $\nu=5$.  For the most general case of p (p$>$2) steps of translation, the fractional charges obey the following equations,
\begin{eqnarray}\label{n=p}
Q_{1,\leftarrow}&=& Q_{2,\rightarrow} = Q_{1,\downarrow} = Q_{2,\uparrow} =\frac{m-(p-1)}{2m+1-(p-1)},   \nonumber\\
Q_{1,\uparrow}&=&Q_{2,\downarrow} =\frac{m+(p-1)}{2m+1-(p-1)}, \nonumber\\
Q_{1,\rightarrow} &=& Q_{2,\leftarrow}= \frac{p}{2m+1-(p-1)},
\end{eqnarray}
here $p\geqq2$. There are $(p-2)$ open tracks sandwiched in between the two dimer tracks. When the number of braiding operations $m$ and the number of translation steps obey the equation $m+1=2p$, the fractional charges $e/3$ are generated by p-step translations in the topological surgery of integral filling state. Because the number of translation steps can not outnumber the integral filling factor, the maximal degeneracy degree is limited by the integer $m$.

\subsubsection{The correspondence between the fractional charge in knot lattice model and that of train track model}

\begin{figure}
\begin{center}
\includegraphics[width=0.42\textwidth]{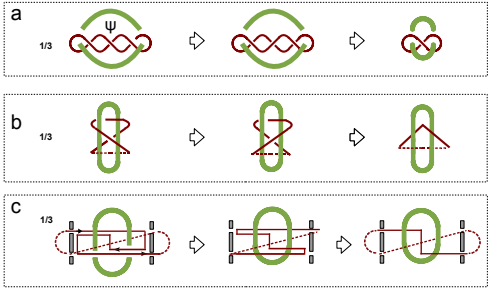}
\caption{\label{fracknot13} (a) The knot with three crossings as initial configuration. $\psi$ is the Majorana fermion operator which flips the crossing state. The connected segment in the same plane can topologically transform into a shorter segment. (b) The knot around flux dimer with respect to the train tracks for $e/3$ and its transformation under flipping operations. (c) The train tracks with respect to $e/3$, and its configuration transformation under flipping operations.}
\end{center}
\vspace{-0.5cm}
\end{figure}

The fractional charge in the train track model here has an exact one-to-one correspondence with the fractional filling state in the knot lattice model of anyon \cite{{SiT2019}}, where the fractional filling factor of anyon is defined as the ratio of the number of Majorana fermion operators $N(\psi)$ to the number of braiding operations $ N(B)$,
\begin{equation}\label{filling}
\nu =\frac{L_{link}}{N{(B)}} =\frac{N(\psi)}{N{(B)}},
\end{equation}
with $L_{link}$ is the linking number, which counts how many flipping operations are needed to bring a multi-crossing knot back to the minimal crossing state. $N{(B)}$ counts the total number braiding for generating the multi-crossing knot out of an uncrossing circle \cite{{SiT2019}}. To show how to map a spiral train track into a knot lattice, the two endings of the flux pair are connected to form a closed loop of magnetic flux tube in Fig. \ref{1+1Dknot1}. Without losing generality, the knot with three crossings as showed in Fig. \ref{fracknot13} (a) is taken as an exemplar configuration. The Majorana fermion operators $\psi$ flips a positive crossing state to a negative crossing state, or vice versa. One flipping operation on the middle crossing point brings the back line to the front, which now can fuse with the other two segments in the front to reduce the multi-crossing knot to the minimal crossing state. Thus by flipping the crossing state in certain location drives the nearest neighboring connected current segments into the same domain so that they fuse into one, which keeps the topology of knot invariant (Fig. \ref{fracknot13} (a)). This knot pattern generated fractional charge $e/3$. The corresponding spiral train track of $e/3$ is showed in Fig. \ref{1+1Dknot1} (b). Flipping the middle crossing point of knot is equivalent to exchange the current segments that form the crossing point in train track pattern Fig. \ref{1+1Dknot1} (b), the current segment on the same side of flux pair can topologically transform into the minimum train track style Fig. \ref{1+1Dknot1} (b), which is exact the same geometric configuration as the minimum knot pattern in Fig. \ref{fracknot13} (a). Note here the source ending and detector ending of the open track is connected to fulfil the boundary condition that they merge into one point at infinity. The projection of the spiral train track of Fig. \ref{1+1Dknot1} (b) into the bottom plane depicts the classical train track around the flux pair in Fig. \ref{1+1Dknot1} (c), where flipping the middle crossing point of the knot in Fig. \ref{fracknot13} (b) brings the middle track segment in Fig. \ref{fracknot13} (c) at the back of the flux pair to the front. The continuously connected track segments on the same side of flux pair fuse into a minimal track segment under topological transformation (Fig. \ref{fracknot13}(c)). This topological operation protocol revealed the rigorous relationship between knot and train track.

\begin{figure}
\begin{center}
\includegraphics[width=0.42\textwidth]{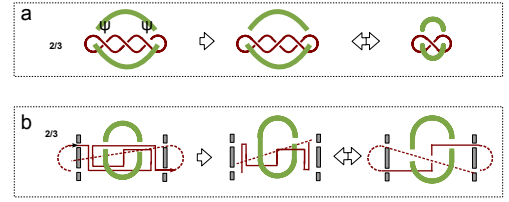}
\caption{\label{fracknot23} (a) The knot transformation for generating $2e/3$, two flipping operations of crossing state are performed on the first and the third crossings. (b) The transformation of train tracks under the two continuous flipping operations.}
\end{center}
\vspace{-0.5cm}
\end{figure}

The fused knot configuration of $2e/3$ is the mirror image of that of $e/3$. Based on the same initial knot with three crossings as that of $e/3$, two flipping operations by Majorana fermion operator on the two crossings away from the middle point bring the two segments at back to the front, then they fuse with middle segment into one (Fig. \ref{fracknot23} (a)).  When it maps into the train tracks for $2e/3$, the first flipping of crossing exchanges the position of the right outmost segment with the left one (Fig. \ref{fracknot23} (a)). A second flipping operation bring the front (back) segment to the back (front) (as shown in Fig. \ref{fracknot23} (a)). It finally leads to the train track of $2e/3$, which is exactly the spatial inversion of the track pattern of $e/3$.

\begin{figure}
\begin{center}
\includegraphics[width=0.42\textwidth]{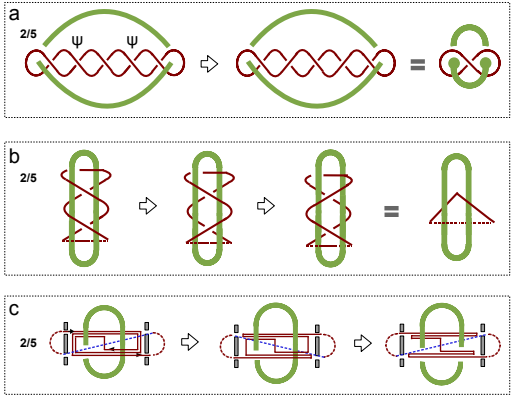}
\caption{\label{fracknot25} (a) The knot with five crossing and its transformation process under crossing flipping operation as a demonstration of $2e/5$ anyon. (b) The corresponding knot around the flux dimer with respect to the trains track of $2e/5$, and its transformation under crossing flipping operations. (c) The train tracks for the anyon with charge of $2e/5$, and the topological transformation under crossing flipping operations.}
\end{center}
\vspace{-0.5cm}
\end{figure}

The topological transformation protocol above can be applied for arbitrary knot lattice. For instance, the knot configuration of $2e/5$ is exactly coincide with the train tracks of $2e/5$. For a knot initially with five crossings in Fig. \ref{fracknot25} (a), two flipping operations are applied to the second and the fourth crossing to connect the decreet segments, transforming it into the minimal crossing state in Fig. \ref{fracknot13} (a). The same topological transformation acted on the spiral train track of $2e/5$ in Fig. \ref{fracknot13} (b) also results in the same minimum track pattern (Fig. \ref{fracknot13} (b)). When the spiral knot lattice is projected into classical train track pattern on the bottom plane (Fig. \ref{fracknot13} (c)), this flipping operation exchanged the two layers of train tracks in the back and that in the front of the flux dimer twice, keeping the track continuous and connected simultaneously. Similarly, the knot of $3e/5$ is reduced to the mirror image of the minimal crossing knot of $2e/5$ after the operation of three Majorana fermions, at the first, the third and the fourth crossing in Fig. \ref{fracknot25} (a). The two exemplar case of $e/3$ and $2e/5$ suggested the topological equivalent between knot and train track is an rigorous and universal relationship that holds for arbitrary fractional charges.

The topological correspondence between knot and train tracks as illustrated by the topological transformations above holds exactly for other serial of fractional charges.  The train track method, as the two dimensional projection of knot in three dimensions, must fulfil the mathematical requirement of avoiding self-crossing. While the knot lattice model is more convenient for constructing a two dimensional knot network than train tracks method as showed in knot lattice model \cite{{SiT2019}}. But the train track method has its own advantage in illustrating the fractional charges in two dimensional system, and provide another method for designing strongly correlated composite fermion. For example, the Laughlins wave function for N particles with a filling factor $\nu=1/(2m+1)$,
\begin{equation}\label{laughlins}
\psi(z_{i}) = \prod_{i<j}(z_{i}-z_{j})^{2m+1}e^{-\sum_{i=1}^{N}\frac{|Z_{i}|^{2}}{4l_{B}^{2}}},
\end{equation}
has a geometric interpretation in this train track theory. Suppose each of the two ending points of the train track of $e/3$ in Fig. \ref{fracknot13} (c) is attached by an electron. Three pairs of track segments must be exchanged to convert the minimal crossing state of $+1$ (Fig. \ref{fracknot13} (c)) to that of  $-1$ in Fig. \ref{fracknot23} (c). Every exchanging operation contributes a $-1$ due to the antisymmetric character of exchanging two electrons. From the picture of the knot for $e/3$ ($2e/3$) in Fig. \ref{fracknot13} (a) ( Fig. \ref{fracknot23} (a)), the three exchanging operations are inevitably in need of mapping the minimal knot with a positive crossing to that with a negative crossing, which offers topological interpretation for Laughlins wave function.

\begin{figure}
\begin{center}
\includegraphics[width=0.4\textwidth]{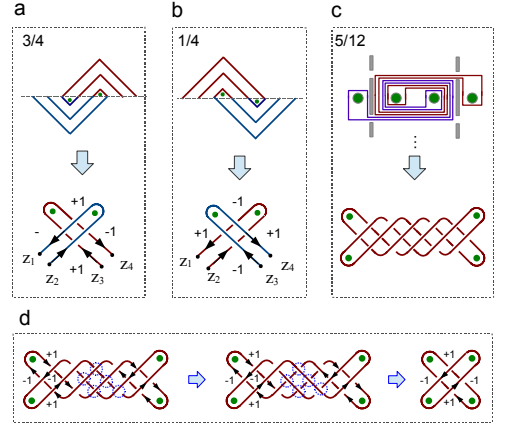}
\caption{\label{knot34} (a) The train tracks for the fractional charge of $3e/4$ and its corresponding knot configuration (b) The train track for $e/4$ and its corresponding knot. (c) The train track with fractional charge $Q = 5e/12$ and its corresponding knot pattern by braiding two current loops. (d) The explicit demonstration of flipping the five crossings (enclosed by the dashed blue circle) to bring the knot pattern back to the original $+1$ or $-1$ crossing state.}
\end{center}
\vspace{-0.5cm}
\end{figure}

\begin{figure}
\begin{center}
\includegraphics[width=0.45\textwidth]{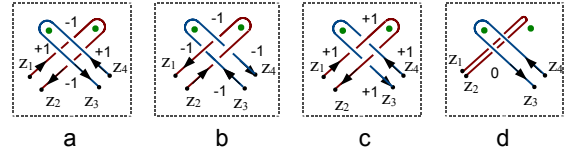}
\caption{\label{Pfaffian} The knot configuration with respect to the three terms in Pfaffian Eq. (\ref {eqpfaffian}) of four crossings, (a) the first term, (b) the second term, and (c) the third term. (d) The fusion of the two segments, $z_{1}$ and $z_{2}$.}
\end{center}
\vspace{-0.5cm}
\end{figure}

The wave function of fractional charges with even denominator can be visualized following the same topological transformation procedure above. For example, the knot with respect to $e/4$ or $3e/4$ is expanded from the winding train track of two double-line tracks as showed in Fig. \ref{knot34}(a)(b), each of them has four crossing sites. Each crossing site is characterized by a number that defines the chirality of crossing. The four ending points of the knot represent four electrons. Exchanging any pair of electrons at the crossing point maps a $+1$ to $-1$ (or vice versa). A collective wave function of this knot state can be constructed in a similar form as Moore-Read state for FQHE \cite{Moore1991},
\begin{equation}\label{moore}
\psi_{MR} = Pf(\frac{1}{z_{i}-z_{j}})\prod_{i<j}(z_{i}-z_{j})^{2m}.
\end{equation}
where $Pf(\frac{1}{z_{i}-z_{j}})$ is the Pfaffian. For the four-fermion system, the Pfaffian includes three antisymmetric terms,
\begin{eqnarray}\label{eqpfaffian}
Pf(\frac{1}{z_{i}-z_{j}})  &=&  \frac{1}{z_{1}-z_{2}} \frac{1}{z_{3}-z_{4}}+
\frac{1}{z_{1}-z_{3}}\frac{1}{z_{4}-z_{2}} \nonumber \\
&+&\frac{1}{z_{1}-z_{4}}\frac{1}{z_{2}-z_{3}}.
\end{eqnarray}
Pfaffian is the square root of determinant, $Pf(M)^2=\det(M)$. However here the Pfaffian equation for the knot configurations of two intersecting loops is summarized as a different form with the following three terms on the right hand side,
\begin{eqnarray}\label{knotfaffian}
Pf_{k}({z_{i}-z_{j}})  &=&  ({z_{1}-z_{2}})({z_{3}-z_{4}})+
({z_{1}-z_{3}})({z_{4}-z_{2}}) \nonumber \\
&+&({z_{1}-z_{4}})({z_{2}-z_{3}}).
\end{eqnarray}
The first term indicates the effect of exchanging the track segments $z_{1}$ and $z_{2}$ ( or $z_{3}$ and $z_{4}$ ) in Fig. \ref{Pfaffian} (a). The second term describe the knot after exchanging $z_{1}$ and $z_{3}$ ( or $z_{4}$ and $z_{2}$ ) (Fig. \ref{Pfaffian} (b)). The third knot in Fig. \ref{Pfaffian} (c) depicts exchanging $z_{1}$ and $z_{4}$ ( or $z_{2}$ and $z_{3}$). When the two track segments $z_{1}$ and $z_{2}$ ( or $z_{3}$ and $z_{4}$ ) are on the same side of the flux, they fuse into one complete segment and contract continuously to zero as Fig. \ref{Pfaffian} (d) illustrated. The same topological fusion process also holds for the other two terms in Eq. (\ref{knotfaffian}). The knot for other fractional charges with even denominator are constructed by the similar procedure. For instance, the knot pattern with respect to $Q = 5e/12$ is generated by braiding two free loop currents three times, each braiding generates four crossings accompanied by a mathematical constraint that single current loop is not twisted (Fig. \ref{knot34}(c)). It takes five crossing flipping (as labeled by the dashed circle) to bring the knot pattern back to the minimal double crossing of $+1$ state (Fig. \ref{knot34}(d)). Therefore the collective wave function of $Q = 5e/12$ can be expressed as
\begin{equation}\label{512}
\psi(z_{i}) = Pf_{k}({z_{i}-z_{j}}) e^{-\sum_{i=1}^{12}\frac{|Z_{i}|^{2}}{4l_{B}^{2}}}.
\end{equation}
This topological transformation protocol can extend to other filling fractions in a straight forward way, providing a new way of constructing collective wave function of strongly correlated electrons in two dimensional system.

\subsection{The irrational charges around flux trimer}

\begin{figure}
\begin{center}
\includegraphics[width=0.45\textwidth]{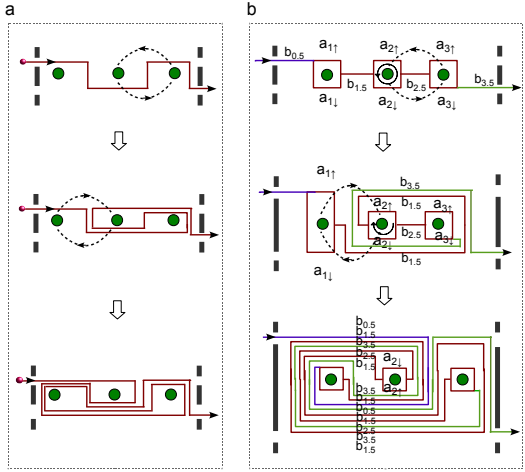}
\caption{\label{trimer} (a) The stacked eclectic current configuration under a pair of braiding operations upon the magnetic flux trimer, i.e., $\sigma^{2}_{(1,2);\circlearrowright}(s_{2})\sigma^{2}_{(2,3);\circlearrowleft}(s_{1})$. (b) The stacked electric currents under the braiding operations $\sigma_{(12);\circlearrowleft}(t_{2})\sigma_{(23);\circlearrowright}(t_{1})$ upon a general currents configuration around the flux trimer.}
\end{center}
\vspace{-0.5cm}
\end{figure}

The train tracks around a flux dimer have output all of the observed fractional charges in FQHE and predicted new fractional serial charges. Thus the two dimensional electron gas in strong magnetic field can be viewed as a gas of free flux dimers surrounded by train track pattern of electron path. However, the FQHE does not exclude the possibility of train tracks around three fluxes. If the finite braiding upon flux dimers above are extended to three magnetic fluxes, for instance, braiding three nearest neighboring fluxes, No. 1, No. 2 and No. 3 in Fig. \ref{trimer}, more complex serial of fractional charges can be produced by combinatoric braiding operations. For example, the single train track that winds around three fluxes showed in Fig. \ref{trimer} (a) are acted by two braiding operations. We first exchange the flux pair $[2,3]$ twice in counterclockwise direction $\sigma^{2}_{(2,3);\circlearrowleft}$, and then exchange the flux pair $[1,2]$ twice in clockwise direction, $\sigma^{2}_{(1,2);\circlearrowright}$. This braiding serial operator, i.e.,
$\hat{B}_{s} = \sigma^{2}_{(1,2);\circlearrowright}\sigma^{2}_{(2,3);\circlearrowleft}$, maps the initial fractional charge serial into a new fractional charge distribution, and keeps the order of three fluxes invariant simultaneously. This leads to a distribution of fractional charges around the three fluxes,
\begin{eqnarray}\label{newQ1231}
Q_{1,\uparrow} &=& \frac{a_{1,\uparrow}}{a_{1,\uparrow} +a_{1,\downarrow} }=\frac{3}{5},\;\;Q_{1,\downarrow} = \frac{a_{1,\downarrow}}{a_{1,\uparrow} +a_{1,\downarrow} }=\frac{2}{5}, \nonumber \\
Q_{2,\uparrow} &=& \frac{1}{5},\;\;Q_{2,\downarrow} =\frac{4}{5}, \;\; Q_{3,\uparrow} =\frac{2}{3},\;\;Q_{3,\downarrow} = \frac{1}{3}.
\end{eqnarray}
Repeating the braiding string operator $\hat{B}_{s}$ on the the charge distribution above maps it to a new charge distribution. As showed in the flux dimer case, the braiding operation modifies the value of effective magnetic field, the combinatoric braiding operator, $\hat{B}_{s} = \sigma^{2}_{(1,2);\circlearrowright}\sigma^{2}_{(2,3);\circlearrowleft}$, indicates the local change of magnetic field strength, it is visualized as two magnetic dipole pointed in antiparallel direction. The change of magnetic field strength drives the electron to redistribute around the flux trimer until it reaches a stable charge distribution.

Every string operator of braiding operations defines a special dancing route for electron. The combined braiding, $\hat{B}_{s} = \sigma^{2}_{(1,2);\circlearrowright}\sigma^{2}_{(2,3);\circlearrowleft}$, leads to irrational charges around the fluxes in the limit of infinite number of actions. We take the fused current tracks as a general initial track distribution. The label $ a_{i,\uparrow}$ (or $ a_{i,\downarrow}$) labels the number of stacked currents above (or below) the $i$th magnetic flux, while $ b_{i.5}$ counts the number of stacked bond current that bridges the flux pair $[i,i+1]$ (Fig. \ref{trimer} (a)). After the action of braiding operator, $\hat{B}_{s} = \sigma^{2}_{(1,2);\circlearrowright}\sigma^{2}_{(2,3);\circlearrowleft}$, the new current stack numbers above (or below) a magnetic flux at time $t_{m}$ can be expressed as the linear combination of that at $t_{m-1}$, which obeys an iterative equation,
\begin{eqnarray}\label{a123m}
a_{1,\uparrow} (t_{m})&=& a_{1,\uparrow}(t_{m-1}) +b_{1.5}(t_{m-1})  +b_{2.5}(t_{m-1})\;\;\nonumber \\
&+&b_{0.5}(t_{m-1})\nonumber \\
a_{1,\downarrow}(t_{m})&=&a_{1,\downarrow}(t_{m-1}) +2b_{1.5} (t_{m-1}) +b_{2.5}(t_{m-1}), \nonumber \\
&+&b_{0.5}(t_{m-1})+2b_{3.5}(t_{m-1})\nonumber \\
a_{2,\uparrow}(t_{m})&=& a_{2,\downarrow}(t_{m-1}) +b_{1.5} (t_{m-1}) +b_{2.5}(t_{m-1}), \;\;\nonumber \\
&+&b_{0.5}(t_{m-1})+b_{3.5}(t_{m-1})\nonumber \\
a_{2,\downarrow} (t_{m})&=& a_{2,\uparrow}(t_{m-1}) +3b_{1.5}(t_{m-1})  +b_{2.5}(t_{m-1}), \nonumber \\
&+&b_{0.5}(t_{m-1})+3b_{3.5}(t_{m-1})\nonumber \\
a_{3,\uparrow}(t_{m})&=& a_{3,\uparrow}(t_{m-1}) +b_{1.5}(t_{m-1})+b_{3.5}(t_{m-1})\nonumber \\
a_{3,\downarrow}(t_{m}) &=& a_{3,\downarrow}(t_{m-1}) +b_{1.5}(t_{m-1})+b_{3.5}(t_{m-1}).
\end{eqnarray}
The bond track number between two neighboring fluxes obeys the equation,
\begin{eqnarray}\label{b1235m}
b_{1.5}(t_{m})&=& 5 b_{1.5} (t_{m-1}) + 2 b_{2.5}(t_{m-1})\nonumber \\
&+&2b_{0.5}(t_{m-1})+4b_{3.5}(t_{m-1}),\nonumber \\
b_{2.5}(t_{m}) &=& 2 b_{1.5} (t_{m-1}) + b_{2.5}(t_{m-1})+2b_{3.5}(t_{m-1}),\nonumber \\
b_{0.5}(t_{m}) &=& b_{0.5}(t_{m-1}), \;\;\; b_{3.5}(t_{m}) = b_{3.5}(t_{m-1}),
\end{eqnarray}
where $b_{0.5}$ and $b_{3.5}$ are not variables, they are constant under arbitrary braiding and has no influence on the track redistribution. Thus we set them as zero for simplicity, i.e. $b_{0.5} = b_{3.5} = 0$. The iterative equations above transform into differential equations when the time step approaches to infinitesimal value. According to differential equation theory, these track distribution variables would finally converge to a stable value. The eigenvectors of the three current stacks above (or below) the fluxes and the two bond track numbers are listed as following:
\begin{eqnarray}\label{abb}
(a_{3,\uparrow}, b_{1.5} , b_{2.5}) &=& (a_{3,\downarrow}, b_{1.5} , b_{2.5}) = ({1}/{2}, 1+\sqrt{2}, 1),\nonumber \\
(a_{2,\uparrow}, a_{2,\downarrow}, b_{1.5}, b_{2.5}) &=&(\frac{2+\sqrt{2}}{4},  \frac{2+3\sqrt{2}}{4}, 1+\sqrt{2},1),\nonumber \\
(a_{1,\uparrow}, b_{1.5} , b_{2.5}) &=& ({1}/{\sqrt{2}}, 1+\sqrt{2}, 1),\nonumber \\
(a_{1,\downarrow}, b_{1.5}, b_{2.5} ) &=& (\frac{1+\sqrt{2}}{2}, 1+\sqrt{2}, 1),
\end{eqnarray}
Here we have eliminated some trivial solutions, in which $b_{1.5} = b_{2.5} =0$. These eigenvectors determines the fractional charge distribution around the three fluxes,
\begin{eqnarray}\label{Q123bb}
Q_{1,\uparrow} &=& Q_{1,\downarrow} = \frac{1}{2},\;\;\;Q_{1,\uparrow}  =\frac{2}{3+\sqrt{2}},\;\;
Q_{1,\downarrow} = \frac{1+\sqrt{2}}{3+\sqrt{2}}, \nonumber \\
Q_{2,\uparrow} &=&  \frac{2+\sqrt{2}}{4(1+\sqrt{2})},\;\; Q_{2,\downarrow} = \frac{2+3\sqrt{2}}{4(1+\sqrt{2})}, \nonumber \\
Q_{3,\uparrow} &=& Q_{3,\downarrow} = \frac{1}{2},
\end{eqnarray}
The fused bond tracks between flux No. 1 and No. 2 also carries irrational charges, $Q_{b1.5} = 1+\sqrt{2}.$ The resistance between flux No. 1 and No. 2 is
\begin{eqnarray}\label{Rx2}
R_{1.5} = R_{0} \frac{1}{1+\sqrt{2}}.
\end{eqnarray}
The Hall resistance around flux No. 1 is determined by
$Q_{1,\downarrow} = Q_{1,\leftarrow} =\frac{1+\sqrt{2}}{3+\sqrt{2}}$,
\begin{eqnarray}\label{Rx3}
R_{H1} = R_{0} \frac{3+\sqrt{2}}{1+\sqrt{2}}.
\end{eqnarray}
The charges around flux No. 2 is the resultant charge of track fusion and splitting. When these train tracks are mapped into knot lattice, it expands a complicate network. However, a measurement of the local resistance around flux No. 2 sill yield irrational conductance that is proportional to the inverse of charges in Eq. (\ref{Q123bb}).

The exemplar braiding string operator above can be generalized to other combinatoric sequence to generates other irrational charges and fractional charges before the system reaches a stable state. From physics point of view, the anisotropic combinatoric braiding sequence can be implemented by anisotropic magnetic field distribution at the spatial scale of elementary magnetic flux quanta. That is partially beyond the current technology level but is still promising for future development.

\section{ Fractional charge current in one dimensional lattice of magnetic fluxes}

\subsection{ Fractional charges of braided currents with space translational symmetry }

\begin{figure}
\begin{center}
\includegraphics[width=0.48\textwidth]{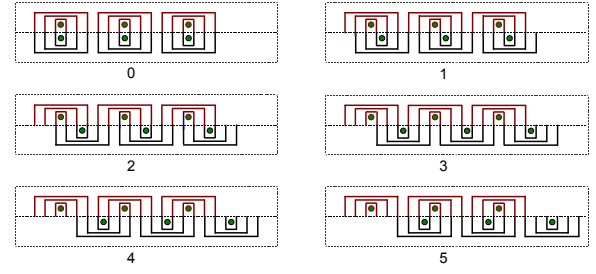}
\caption{\label{transymmetry} (0) A chain of flux dimer oriented in the vertical direction. Each flux is surrounded by three track loops. (1-5) The flux dimer is cut along their border line to generate two chains of semicircle around single flux. The upper semi-circles are collectively translated to the left by $p$ (p=1,2,3.4,5) steps and reunite with new contacting points. This generated different train track patterns under different braiding modes.}
\end{center}
\vspace{-0.5cm}
\end{figure}

The braided electric current around one flux pair demonstrates the main character of FQHE. Similar to the composite fermion theory of FQHE, the two dimensional electron gas in strong magnetic field can be well approximated by a gas of weakly connected train tracks around flux pairs. When many flux pairs are confined in a one dimensional chain with strong neighboring interactions, we apply the topological surgery method and translation operation to construct an one dimensional quantum gas of flux pair. The initial state of the dimensional gas of fractionally charged anyons is set as concentric current circles around flux pairs arranged in vertical direction, which represents the integral filling state. The integral charge state $\nu=1$ is represented by two layers of concentric circles and $\nu=2$ is illustrated by three layers of concentric circles in Fig. \ref{transymmetry}-(0). The concentric circles are first cut into two sets of concentric semicircles around single flux along the borderline between the two fluxes. The set above the cutting line is dyed blue, while the other set is dyed red in Fig. \ref{transymmetry}. The cutting points are placed on a regular one dimensional lattice with the lattice spacing equals to the distance between the nearest neighboring circles. The semicircles above the cutting line together with the flux they surround are translated to the right hand side by one step and then docked the cutting points in their new locations with the semicircles below (Fig. \ref{transymmetry}-(1)). This operation fuses isolated circles into a continuous train track of anyon with $3e/5$ that is periodically distributed along the cutting line. This translation operation is equivalent to braiding a simple track twice in clockwise direction periodically. Further more, two steps of translation generates a one dimensional lattice of fractional charge $3e/4$, representing by the braiding the double-line around flux dimer in clockwise direction (Fig. \ref{transymmetry}-(2)). Three steps of translation generates three layers of open track winding through the one dimensional flux lattice (Fig. \ref{transymmetry}-(3)), indicating a conducting state of integral charges. While four steps of translation drives the train track back to the fractional charge $e/4$, which matches the tracks generated by braided the double line around flux dimer in counterclockwise direction (Fig. \ref{transymmetry}-(4)). Five steps of translation generates the mirror image of that of one-step translation (Fig. \ref{transymmetry}-(5)). The translation operation showed by the exemplar train track pattern above offers an explicit protocol for topological paths fusion and applies for other fractional charge serial.

\begin{figure}
\begin{center}
\includegraphics[width=0.45\textwidth]{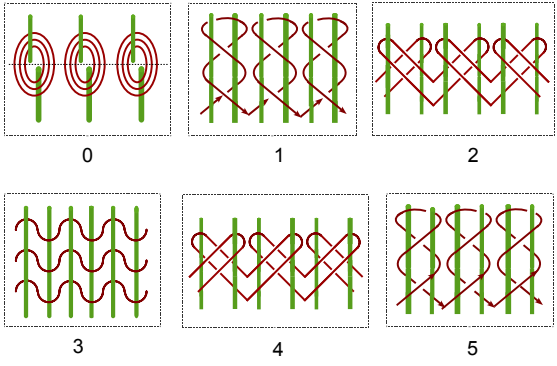}
\caption{\label{1+1Dknot} The two dimensional knot lattice as an expansion of the one dimensional lattice of train tracks by $p$ (p=1,2,3.4,5) steps of translation operations in Fig.\ref{transymmetry}.}
\end{center}
\vspace{-0.5cm}
\end{figure}

As showed before the winding train track around one flux dimer is the projection of an one dimensional knot lattice, similarly the one dimensional lattice of winding train tracks pattern is also the projection of a two dimensional knot lattice. For instance, the initial train tracks of integral filling state in Fig. \ref{transymmetry}-(0) is the projection of dimer current lattice in Fig. \ref{1+1Dknot}-(0). The one dimensional lattice of $2e/5$ corresponds to the two dimensional knot lattice in Fig. \ref{1+1Dknot}-(1), whose mirror image is the one dimensional lattice of $3e/5$ as showed in Fig. \ref{1+1Dknot}-(5). The train tracks generated by braiding dimer current is the projection of double-line knot lattice in Fig. \ref{1+1Dknot}-$(2) \& (4)$. While the train tracks of three layers of simple train track also corresponds to three continuous simple current in two dimensional knot lattice (Fig. \ref{1+1Dknot}-$(3)$). The weak interaction between neighboring fractional charges $2e/5$ is clearly illustrated by the single connection at the bottom of Fig. \ref{1+1Dknot}-(1), so does the fractional charge $3e/4$ as showed in Fig. \ref{1+1Dknot}-$(2)(4)$.

Every semi-circle arc connecting two cutting points is equivalent to a directed line in Feynman diagram, which indicates the generation of an electron at one point and its annihilation at the other point. The train track can be viewed as the continuous connection of a serial of  discrete Feynman diagrams in space. Thus each train track pattern in Fig. \ref{transymmetry} can be well described by Hamiltonian. For instance, the Hamiltonian equation for Fig. \ref{transymmetry} - (0) reads,
\begin{eqnarray}\label{Hamitrans0}
H_{0}&=& \sum_{r,j}t\;c^{\dag}_{r2(n+1)+(j+\frac{1}{2})}c_{r2(n+1)+(j-\frac{1}{2})}e^{i{\phi_{r2(n+1),\uparrow}}} \nonumber\\
&+& \sum_{r,j}t\;c^{\dag}_{r2(n+1)+(j-\frac{1}{2})}c_{r2(n+1)+(j+\frac{1}{2})}e^{-i{\phi_{r2(n+1),\downarrow}}}\nonumber\\
&+&h.c.
\end{eqnarray}
where $n$ is the integral filling factor of isolated concentric circles before the topological path fusion. $h.c.$ represents the Hermitian conjugate of the terms above. $t$ is the hopping rate coefficient. $j[<2(n+1)]$ labels the location of cutting points between the nearest neighboring fluxes. This Hamiltonian represents the train tracks with an integral filling factor $\nu = n$. The Hamiltonian $H_{0}$ is the initial state for translation operations. The translation operator $\hat{T}_{p}$ maps the location index of all cutting points $p$-steps forward in the right hand side, i.e.,
\begin{eqnarray}\label{HatTp}
\hat{T}_{p}c^{\dag}_{r}=c^{\dag}_{r+p},\;\;\;\hat{T}_{p}c_{r}=c_{r+p},\;\;\;\hat{T}_{p}\phi_{r}=\phi_{r+p}.
\end{eqnarray}
When $\hat{T}_{p}$ acts only on the red semi-circle arcs below the cutting line, it generates the Hamiltonian for fractionally charged anyons. For example, one-step translation to the right hand side derives the Hamiltonian,
\begin{eqnarray}\label{Hamitrans1}
H_{1}&=& \hat{T}_{1}H_{0}\nonumber\\
&=&\sum_{r,j}t\;c^{\dag}_{r2(n+1)+(j+\frac{1}{2})}c_{r2(n+1)+(j-\frac{1}{2})}e^{i{\phi{r2(n+1),\uparrow}}} \nonumber\\
&+& \sum_{r,j}t\;c^{\dag}_{r2(n+1)+(j+\frac{1}{2})}c_{r2(n+1)+(j+\frac{3}{2})}e^{-i{\phi({r2(n+1)+1,\downarrow})}}\nonumber\\
&+&h.c.
\end{eqnarray}
This Hamiltonian describes a chain of $3e/5$ charged anyons for n = 2. It is equivalent to braiding the neighboring flux pairs twice in clockwise direction. For a general case, p-step translation generates the Hamiltonian for other train tracks in Fig. \ref{transymmetry}, $H_{p}= \hat{T}_{p}H_{0}$,
\begin{eqnarray}\label{Hamitransp}
H_{p}&=& \sum_{r,j}t\;c^{\dag}_{r2n+j+\frac{5}{2}}c_{r2n+j+\frac{3}{2}}e^{i{\phi{r2(n+1),\uparrow}}} \nonumber\\
&+& \sum_{r,j}t\;c^{\dag}_{r2n+j+\frac{3}{2})+p}c_{r2n+j+\frac{5}{2}+p}e^{-i{\phi({r2(n+1)+p,\downarrow})}}\nonumber\\
&+&h.c.
\end{eqnarray}
The Hamiltonian has translation symmetry with respect to the major index $r$, which labels the periodic distribution of fluxes. The Fourier transformation applies for the index $r$ but not for the inner index $j$, because there is no translational symmetry for different semi-circle layers. This Hamiltonian describes the motion of electron in flux lattice before the stacked track layers fuse into one bundle. When the track stack around each flux are confined into one bundle, the internal index $j$ that labels the different layers in the hopping operator becomes redundant and could be removed out of the Hamiltonian. While the phase factor after track fusion is now fractional number, which counts the fractional fluxes enveloped by the square track with anisotropic weight. As a result, the Hamiltonian for periodical train tracks could be approximated by the
Harper Hamiltonian for one dimensional flux lattice \cite{Harper1955},
\begin{eqnarray}\label{harperHami}
H_{hp}=\sum_{r,j}t_x\;c^{\dag}_{r+1,j}c_{r,j}+t_y\;c^{\dag}_{r,j+1}c_{r,j}e^{-i2\pi{\phi}r}
+h.c.
\end{eqnarray}
The rational number ${\phi}$ in Harper Hamiltonian is essentially the fractional filling factor $\nu$ of FQHE. Before the track fusion, a moving electron along these winding paths feels an integral flux. The electric currents in the nearest neighboring tracks are always pointed in opposite direction. As all know, two parallel currents in opposite direction repel each other, otherwise they attract each other. Thus track fusion drives the system into a low energy state. The electron in the fused tracks flow in the same direction by splitting itself according to the weight distribution on the four edges around the flux. This phenomena can be equivalently realized by cutting the integral flux into fractional flux and keeping the electron around the four edges as an integral charge. As a result, a fractional flux is raised up to the phase factor of hopping terms in Harper Hamiltonian. In another case, if the flux number ${\phi}$ in Harper Hamiltonian is not a rational number, it extends into the Hofstadter model \cite{Hofstadter1976}, in which the energy spectrum shows a fractal structure. According to the correspondence theory between knot on torus and train track around flux pair in the first section, a rational filling factor leads to a closed curve on torus and a well connected winding track around flux pair, while an irrational filling factor leads to an endless open trace on torus and winding track generated by a braiding operation over an arbitrary angle $\theta\neq{n\pi},n = 1,2,\cdots$. Therefore, the fractal structure in Hofstadter model corresponds to the chaotic pattern of electron path on torus and the incomplete winding track under an irrational angle of rotation. These chaotic path results in the classical transportation behavior of electron, while those complete knotted path lead to the Hall plateau in FQHE.

\subsection{Fractional charges of the braided loop currents around magnetic flux clusters}

\begin{figure}
\begin{center}
\includegraphics[width=0.4\textwidth]{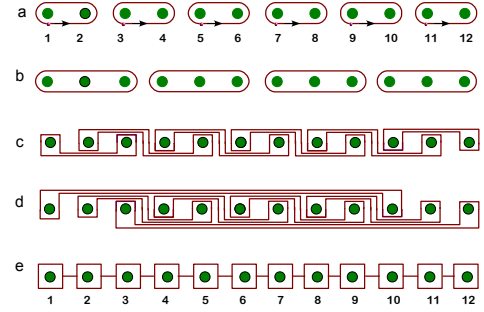}
\caption{\label{dimerc}(a) A chain of flux dimers, each of which is surrounded by an electric current loop (represented by the red circle). (b) The chain of flux trimers. Every electric loop current enclose three fluxes. (c) The stack electric currents around the lattice of magnetic fluxes generated by one counter-clockwise braiding over the nearest neighboring fluxes. (d) The train track generated by one more braiding counterclockwise braiding over the first and the last flux in the lattice above. (e) A scheme representation of the most general current track distribution around the one dimensional flux lattice. Each edge are assigned with different weight, which counts the number of track that fused there.}
\end{center}
\vspace{-0.5cm}
\end{figure}

Besides the minimal closed current that winds around two nearest neighboring fluxes, a moving electron can also be captured by a cluster of fluxes that enclosed more than two fluxes to form a closed electric loop. In an one dimensional lattice of magnetic fluxes, in which every two nearest neighboring magnetic fluxes combines together as a dimer (Fig. \ref{dimerc} (a)), braiding the two magnetic flux inside a dimer does not change the energy of flux lattice. The free dimer state is an insulating state, because electrons are confined in local flux pair, failed to move freely across whole space. When the flux dimers covering the whole one dimensional lattice are braided synchronously in the clockwise direction, it reproduces the same train tracks as that generated by two steps of translation in Fig. \ref{transymmetry} - (2). This braiding operation over the nearest neighboring dimers transforms the insulating state into a conducting state of fractional charges, generated many parallel (or anti-parallel) electric currents around fluxes and provided effective transportation channel for electrons after track fusion. According to electromagnetic field theory, parallel currents in the same orientation attract each other, while anti-parallel currents repel each other. The total energy of the initial free dimer lattice of minimal loops (Fig. \ref{dimerc} (a)) is the sum of anti-parallel current segments in the lattice,
\begin{eqnarray}\label{Hami02}
H_{0,[2]}=\sum_{i=1}^{6}\frac{-\mu_{0}}{2\pi}l_b\ln(d)I_{i+1/2,j}I_{i+1/2,j+1},
\end{eqnarray}
Here $l_b$ is the lattice spacing between two neighboring fluxes. If the finite flux lattice is grouped into triple clusters (Fig. \ref{dimerc} (b)), the total energy is higher than the dimer covering state, because the summation of energy terms grows up from six to eight.
\begin{eqnarray}\label{Hami03}
H_{0,[3]}=\sum_{i=1}^{8}\frac{-\mu_{0}}{2\pi}l_b\ln(d)I_{i+1/2,j}I_{i+1/2,j+1}.
\end{eqnarray}
Therefore the lattice of loop currents enclosing larger flux cluster has higher energy. However the neighboring fluxes that is bridged over by many pairs of parallel currents has lower energy than the unbraided loop dimer. Therefore, braiding operation drives the system to a state with lower energy. Since a multi-layer track stack with even number of electric current tracks always has a higher potential energy. The electrons on the bond prefer to gather around the core region of magnetic fluxes which admits a pair of odd number of current tracks (Fig. \ref{dimerc} c). For instance, after one braiding over magnetic flux pair $(2;3)$, there are one current track above the magnetic flux No, 2 and three current tracks below. An inverse distribution is gathered around magnetic flux No. 3 (Fig. \ref{dimerc} (c)). This braiding maps the high energy function of free dimer state into a lower energy state (Fig. \ref{dimerc} (c)) with following Hamiltonian,
 \begin{eqnarray}\label{Hami1}
H_{1}= - \frac{\mu_{0}}{2\pi}l_b\sum_{r>l=1;i}^{4}(-1)^{r-l}\ln[(r-l)d](I_{i,l}I_{i,r}).
\end{eqnarray}
Braiding operation is an effective way of driving the train tracks to the ground state.

The current channel can be viewed as the central orbital along which the probability distribution of electron cloud reaches the maximal value. Since an electron in one loop is not distinguishable from the one in its neighboring loop, the measurement of electron probability should count every track around the flux. Therefore, this splitting current tracks in fact divide one charge into four possible tracks, each track carries a fractional charge. The upper charge is $Q_{2,\uparrow}$  = 1/4 and the down charge is the sum of three fractional charges, $Q_{2\downarrow}$  = 3/4. Since there is no new charge generated or annihilated, the total electric charge must be conserved, this conservational law imply a topological invariant equation, $ Q_{i\uparrow} + Q_{i,\downarrow}  =  Q_{(i+j)/2}. $Here (i,j) is a pair of nearest neighboring magnetic fluxes upon which the braiding operation is performed. A braiding operator always contributes a pair of anti-parallel current to the bond, while an inverse braiding cancels a pair of anti-parallel currents. For a sequence of braiding operations,
\begin{eqnarray}\label{bra}
 B_{ij,n} = \sigma_{ij} (t_1)\sigma_{ij}^{-1}(t_{2})...\sigma_{ij}(t_{n}) = \sigma_{ij}^{p},
\end{eqnarray}
 the ultimate index of braiding operator $p$ multiply by two equals to the denominator in the fractional charge equation with even denominator, i.e, $Q=(m\pm1)/2m$, $m=2p$. Here $t_{i}$ indicates the time when the braiding operation is performed.  $\sigma_{ij}$ is the braiding operator on (ij) with $\sigma_{ij}^{-1} $ as its inverse operator, and their product fuses into unity,   $\sigma_{ij}\sigma_{ij}^{-1} =1$.

The braiding operation could be performed on two magnetic fluxes far separated, it induces a different fractional charge sequence from that generated by braiding the nearest neighboring flux pairs. For example, braiding the first magnetic flux and the last magnetic flux generates a long range current bridge covering the whole chain (Fig. \ref{dimerc} (d)). This newly added current reset the charge splitting ratio around each magnetic flux core (Fig. \ref{dimerc} (d)). The upper fractional charge at the second magnetic flux is increased to $Q_{2,\uparrow}$  = 3/4 and the down charge is reduced to $Q_{2,\downarrow}$  = 1/4. While the fractional charge at the 3rd magnetic flux becomes ($Q_{3,\uparrow}$ = 5/6, $Q_{3,\downarrow}$ = 1/6). While the flux covering the lattice sites from $i=4$ to $i=9$ are sandwiched in between two fractional charges ($Q_{i,\uparrow}$ = 3/8, $Q_{i,\downarrow}$ = 5/8) (Fig. \ref{dimerc} (d)). The denominator of these fractional charge after braiding is still an even number which counts the number of currents on the bond between two neighboring fluxes. This exemplar braiding operation suggests long range braiding causes charge fluctuation in large spatial scale. The spatial range of topological correlation is proportional to the distance between the most far separated two fluxes in the braiding operator. Every long range braiding operator can be exactly express as the product of braiding operators over the nearest neighboring fluxes. In order to map the initial ordering of magnetic fluxes, $[1,2,3,...,12]$, to the final ordering $[12,2,3,...,1]$, we first bring the first flux to the last position, $[2,3,...,12,1]$, then bring the 12 flux to the first position, $[12,2,3,...,11,1]$. This mapping can be realized by following mapping sequence,
\begin{eqnarray}\label{bralong}
 B_{1,12}& =& \sigma_{1,2}(t_1)\sigma_{2,3}(t_{2})\cdots\sigma_{11,12}(t_{12})\sigma_{10,11}(t_{13})  \nonumber\\
&&\sigma_{9,10}(t_{14}) \sigma_{8,9}(t_{15}) \cdots \sigma_{1,2}(t_{22}).
\end{eqnarray}
This mapping sequence generates the same topological pattern of current tracks as that under one straightforward operation between flux [1,12]. Note here the operation in the sequence above must be kept in the same direction (either clockwise or counterclockwise) as that of one step braiding $ B_{1,12}$. Even though replacing one braid operator by its inverse operator also exchange the position of the nearest neighbors, it would map into a completely different topological pattern that mismatches the output of $B_{1,12}$.

Braiding string operator can be implemented by serial manipulation of local magnetic field strength. Thus every electromagnetic energy corresponds to the eigen-energy of certain magnetic field strength. The initial free dimer state is the highest excited state which has the maximal eigen-energy. The second highest excited state is generated by only one braiding upon only one pair of the nearest dimers loops  (Fig. \ref{dimerc} b). For a magnetic flux chain of $N$ dimers,  the second highest excited has $(N-1)$ fold degeneracy. The topological eigenstate shows only a pair of local fractional charged states, ($Q_{u,i}$ = 1/4, $Q_{d,i}$ = 3/4) and  ($Q_{u,j}$ = 3/4, $Q_{d,j}$ = 1/4). While other unbraided dimer loops remains half-charged state.  The third highest excited state are created by two continuous braiding operations over the same pair of dimer loops, which generates fractional charges, ($Q_{i,\uparrow}$ = 3/8, $Q_{i,\downarrow}$ = 5/8) or ($Q_{i,\downarrow}$ = 3/8, $Q_{i,\uparrow}$ = 5/8). The example above suggested that fractional charges are only generated by finite number of braiding. In the case of infinite number of random braiding over the whole flux lattice, every flux is wrapped by large number of current segments. The total number of upper currents almost equal to that of down currents around each flux, correspondingly one electron splits into two approximately equal charges after track fusion. Therefore, those fractional charged states finally converge to half charged state, which is exactly the ground state of this loop dimer chain model. Note that this half-charge state is a conducting half-charges state, which is completely different from initial free dimer state, where electrons are localized around the flux pair and showed an insulating state. From the point view of energy, the total electromagnetic energy of current stacks in the flux lattice would finally reach a fixed point as suggested by the simplest energy sequence of braiding one flux pair (Fig. \ref{energy}). That fixed point is the ground state of this loop dimer chain model. In practical physical system, it is a technological challenge to manipulate single magnetic flux, but electrons could move around magnetic flux in different possible paths. It is always possible to find a winding path of electron that matches certain track generated by braiding fluxes. At finite temperature, the reduced mean free path of electron limited the maximal length of the winding path and exclude the existence probability of fractional charges. As a result, the partition function term with respect to the winding path in Fig. (\ref{dimerc} (c)) is computed by substituting energy function equation (\ref{Hami02}) into partition function,
\begin{eqnarray}\label{partition1}
 Z_{1}& =&\exp[\frac{\mu_{0}a}{2\pi{k_{b}T}}][d^{6}\prod_{r>l=1}^{4}[(r-l)d]^{(-1)^{r-l}}].
\end{eqnarray}
The denominator and numerator in the fraction formulation of fractional charges now appears as the upper limit of the product equation in partition function terms above. The Partition function of a general winding path in one dimensional lattice of fluxes can be computed along the same route above.

\begin{figure}
\begin{center}
\includegraphics[width=0.4\textwidth]{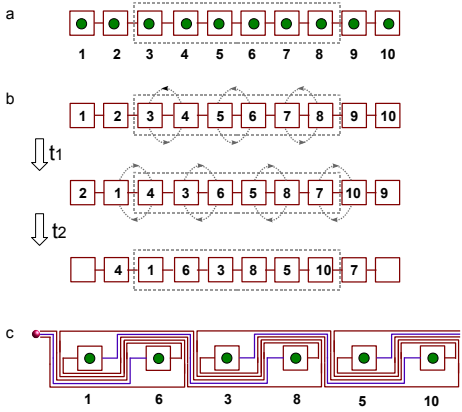}
\caption{\label{tripletrain} (a) A general train track distribution in one dimensional lattice as the initial state of braiding operations. (b) A collective braiding in counterclockwise direction is first performed(as labeled by $t_{1}$), followed by a collective clockwise braiding as labeled by $t_{2}$. (c) The resultant train track pattern generated by the combined braiding above. }
\end{center}
\vspace{-0.5cm}
\end{figure}

Homogeneous braiding operations indicates a homogeneous application of an effective magnetic field. If braiding operations are performed in clockwise and counterclockwise directions alternatively, it is equivalent to the action of an alternative magnetic field. We first perform a counterclockwise braiding over the flux lattice in Fig. \ref{tripletrain} (a) which corresponds to the fused train tracks after braiding the flux dimer lattice in Fig. \ref{dimerc} (e). Then translate the braiding centers to the left by one step and perform a clockwise braiding in their new locations in Fig. \ref{tripletrain} (b). Repeating this hybrid braiding operation on the flux lattice after infinite number of times generates a train track pattern carrying irrational charges as that showed in Eq. (\ref{Q123bb}). Fig. \ref{tripletrain} (c) shows the winding track pattern of fractional charge states before it reaches the ground state.

\section{Fractional charges in two dimensional lattice of magnetic fluxes}

\subsection{Fractional charges generated by periodical braiding of two dimensional flux lattice}

When the radial center of magnetic flux tubes are placed on a two dimensional square lattice, the electron goes around the pair of flux tubes along double helix paths that expands into three dimensional space, depicting a three dimensional knot lattice. The projection of these knot curves in three dimensional multi-connected domain to its two dimensional boundary surface reduced many degrees of freedom. Each double helix path can be represented by a collective state of Ising spin along the flux tube \cite{{SiT2019}}. For example, for the knot showed in Fig. \ref{1+1Dknot1} (c), the knot state $|\psi\rangle=(-1,-1,0,-1,0)$ project the same train track curve to two dimensional boundary as $|\psi\rangle=(-1,-1,-1,0,0)$ and $|\psi\rangle=(-1, 0, -1, 0, -1)$, and so on. However the topology of the knot curve in three dimensions are preserved in the two dimensional boundary.

\begin{figure}
\begin{center}
\includegraphics[width=0.48\textwidth]{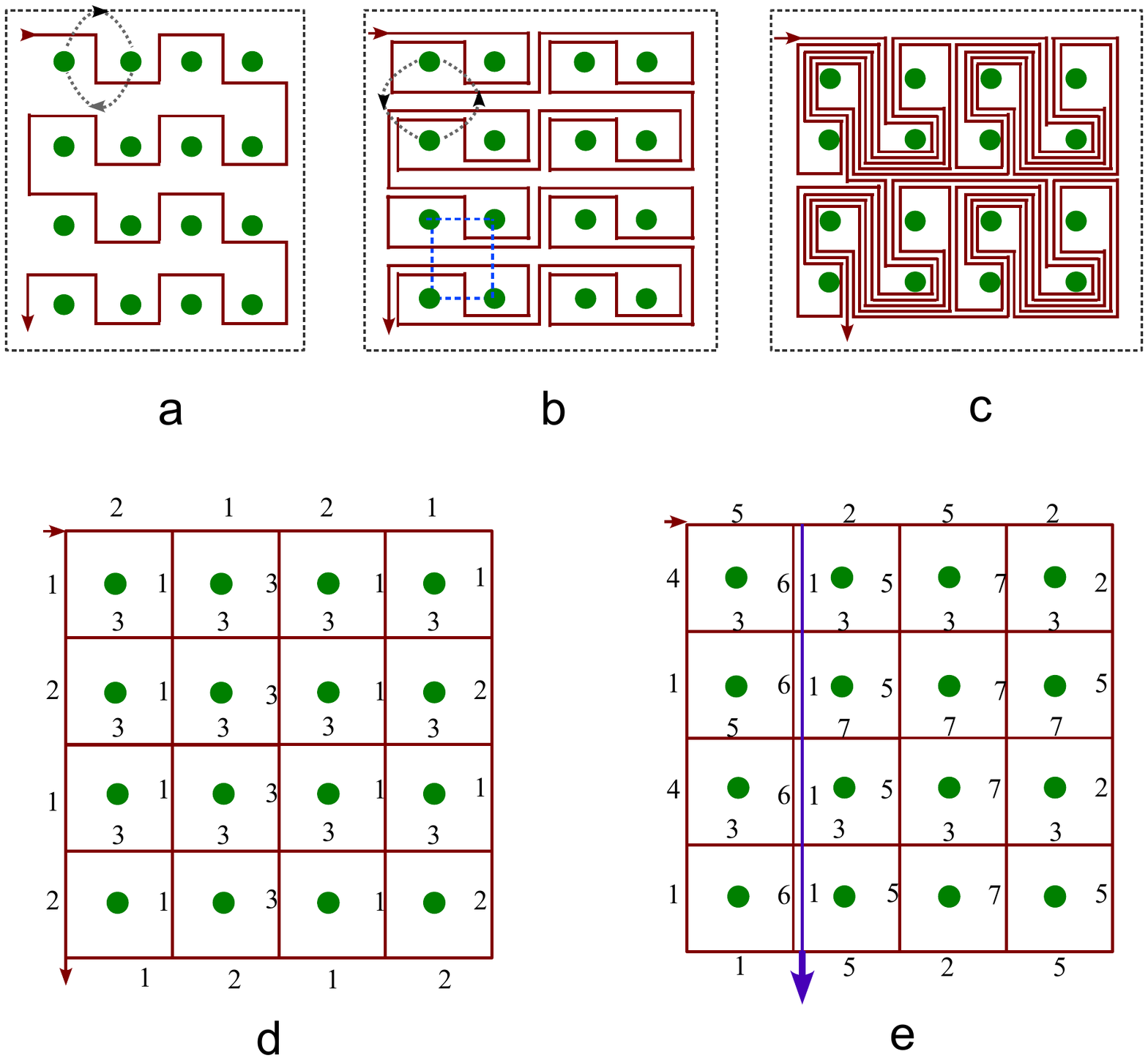}
\caption{\label{2Dfluxlattice} (a) An electron passes through the two dimensional flux lattice following semi-circle tracks with opposite chirality. (b) The train track pattern generated by a collective braiding over flux pairs in horizontal direction. (c) Take the train track in (b) as initial state, another collective braiding on flux pairs in the vertical direction is performed, it generates a multilayered train tracks that periodically distributed in two dimensional square lattice. (d) The number of layers of track segments on the edges around every flux with respect to the track pattern in (b). (e) The distribution of layer numbers on edges with respect to the track pattern in (c).}
\end{center}
\vspace{-0.5cm}
\end{figure}

A collective braiding of electron path around flux pairs in two dimensional lattice also generate similar fractional charge serial as that of one dimensional case as well as single flux pair, but the charge distribution on the edge shows a different character from that in the bulk. An arbitrary path curve divides the two dimensional space into two separated domains. If there exists at least one flux that is located in the opposite domain of other fluxes, the continuous path of an electron can be pushed around to wind through all fluxes under braiding operations. An alternative winding path in two dimensional square lattice is showed in Fig. \ref{2Dfluxlattice} (a) as an initial state. After a collective braiding operation on all of the flux dimers oriented in horizontal direction, the initial single track transforms into a periodical distribution of stacked track in two dimensions as Fig. \ref{2Dfluxlattice} (b) showed. These folded track segment fused into one bundle bond when the perpendicular distance between neighboring track segments is below a critical value. Each bond still keeps its original weight which is proportional to its original total number of stack layers. The total number of stack tracks passing through the line of cross section between the two nearest neighboring fluxes (as that represented by the blue dashed lines in Fig. \ref{2Dfluxlattice} (b)) is labeled in Fig. \ref{2Dfluxlattice} (d). When an elementary charge is input from the upper left corner (showed by the red arrow in Fig. \ref{2Dfluxlattice} (a)), it splits into two fractional charges, $2e/3$ goes into the horizontal bond, with the other $e/3$ flowing into the vertical bond (Fig. \ref{2Dfluxlattice} (d)). At each crossing node where four bonds meet, the four electric currents obey the Kirchhoff's Law in electric circuit theory, i.e., the sum of all currents flowing into one node is equal to the sum of currents flowing out of that node, which is essentially the conservation law of charges. The fractional current  $2e/3$ and $e/3$ alternatively distributed along the edges of the square lattice in Fig. \ref{2Dfluxlattice} (d). In the bulk region, the elementary charge keeps an integral value $e$ along the horizontal channels, but a strip of $e/3$ exists alternatively along different vertical channels (Fig. \ref{2Dfluxlattice} (d)). This fractional charge density wave is an exotic phenomena on finite lattice.

One more periodical braiding on the winding path pattern in \ref{2Dfluxlattice} (b) maps it into a new lattice of heavily stacked currents in Fig. \ref{2Dfluxlattice} (c), where the boundary current shows more difference from the edge current than that above. After the collective braiding over flux dimers oriented in vertical direction, it maps the train tracks of $2e/3$ and $e/3$ in Fig. \ref{2Dfluxlattice} (b) into the multilayered train track of fractional charges of $4e/9$, $5e/9$, $3e/4$, $e/4$, and so on (Fig. \ref{2Dfluxlattice} (c) (e)). The $4e/9$, $5e/9$ are only located at the upper left corner (Fig. \ref{2Dfluxlattice} (e)). While $5e/7$ and $2e/7$ anyon forms convective flow on the upper boundary.  $3e/4$ and $e/4$ run on the left boundary, and fuse into another channel of $4e/5$ and $e/5$ anyons. On the right boundary, $3e/5$ and $2e/5$ anyon flow to the middle and fuse into anyon pairs of $5e/7$ and $2e/7$, which also form convective current on the bottom boundary (Fig. \ref{2Dfluxlattice} (e)). Integral charge runs in most channels of the bulk region except the channel that is connected to the output detector on the bottom boundary (as represented by the arrow at the bottom of Fig. \ref{2Dfluxlattice} (e)). At the node that extends to output detector, the elementary charge splits into two $e/7$ and one $5e/7$. The detector only collected a $e/7$, which is generated by first fusion of two fractional charges, $5e/7$ and $2e/7$, and then splitting on the upper boundary. As a result, the Hall resistance measured by the detector reads $R_{H} = (R_{0}{7})/{e}.$ This fractional charge distribution is the result of the special initial state in Fig. \ref{2Dfluxlattice} (a) and the combinatorial braiding operations. It suggests the Hall resistance has a strong dependence on initial state and magnetic field strength.

\begin{figure}
\begin{center}
\includegraphics[width=0.45\textwidth]{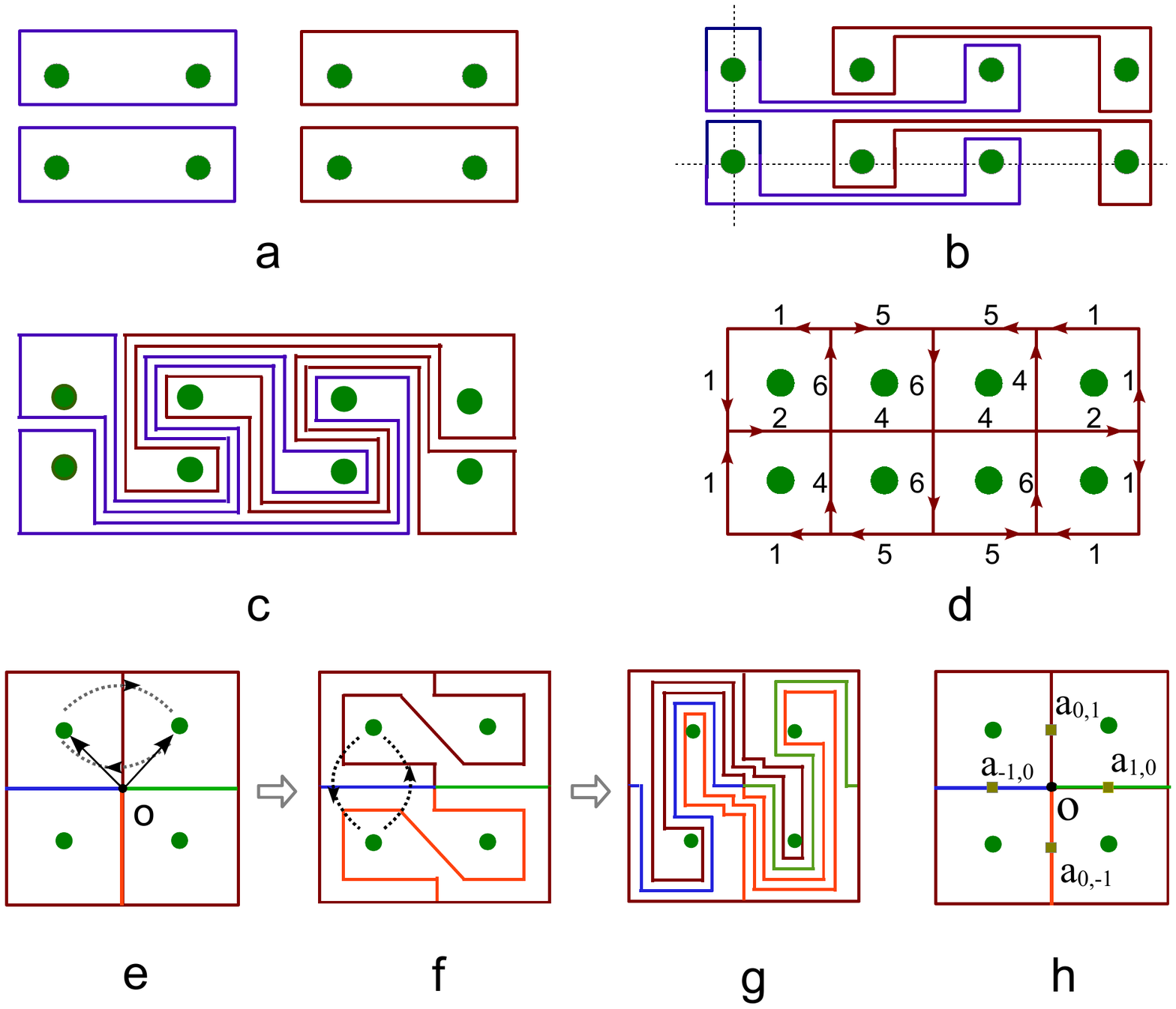}
\caption{\label{2Dgeneral} (a) Four loop currents around four flux dimers oriented in horizontal direction. (b) The stacked train track after one counterclockwise braiding on the two fluxes in the middle of each row. (c) The stacked train track after one more counterclockwise braiding on the two fluxes in vertical ordering in the middle of the two rows. (d) The weight distribution of fused current tracks on the edges around every fluxes. (e) A general train track network around four fluxes, with the four bonds meeting at the central node dyed in different color. (f) The train track distribution after a clockwise braiding on the two flux dimer in horizontal ordering simultaneously. (g) Based on the track pattern of (f), another collective counterclockwise braiding is performed on two flux dimer in vertical ordering. (h) the weight distribution of multilayered train track distribution on the same square network as before after track fusion, each bond now is assigned with new weight, as labeled by $a_{ij}$.}
\end{center}
\vspace{-0.5cm}
\end{figure}

The different fractional charge distribution between edge and bulk also exist on a dimerised flux lattice covered by loop currents. For instance, for the four horizontal flux dimers covering eight fluxes in Fig. \ref{2Dgeneral} (a), a braiding operation on the horizontal pair of fluxes in the middle in counterclockwise direction followed by a braiding operation on the vertical pair of fluxes in the middle also in counterclockwise direction leads to the periodically distributed train track pattern in Fig. (\ref{2Dgeneral}) (c). The weight of each fused bundle bond is labeled in Fig. (\ref{2Dgeneral}) (d). This train track results in half charge on the left and right boundary, and fractional charge $(1/6,5/6)$ and $(4/5,1/5)$ on the upper and bottom boundary. fractional charges with even denominator. The collision process between different fractional charges in the bulk fulfils the conservation law of charge and mass.

The two fundamental path configurations, the open path and loop path, may combine together to form a more general stack tracks in two dimensional lattice. For the most general track distribution in two dimensions as showed in Fig. \ref{2Dgeneral} (e), a clockwise braiding on the horizontal flux pair followed by a counterclockwise braiding on the vertical flux pair as showed in Fig. \ref{2Dgeneral} (e)(f) maps the initial simple track distribution into unit quadramer of crowded stacked tracks in two dimensions (Fig. \ref{2Dgeneral} (g)). After the train track fusion, the weight of the four bonds out of the origin point $o$ is the linear sum of the initial weight distributions, as governed by the following difference equations,
\begin{eqnarray}\label{genea0101}
a_{_{[0,1]}} (t)&=& 3 a_{_{[0,1]}}(t-1) + 2 a_{_{[0,-1]}}(t-1) + a_{_{[-1,0]}}(t-1) \nonumber \\
&+& a_{_{[1,0]}}(t-1),\;\;\nonumber \\
a_{_{[0,-1]}} (t)&=& 2 a_{_{[0,1]}}(t-1) + 3 a_{_{[0,-1]}}(t-1)+ a_{_{[-1,0]}}(t-1)  \nonumber \\
&+& a_{_{[1,0]}}(t-1),\;\;\nonumber \\
a_{_{[-1,0]}} (t)&=& a_{_{[0,1]}}(t-1) + a_{_{[0,-1]}}(t-1) + a_{_{[-1,0]}}(t-1) , \nonumber \\
a_{_{[1,0]}} (t)&=& a_{_{[0,1]}}(t-1) + a_{_{[0,-1]}}(t-1) + a_{_{[1,0]}}(t-1) .\;\;\;\;\;\;
\end{eqnarray}
Here $t$ indicates the step of operation. The subscribe index of $a_{_{[i,j]}}$ labels the location of the bond. The weight on bonds determine the distribution of fractional charges around each crossing node in square lattice, which were mapped from one serial to another under braiding operations. The stable distribution of fractional charges is derived from the eigenvectors of the braiding operation matrix as follows,
\begin{eqnarray}
B=\begin{pmatrix}
3 & 2 & 1 & 1\\
2 & 3 & 1 & 1 \\
1 & 1 & 1 & 0 \\
1 & 1 & 0 & 1\\
\end{pmatrix}.
\end{eqnarray}
The weight on four bonds reach a stable distribution after infinite number of rounds of braiding,
\begin{eqnarray}\label{gene0101eigen}
a_{_{[0,1]}} = a_{_{[0,-1]}} = 1+\sqrt{2},\;\;\; a_{_{[-1,0]}} = 1, \;\;\;a_{_{[1,0]}} = 1.
\end{eqnarray}
Thus the ground state of this local flux quadramer is irrational charges located on the two vertical bonds (Fig. \ref{2Dgeneral} (h)).

The eigenenergy of these multi-layer current stacks can be calculated by the interaction between vectorial currents. Unlike the uniformly oriented dimers in one-dimensional flux lattice, electric currents on two dimensional lattice are oriented into two perpendicular directions. The interaction between the vertical and horizontal currents is governed by the electromagnetic coupling equation,
\begin{eqnarray}\label{Hami2D}
H_{2D}=\sum_{i,j}-\frac{\mu_{0}a^2}{4\pi}\ln(r_{ij})I_{i}I_{j}[\textbf{e}_{j}\times(\textbf{e}_{i}\times{\textbf{e}_{r}})],
\end{eqnarray}
where $\textbf{e}_{i}$ indicates the unit orientation vector of the current on the $i$th bond, $\textbf{r}_{ij}$ is relative position vector directed from the $i$th current to the $j$th current. $\textbf{e}_{r}$ is the unit orientation vector of $\textbf{r}_{ij}$. The eigenenergy strongly depends on the orientation of current segments on each bond. The collective orientation of the current segments in the square network is highly simplified after track fusion, because the nearest neighboring anti-parallel currents turns into parallel current. For an explicit distribution like Fig. \ref{2Dfluxlattice} (d) or Fig. \ref{2Dgeneral} (d), the eigenenergy of the square network of currents is a straightforward summation. In this real space representation, the energy of bulk currents and edge currents could be performed separately, provided a clear vision on the correspondence between bulk state and edge state. In the continuous limit, the edge fluxes surrounded by electric paths behaves like composite particles running along a closed loop chain, governed by the  Calogero-Sutherland model from dimension reduction of composite fermion model in two dimensions \cite{Yu2000}. Since the bulk network transforms synchronously together with the edge network under a global braiding operation or the action of a global magnetic field, the information of bulk charge is deducible from the edge charges. This correspondence is independent of impurity on the edge, because the electron can always circumvent the impurity to fuse into neighboring tracks as long as the fluxes lattice is perfectly in order. If a magnetic impurity is introduce into the flux lattice, it would change the local distribution of fluxes around it. In that case, the same braiding operation sequence as above would generate a different fractional charge distribution especially at finite temperature. The influence of impurity on the fractional conductance is also observed in experiment \cite{Stormer1999}.

\subsection{Fractional charges generated by translation of winding paths in two dimensional space}

\begin{figure}
\begin{center}
\includegraphics[width=0.4\textwidth]{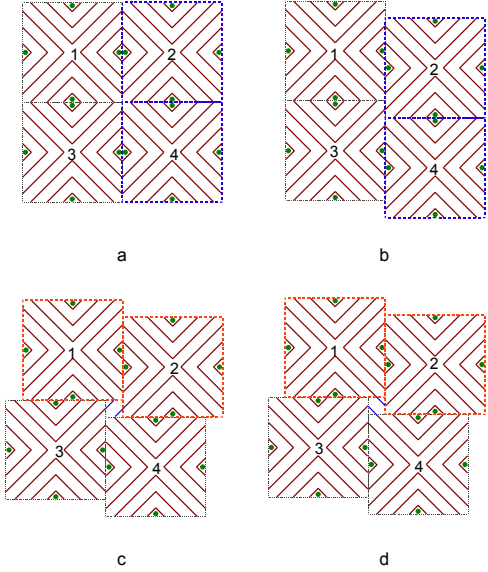}
\caption{\label{braidflux2D} (a) A square lattice of many flux pairs, each of them is surrounded by three layers of square curves. (b) The train track pattern under one step translation of unit cells No. 2 and No. 4 in the vertical direction. (c) The train track patter under one step of translation of unit cells No. 1 and No. 2 to the right hand side. The two parallel blue lines bridge over the unit cells (No. 1,No. 3) and (No. 2, No. 4) respectively. (d) The two parallel blue lines bridge over the unit cells (No. 1,No. 2) and (No. 3, No. 4) respectively after the combined translation.}
\end{center}
\vspace{-0.3cm}
\end{figure}

The periodical braiding operation in two dimensions is not the only method for visualizing fractional charges in two dimensional flux lattice. As showed in previous section of correspondence between integral charge and fractional charge, the fractional charge can be generated by translation operation on their mother tracks around a pair of fluxes. This topological translation surgery was already applied to construct one dimensional chain of anyon. Its extension to two dimensional flux lattice is straightforward. We take the initial train track of $\nu = 2$ as an example to show the translational surgery method for constructing two dimensional train tracks of fractional charges. The initial track pattern is three layers of concentric circles around a pair of fluxes as showed in Fig. (\ref{braidflux2D}) (a). The square unit cell contains four fluxes located on the middle point of its four edges (as labeled by the green disc in Fig. (\ref{braidflux2D}) (a)). Each flux within one unit cell combines with another flux in the nearest neighboring unit cell to form a flux pair, around which is three layers of concentric square. The three layers of concentric semi-square curves intersect with the boundary of the unit cell, pinning down the docking sites for connecting tracks in another unit cell. Translating the column of two unit cells No. 2 and No.4 downward by one step and docking the tracks in their new locations one by one generates open strips of periodical winding tracks, separated by isolated concentric squares around flux pairs (Fig. (\ref{braidflux2D}) (b))). In order to break the flux pair oriented in vertical direction in Fig. \ref{braidflux2D}) (b)), we group the unit cells No. 1 and No. 2 together and translate the whole row to the right hand side by one step. Most docking sites on the boundary naturally meet their next neighboring site except the four sites on the corners (as showed by the empty square hole in the middle region where four unit cells meet in Fig. \ref{braidflux2D} (c)). Since the train track construction forbids self-crossing, there are only two possible ways to dock the four sites. Fig. \ref{braidflux2D} (c) shows the first way that connects unit cells (No. 1, No. 3) and (No. 2, No. 4). The other way connects unit cells (No. 1, No. 2) and (No. 2, No. 3) as showed in Fig. \ref{braidflux2D} (d). In the end, a periodical distribution of winding train tracks around the flux pairs is constructed over the whole two dimensional space. In the bulk region, there are five tracks passing through the border region between neighboring fluxes, interwind by the solo track between the nearest neighboring fluxes.

\begin{figure}
\begin{center}
\includegraphics[width=0.42\textwidth]{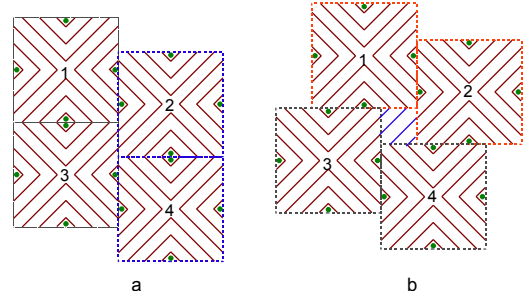}
\caption{\label{fluxdimer2d} (a) With same initial track lattice of three layers of squares around flux pair as Fig. \ref{braidflux2D} (a), a different train track pattern under two-step translation of column in vertical direction is obtained. (b) The train track patter under two-step of translation of rows to the right hand side. There are four parallel blue lines bridge over the unit cells (No. 1,No. 3) and (No. 2, No. 4) respectively, which may also be replaced by another four parallel blue lines bridge over the unit cells (No. 1,No. 2) and (No. 3, No. 4).}
\end{center}
\vspace{-0.3cm}
\end{figure}

An odd number of steps of translation generates a lattice of open tracks around flux pairs, while an even number of steps of translation generates closed current loops that interwind each other with two fluxes enclosed inside. As Fig. \ref{fluxdimer2d} shows, with the same initial track state as Fig. \ref{braidflux2D} (a), the column of unit cells, No. 2 and No. 4, are combined together to translate downward by two steps, generated entangled loop currents that enclose a pair of fluxes respectively (Fig. \ref{fluxdimer2d} (a)). Then the row of unit cells in Fig. \ref{fluxdimer2d} (a), No. 1 and No. 2 together, is translated to the right by two steps. This leaves a blank square where two unconnected docking sites of each corner of the four unit cells meet (Fig. \ref{fluxdimer2d} (b)). Four parallel track segments are added to bridge over the gap and connect neighboring unit cells aligned in the diagonal direction, (No. 1, No. 2) and (No. 3, No. 4), or in the off-diagonal direction, (No. 1, No. 3) and (No. 2, No. 4) in Fig. \ref{fluxdimer2d} (b). The whole flux lattice is covered by winding loop currents.

\begin{figure}
\begin{center}
\includegraphics[width=0.44\textwidth]{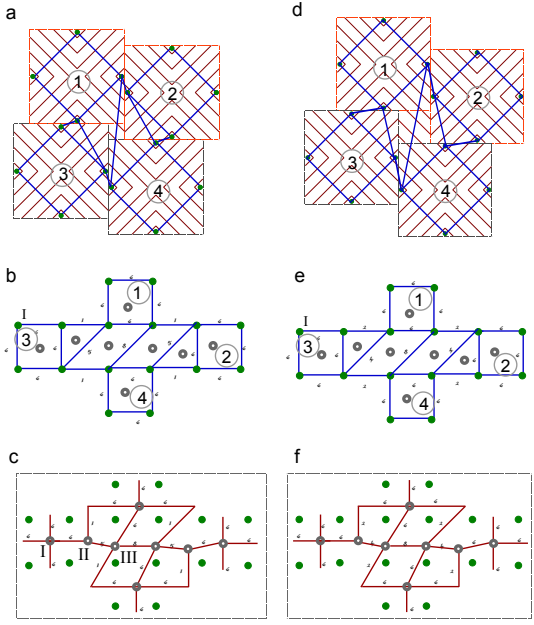}
\caption{\label{flux2dnet} (a) The network of cross sectional bonds that connects the nearest neighboring fluxes in the winding track lattice generated by one step of translation, as indicated by the blue lines. Green disc represents flux. (b) The network of the cross sectional bonds in (a) is reformed into a regular square lattice. (c) The dual network of fused current around flux, with each node (represented by the black annulus) placed as the center of plaquette. (d) The network connecting fluxes of the winding track lattice generated by two steps of translational surgery. (e) The reformed of network of fluxes in (d). (f) The dual network of fused currents after two steps of translation.}
\end{center}
\vspace{-0.3cm}
\end{figure}

The fractional charge generated by this topological translation surgery method is read out from the network of fused current tracks, that is the dual network of that connecting magnetic fluxes. We first connect the nearest neighboring fluxes by cross sectional bonds (represented by the blue bonds in Fig. \ref{flux2dnet} (a) (d)) to construct a periodical network of complex unit cell, . The number of tracks that is cut by the cross sectional bond is labeled as its weight. By placing all fluxes on a square lattice and keeping the topology of network invariant in the mean time, it yields a brief network composed of square and triangle plaquette (Fig. \ref{flux2dnet} (b)(e)). The center of each plaquette is labeled as a dual node, which is connected to its nearest neighboring dual node by fused electric current, perpendicular to the cross sectional bond with the same weight factor (Fig. \ref{flux2dnet} (c)(f)). Following the construction procedure above, we derived the fused current network generated by one step translation (Fig. \ref{flux2dnet} (c)) and two-step translation (Fig. \ref{flux2dnet} (f)). There are three different types current distributions in this exemplar finite flux lattice, (1) four identical currents (each of which has a weight factor 6) meet at node type I; (2) three currents meet at node type II; (3) four anisotropic current meet at node type III. For case (1), two integral charge collide at the node and run out in the other two routes still with integral charge. For case (2), in the weigh distribution around node type II in Fig. \ref{flux2dnet} (c) with respect to one step translation, one charge splits into two fractional charges, $1/6$ and $5/6$, while around the same node type II in Fig. \ref{flux2dnet} (f) with respect to two step translation, one charge splits into fractional charge $1/3$ and $2/3$. For case (3), the node is surrounded by four different weights, (1,5,6,8) with respect to one step translation in Fig. \ref{flux2dnet} (c), and $(2,4,6,8)$ with respect to one step translation in Fig. \ref{flux2dnet} (f). In this case, we first normalize the weight factor of each current $a_{i,\alpha}$ by the half of their sum around the $i$th node,i.e.,
\begin{eqnarray}\label{q1=1234}
Q_{i,\alpha}=\frac{2a_{i,\alpha}}{a_{i,\uparrow}+a_{i,\downarrow}+a_{i,\leftarrow}+a_{i,\rightarrow}},\alpha=\uparrow,\downarrow,\rightarrow,\leftarrow.
\end{eqnarray}
Then $(1,5,6,8)$ is normalized as $(1/10,5/10,6/10,8/10)$ = $(1/10, 1/2, 3/5, 4/5)$, and $(2,4,6,8)$ is normalized as $(1/10, 2/5, 3/5,4/5)$. In the two cases above, the four currents carries anisotropic fractional charges, the only difference is the half charge 1/2 on one branch generated by one-step translation and a 2/5 charge on one branched generated by two steps of translation. This fractional charge distribution exist in the bulk region of winding track lattice, showing a strong dependence on construction rule of track fusion, i.e., the construction of flux network. Only the nearest neighboring fluxes are connected in the case above. If the next nearest neighboring fluxes are also connected, it would induce the current fusion between the next nearest neighboring fluxes, resulting in more fractional charges splitting at a node where more fused currents meet.

This topological translational surgery method has advantage of constructing a Hamiltonian for periodical track lattice, for instance, the Hamiltonian for train tracks undergoing a general p-step translation in two dimensions is expressed as $H^{[2d]}_{p}= \hat{T}_{p}H^{[2d]}_{0}$,
\begin{eqnarray}\label{Hamitransp2D}
H^{[2d]}_{p}&=& \sum_{x,y,j}t\;c^{\dag}_{x2n+j+\frac{5}{2},y2n+j+\frac{5}{2}}\nonumber\\
&&c_{x2n+j+\frac{5}{2},y2n+j+\frac{5}{2}}e^{i{\phi{r2(g+1),\uparrow}}} \nonumber\\
&+& \sum_{x,y,j}t\;c^{\dag}_{x2n+j+\frac{3}{2}+p_{x}, y2n+j+\frac{3}{2}+p_{y}} \nonumber\\
&&c_{x2n+j+\frac{5}{2}+p_{x}, y2n+j+\frac{5}{2}+p_{y}}e^{-i{\phi({r2(n+1)+p_{x}+p_{y},\downarrow})}}\nonumber\\
&+&h.c.
\end{eqnarray}
Translations in opposite direction generated braiding operations in opposite direction. The accumulation of phases is recorded by complex phase in the hopping term when an electron propagates along a winding track. A simplified version of this Hamiltonian can be approximated by Hofstadter model in two dimensional lattice \cite{Hofstadter1976}. The fractal structure of energy spectrum of Hofstadter model can be explained by the closed knot path and open chaotic path in two dimensional magnetic flux lattice in real space, disclosed interference pattern of winding paths in real space and the corresponding fractional charges.

\subsection{ The train track of fractional charges derived from the full vacuum states of two dimensional knot lattice}

\begin{figure}
\begin{center}
\includegraphics[width=0.45\textwidth]{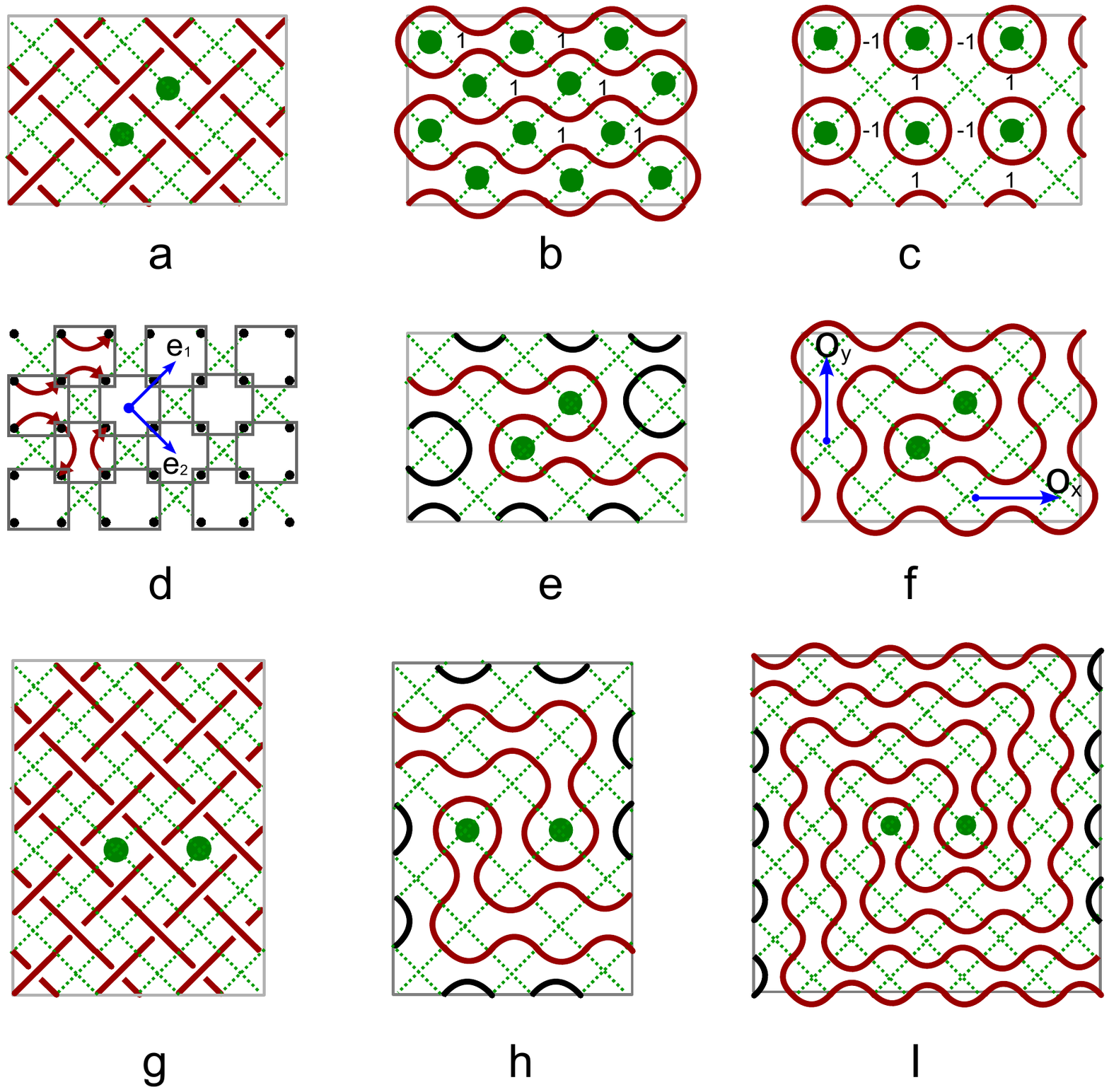}
\caption{\label{Hallknot2D} (a) A local knot lattice pattern around two nearest neighboring fluxes that penetrate through the center of two plaquette. The flux is represented by green disc. Bold red lines represent electron path. Dashed green lines draw the square lattice in space.(b) The ferromagnetic phase of vacuum state. (c) The anti-ferromagnetic phase of vacuum state. (d) A two dimensional lattice of unit cell (showed as the square enveloped two vacuum arcs) constructs one general winding path pattern in two dimension. (e) The winding string of vacuum states for fractional charge $1/3$ is designed route for on square lattice. (f) The winding string of vacuum states for $2/5$ is a vortex pattern. (g) The local knot lattice pattern around two next nearest neighboring fluxes.(h) The winding string of vacuum states for fractional charge $1/4$ is designed as two paths surrounding two fluxes respectively but avoiding each other. (I) The winding string of vacuum states for fractional charge $3/8$. }
\end{center}
\vspace{-0.3cm}
\end{figure}

Here we developed a new method to derive the train tracks of fractional charges from the vacuum states of two dimensional knot lattice of electron paths \cite{{SiT2019}}. A continuous self-avoiding electron path snaking through two dimensional lattice of magnetic fluxes represents a collective vacuum state of two dimensional knot lattice. In a local snapshot of square lattice of knots around a pair of fluxes (as showed in Fig. \ref{Hallknot2D} (a)), the flux locates at the vertex of square lattice surrounded by square plaquette, in which two currents meet to form an over-crossing, an under-crossing or a vacuum state. In the train track theory, a curve is forbidden to intersect with itself everywhere. While in the knot lattice model, this self-avoiding rule is naturally implemented by confining the block spin 1 of current crossing state to vacuum state, i.e., $S = 0$, while the over-crossing (under-crossing) state corresponds to $S=1$ (S=-1) \cite{{SiT2019}}. A collective vacuum state of the knot lattice depicts an unfused train track in two dimensional space, as showed by the exemplar pattern in Fig. \ref{Hallknot2D} (b)(c)(e)(f). Unlike the conventional spin 1, here the vacuum state of block spin 1 ( i.e., $|S=0\rangle = |O\rangle $) is a vector composed of two orthogonal components,
\begin{eqnarray}\label{oxy}
|O\rangle  =  O_{x}|{x}\rangle + O_{y}|{y}\rangle, \;\;\;O_{\mu}O_{\nu}=\delta_{\mu\nu}.
\end{eqnarray}
Every vacuum state carries two turning arcs that bend the electric current into its perpendicular direction (Fig. \ref{Hallknot2D}(b)). $O_{x}$($O_{y}$) denotes two turning arcs oriented in X-direction (Y-direction). These two turning arcs connect the current in one direction but disconnect that in its perpendicular direction. In the Hilbert space of vacuum states, the two vacuum components are equivalent to classical Ising spin, ($O_{x} = +1$, $O_{y} = -1$) or vice versa. Every train track in two dimensional lattice exactly corresponds to a spatial distribution of the binary value. In the ferromagnetic phase, all vacuum arcs in the bulk are oriented in the same direction, those perpendicular to that direction only exist on the edge to fulfill the conservation law of mass and charge (Fig. \ref{Hallknot2D} (b)). In the anti-ferromagnetic phase, the nearest neighboring vacuum arc pairs are oriented into perpendicular directions, demonstrating a lattice of isolated current loops (Fig. \ref{Hallknot2D} (c)). A general winding path is constructed by translation and rotation operations on the unit cell of vacuum arc pairs, which is essentially a quadramer enveloping four lattice sites (as showed by the black square in Fig. \ref{Hallknot2D} (d)).

The total energy of a general winding path is the sum of the potential energy of interacting electric current pair within one unit cell. The orientation of the two current segments is either parallel or anti-parallel to each other. The potential energy in each unit cell is governed by the product of two current tensors,
\begin{eqnarray}\label{E0}
\epsilon_{r} = u(I_{r,a}^{\alpha}I_{r,b}^{\bar{\alpha}}),\;\;\alpha=1;\bar{\alpha}=2;\;\; a,b=\leftarrow,\rightarrow,\uparrow,\downarrow.
\end{eqnarray}
where $u = -\frac{\mu_{0}l_{i}}{2\pi}\ln[d]$ denotes the potential energy between two parallel electric currents. $\alpha(\bar{\alpha})$ denote the inner location of the two current segments. $a(b)$ denote the orientation of the current along the vacuum arc. $r$ denotes the central location of the unit cell. The total energy of free quadramers is $H_{1}=\sum_{r} \epsilon_{r} $, which admits an eigenstate with a random orientation of current segment within every unit cell. When we study the current flow from one unit cell into another unit cell, the conservation law of electric charge and mass confined the two docking currents that belongs to the two nearest neighboring unit cells respectively into the same direction. The coupling terms describing two counter-propagating current segment should be eliminated. To exclude the coupling terms like
\begin{eqnarray}\label{Ilrud} I_{r,\leftarrow}^{\alpha}I_{r+1,\rightarrow}^{\beta}, \; I_{r,\rightarrow}^{\alpha}I_{r+1,\leftarrow}^{\beta},\;
I_{r,\uparrow}^{\alpha}I_{r+1,\downarrow}^{\beta}, \; I_{r,\downarrow}^{\alpha}I_{r+1,\uparrow}^{\beta},\;
\end{eqnarray}
the directed electric current is viewed as Feynmann diagram in quantum field theory and expressed by the hopping operator of fermions, for example,
\begin{eqnarray}\label{I=cc+}
I_{r,\rightarrow}^{\alpha} = c_{ij,\alpha}c_{i+1,j,\alpha}^{\dag};\;\;I_{r,\uparrow}^{\alpha} = c_{ij,\alpha}c_{i,j+1,\alpha}^{\dag}.
\end{eqnarray}
Because the product of two identical fermion operators at the same lattice site is exactly zero, $c_{ij,\alpha}c_{ij,\alpha}=0$, the counter-propagating currents meet at the same site naturally becomes zero. The fermion operator provides a natural expression of the Hamiltonian of coupled quadramers,
\begin{eqnarray}\label{H2}
H_{2} &=& u \sum_{<r,r'>} \epsilon_{r}\epsilon_{r'}\nonumber\\
&=& I_{r,a}^{\alpha}I_{r,b}^{\bar{\alpha}}I_{r',a'}^{\beta}I_{r',b'}^{\bar{\beta}}\delta_{aa'}\delta_{bb'}
\end{eqnarray}
where $r$ indicates the center of plaquette. The eigenstate of Hamiltonian Eq. (\ref{H2}) corresponds to many continuously oriented winding paths. There exists topological correlation between the two current segments in one unit cell, even though they are nearest neighboring current segments. This is because the upper current propagates along a continuous path would finally turns back into its bottom partner within the same unit cell, only continuously oriented pattern survived since any counter-propagating current naturally is zero.

The vortex path with double flux core can generates a serial of fractional charges after topological path fusion (as showed in Fig. \ref{Hallknot2D} (e)(f)(g)(h)). The magnetic fluxes are placed on the vertex of square lattice. Every hopping current gains a phase factor when it goes around a flux. If we represent the two vacuum components ($O_{x}$ and $O_{y}$) by current operators, i.e., ($O_{x} = I_{x}I_{x}$ and $O_{y}=I_{y}I_{y}$), then $I_{x} = 1$ and $I_{y} = e^{i\pi/2}=i$. Whenever the vacuum state transforms from a X-state to a Y-state or vice versa,  the vacuum arc gains a phase increment of $\pi/2$. Setting the flux as origin point and recording the sequence of vacuum state along the vortex path, it maps one of the two spiral arms of vortex path into a sequence of binary code, for instance, $(+1-1+1+1)$  corresponds to Fig. \ref{Hallknot2D} (e) and $(+1-1+1+1-1-1)$ corresponds to Fig. \ref{Hallknot2D} (f). Every flipping point from $+1$ to $-1$ is a kink excitation quantified by a gauge potential equation, $a = \partial_{y}O_{x}-\partial_{x}O_{y}$. These kink configuration is a topological excitation, whose population number is invariant during topological transformation and topological fusion. A continuous path can only fuse with the current segments on the same side of the flux under topological transformation, since each magnetic flux exists as a forbidden tubular hole in space. Topological fusion induced a high degeneracy of fractional charge state, because topological transformation shortens or elongates the distance between current segments continuously to keep the topological character but inevitably change the coupling potential energy between current. The topological fusion of the red path in Fig. \ref{Hallknot2D} (e) generates fractional charge $1/3$. While fractional charge $2/5$ is derived from Fig. \ref{Hallknot2D} (f). If the two magnetic flux are placed on the next nearest neighboring sites in knot lattice of Fig. \ref{Hallknot2D} (g), fractional charges with even denominator are generated after topological fusion. For instance, the winding track in Fig. \ref{Hallknot2D} (h) and Fig. \ref{Hallknot2D} (I) generates the fractional charge $1/4$ and $3/8$ respectively.

\begin{figure}
\begin{center}
\includegraphics[width=0.45\textwidth]{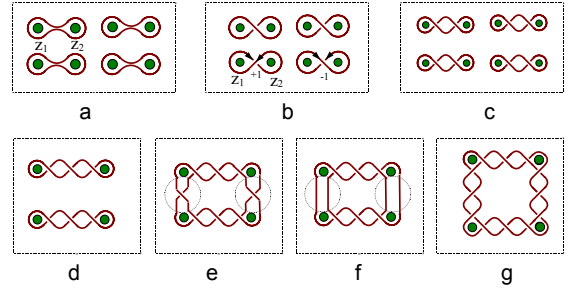}
\caption{\label{dimer2d}  (a) The square lattice of unbraided loop currents around flux pairs. (b) The knot lattice of composite fermion under one braiding operation. (c) The knot lattice of composite boson under two braiding operations. (d) The knot lattice for free anyon with fractional $e/3$. (e) The knot lattice of coupled anyons by a fermion. (f) The knot lattice of coupled anyons by a boson. (g) The knot lattice of collectively coupled anyon of $e/3$.}
\end{center}
\vspace{-0.3cm}
\end{figure}

Picking the closest current segments to the double flux core and stretching them upward to the third dimension maps a winding path of vacuum states into a knot lattice \cite{{SiT2019}}, which is constructed by the two simple elementary knot configurations, the fermionic knot with a current penetrating through their border region and . bosonic knot with current avoiding their border region (Fig. \ref{dimer2d} (a)). Fixing one of the two fluxes enveloped in the composite boson path and flipping the other flux into opposite direction transforms the bosonic winding path into a fermionic winding path (Fig. \ref{dimer2d} (b)). These two fundamental knot patterns are the mother state of all of other fractionally charged states. Two braiding operations generates the boson path with half charge $\nu = e/2$ in Fig. \ref{dimer2d} (c), since a flipping on one of the two crossings brings it back to the bosonic mother pattern in Fig. \ref{dimer2d} (a). In a knot pattern with odd number of crossings (such as Fig. \ref{dimer2d} (b)(d)), exchanging the location of two fluxes within one loop maps the positive crossings to negative ones. The knot in Fig. \ref{dimer2d} (d) has a fractional charge $Q = e/3$ or $Q = 2e/3$. Since it takes three braiding operations (or exchange operation of the ending points) to map a $e^{i\pi} = -1$ crossing state to $+1$ crossing state, each braiding operator contributes a fractional phase $\pi/3$. The statistical character of the knot current with odd number $(2m+1)$ of crossings matches the Laughlin wave function,
\begin{eqnarray}\label{psidimer}
\psi(z_{1},z_{2}) =  (z_{1}-z_{2})^{2m+1}e^{-\frac{|z_{1}|^{2}+|z_{2}|^{2}}{4l_{B}^{2}}}£¬
\end{eqnarray}
where $z_{i} (i=1,2)$ labels the location of the two cutting ends of current in Fig. \ref{dimer2d}, which represents the point particle---electron. Exchanging the two electrons is equivalent to flipping one flux within the flux pair. The number of flipping operations is proportional to the distance between the two fluxes. The effective magnetic field strength decreases linearly with respect to the number of flipping operation. Under infinite number of braiding operations, it converges to the limit of half-charge state when the effective magnetic field strength reaches zero. For the knot lattice formed by an even number of flux flipping operation (such as Fig. \ref{dimer2d} (c)), it takes exactly $m/2$ (m is even) steps of braiding on the $m$ crossings to map the highly braided knot current back to the mother state of dimer covering in Fig. \ref{dimer2d} (a), indicating a half filling factor $\nu = 1/2$. Thus the half-charge anyon, ${Q={e}/{2}}$, exists at different values of magnetic field strength, revealing a highly degenerated topological state.

The strongly interacted composite boson or fermions can be constructed from flux dimer coverings on two dimensional lattice. The same dimer covering pattern on square lattice may be implemented by either bosonic loop (Fig. \ref{dimer2d} (a)) or fermionic knot path (Fig. \ref{dimer2d} (b)). The potential energy of every knot path is counted by electromagnetic potential equation (\ref{HamiE}). For a $N_{x}\times{N_{y}}$ flux lattice,  the total energy of the bosonic simple loop in Fig. \ref{dimer2d} (a) is counted as $E_{0,a} = {N_{x}N_{y}}E_{0}/{2}$, where $E_{0}$ is the electromagnetic potential of two antiparallel currents. There are many different dimer covering patterns with respect to the same eigenenergy $E_{0,a}$, expanded Hilbert space of many degenerated states. The degree of degeneracy can be counted by Kasteleyn's counting method \cite{Kasteleyn1961}. Here each flux at the $i$th site can be viewed as a fermionic object, represented by Grassmann variable $\eta_{i}$, the number of all possible dimer coverings $Z$ is given by the Pfaffian of the matrix $M$, $ Z = Pf [M]$, which is also the partition function of interacting Grassmann variables, $Z = \int[D\eta] \exp[ \sum_{i<j}M_{ij}\eta_{i}\eta_{j}]$. For a $N_{x}\times{N_{y}}$ flux lattice, the degree of degeneracy of free dimer covering reads,
\begin{eqnarray}\label{k}
Z=\left [\prod_{k=1}^{N_{y}}\prod_{j=1}^{N_{x}} (2\cos[\frac{\pi j}{N_{x}+1}]+2 i
\cos[\frac{\pi k}{N_{y}+1}]) \right ]^{1/2}.
\end{eqnarray}
The high degeneracy of free dimer state is reduced by introducing interacting channels between two composite particles. There are six vacuum states (or zero crossing states) around each flux dimer. Every vacuum state can map into a positive crossing, a negative crossing or a perpendicular vacuum state to itself to bridge the neighboring flux dimers (as shown in Fig. \ref{dimer2d} (e)(f)). From the energy point of view, two crossing currents which are mutually perpendicular to each other contributes a zero energy, which is lower than the electromagnetic energy of two vacuum currents. Thus free composite particles prefer being connected by crossing states to reduce the total energy. The number of crossings that connects neighboring fluxes determines the fractional charges. For example, Fig. \ref{dimer2d} (g) illustrates the knot lattice of many interacting anyons with fractional charge $Q = e/3$, where fluxes are connected by one dimensional knot lattice with three crossings. The distance between neighboring fluxes increases with respect to a decreasing magnetic field strength.

\begin{figure}
\begin{center}
\includegraphics[width=0.45\textwidth]{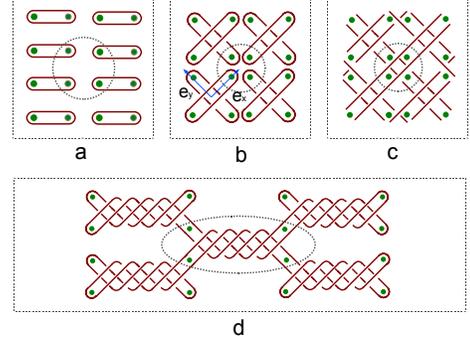}
\caption{\label{2Devenknot}(a) A homogeneous dimer covering of flux lattice by loop current. (b) The knot lattice pattern after braiding the nearest neighboring flux dimers, formed by square lattice of free pairs of two crossing loop currents.  (c) The four free crossings are coupled to one another by one block crossing that bridges them. (d) The knot lattice of coupled anyons with fractional charge $Q = 5e/12$.}
\end{center}
\vspace{-0.3cm}
\end{figure}

When the fluxes are connected by an even number of current crossings, it constructs an knot lattice of interacting composite bosons, that carries fractional charge with an even denominator. Take a dimer covering of bosonic free loops as the initial state (Fig. \ref{2Devenknot} (a)), fix the left flux within the dimer and braid the right flux with its vertical neighbor, it generates a knot of double currents with four crossings (Fig. \ref{2Devenknot} (b)), representing a lattice of fractionally charged states with even denominator as described by the following collective wave function,
\begin{eqnarray}\label{2Dfreefaffian}
\Psi_{o} &=& {\prod}_{i = 2le_{x}+2n{e}_{y},}P_{f{k};i} (z_{i} - z_{i\pm\vec{e}}),  \nonumber \\
&&P_{f{k};i} (z_{i} - z_{i\pm\vec{e}})  =  ( z_{i-e_{x}} - z_{i+e_{x}}) (z_{i-e_{y}}-z_{i+e_{y}})\nonumber \\
&+&(z_{i-e_{x}} - z_{i-e_{y}}) (z_{i+e_{y}} - z_{i+e_{x}}) \nonumber \\
&+&(z_{i-e_{x}}-z_{i+e_{y}}) (z_{i+e_{x}}-z_{i-e_{y}}).
\end{eqnarray}
To introduce interaction between free composite fermions with four crossings, the vacuum state in the middle cell where four dimers meet (as enclosed by the dashed circle in Fig. \ref{2Devenknot} (b)) is replaced by crossing state of double currents. This generates a square lattice of block Ising spin $S = \pm1$, where each state is characterized by four crossings of single line. This knot pattern of Fig. \ref{2Devenknot} (c) represents the collective state of interacting anyon with fractional charge $Q = e/4$ or $Q = 3e/4$, governed by the Pfaffian equation,
\begin{eqnarray}\label{2Dcouplefaffian}
\Psi_{1} = {\prod}_{i = le_{x}+n{e}_{y}}P_{f{k};i} (z_{i} - z_{i{\pm}e_{x}{\pm}{e}_{y}}),
\end{eqnarray}
Braiding each block spin in the knot lattice of $Q = e/4$ two more times generates the fractional charge state $Q = 5e/12$ and $Q = 7e/12$, which depicts the knot lattice of entangled double helix in Fig. \ref{2Devenknot} (d). Its corresponding wavefunction is constructed as,
\begin{eqnarray}\label{2Dcouplefaffian6}
\Psi_{m} = {\prod}_{i = le_{x}+n{e}_{y}}P_{f{k};i} (z_{i} - z_{i{\pm}e_{x}{\pm}{e}_{y}})^{m},
\end{eqnarray}
where $m = 3$ indicates the times of braiding over the current loop. Repeating the same operation generates a serial fractional charges with even denominator. If the coupling subunit (indicated by the crossing state within the dashed circle in Fig. \ref{2Devenknot} (d)) are replaced by knot with a different fractional charge from the unit cell of the lattice, it results in a hybrid fractional charge serial. Because all current tracks must be closed on the boundary, a gapless vacuum current always exist on the boundary of a finite knot lattice.

\section{The fractionally charged anyon in three dimensional lattice of magnetic fluxes}

\begin{figure}
\begin{center}
\includegraphics[width=0.42\textwidth]{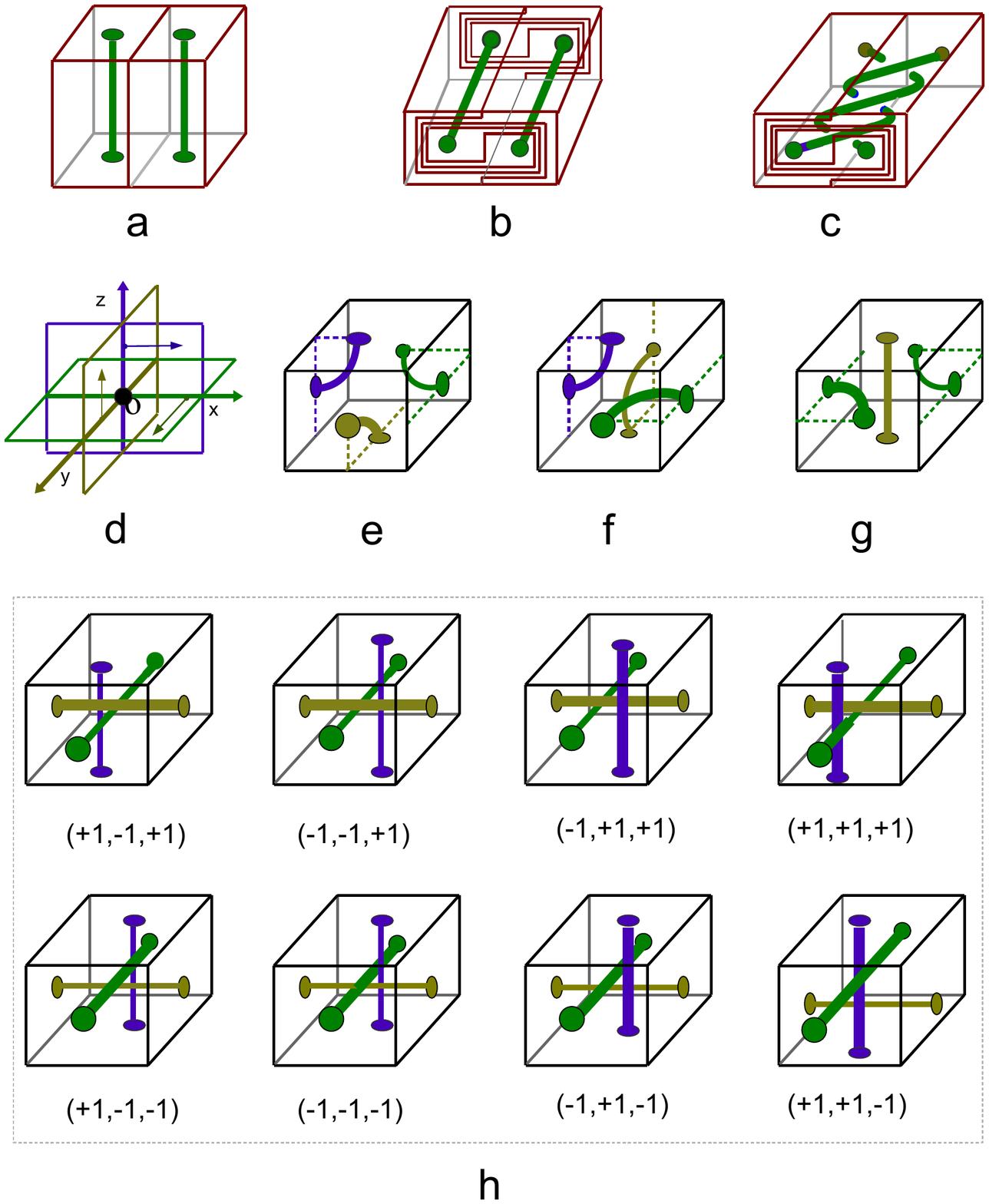}
\caption{\label{3DdualEM}(a) Flux tubes are surrounded by three dimensional network of electron paths denoted by the red edges of the cubic. (b) The electron paths in different layers perpendicular to the homogeneous magnetic fluxes are braided simultaneously. (c) The braided electron path and knotted magnetic flux tubes in an inhomogeneous magnetic field with one ending of the path fixed to a plaquette center. (d) A general case of magnetic fluxes with three projected components in X, Y and Z direction which are represented by the green, brown and blue axis. (e)(f)(g) The three vacuum states of  non-intersect magnetic flux tubes. (h) The eight spin states represented by three intersecting magnetic flux tubes.}
\end{center}
\vspace{-0.3cm}
\end{figure}

The two dimensional fluxes lattice is the projection of three dimensional array of flux tubes to the ground plane (as shown in Fig. \ref{3DdualEM} (a)). The fractional Hall conductance characterize the transportation capability of the braided paths of electrons gas confined in the the ground plane. Usually the thickness of the electron gas layer is too small to demonstrate the physical effect of a tilted flux tube. The top layer and bottom layer of the 2D electron gas are braided simultaneously, as shown in Fig. \ref{3DdualEM} (b). However, in a strong magnetic field with high gradient, the top layer feels a different magnetic field strength from that of bottom layer. As a result, the current between two fluxes in top layer is braided for different period from the bottom layer. If the bottom layer is located in the plane with the strongest magnetic field strength, the trajectory of an electron is confined in such a small radius that it cannot wind around the two fluxes except penetrating through their border region. This case is equivalent to fixing the endings of fluxes to the bottom layer (showed by Fig. \ref{3DdualEM} (c)), where a magnetic monopole is located at the center of a serial of concentric spheres. Then the train tracks of electron paths in the top layer has an one-to-one correspondence with the knot structure of braided magnetic flux tubes in the bulk (showed by Fig. \ref{3DdualEM} (c)). The fractional conductance of the train tracks in top layer can be extracted out of the knot state of the braided magnetic fluxes. The braided magnetic field lines is not only a theoretical proposal. In fact, a helical magnetic field is found in the Sun's surface, illustrated by the eruption of coronal mass ejections \cite{Seehafer1990}.

\subsection{Fractional charges generated by winding tracks in three dimensional knot lattice of magnetic fluxes}

One arbitrary knot of magnetic fluxes in three dimensions is constructed by combinatorial connection of ten elementary crossing states showed in Fig. \ref{3DdualEM} (e)(f)(g). The three dimensional space is divided into a lattice of many identical unit cubic. A local Cartesian coordinate system is set up inside each unit cubic with the origin point located exactly at its center. As showed in Fig. \ref{3DdualEM} (c), the X-axis segment (indicated by the green line) divides the X-Z plane into two separated domains, so do the Y-axis (indicated by the yellow line) and Z-axis (indicated by the blue line). There are 12 separated domains in total, each of which is dyed by a color. The three axis intersect with the center of the six faces of the cubic. Each intersecting point is a docking site of one magnetic flux segment with another one in its neighboring cell. Six face centers are bridged by three magnetic flux segments, since each segment has only two endings. The segment connecting two opposite faces are defined as crossing line, while the arc connecting the nearest neighboring faces are defined as vacuum arc. For the first case that all of the three segments are vacuum arcs, there are only two independent configurations as showed in Fig. \ref{3DdualEM} (e)(f). The other configurations are derived by three dimensional rotation transformation. For the second case of two vacuum arcs and one crossing line, there is only one independent configuration that two vacuum arcs lie in the same plane with a crossing line perpendicularly penetrating through the center of the plane (Fig. \ref{3DdualEM} (g)). For the third case that all of the three flux segments are crossing lines, there are eight independent configurations in total as Fig. \ref{3DdualEM} (h)showed, in which each segment is parallel to one axis but avoid crossing with one another. The location of the flux segment parallel to Y-axis is determined by its projection to X-axis and Z-axis, which are labeled by $\hat{P}Y^{x}$ and $\hat{P}Y^{z}$ correspondingly, with $\hat{P}$ is a projection operator. The location of other flux segments are determined in the same way. Then the crossing states can be characterized by a relative location vector,
\begin{eqnarray}\label{Svector}
S_{i} &=& (S_{x},S_{y},S_{z})\nonumber\\
&=&  (\hat{P}Y^{x} - \hat{P}Z^{x}, \hat{P}Z^{y} - \hat{P}X^{y}, \hat{P}X^{z} - \hat{P}Y^{z}),
\end{eqnarray}
where $i$ locates the center of the unit cubic. A normalized vector $S_{i}$ is equivalent to an Ising spin in three dimensions, $S_{\alpha} = \pm1$. The Ising spin values of the eight crossing states are listed in Fig. \ref{3DdualEM} (h). The full vacuum state is represented by self-avoiding arcs in Fig. \ref{3DdualEM} (e). For instance, the green arc in X-Y plane of Fig. \ref{3DdualEM} (e) connects the two face centers at (x = +1) and (y = -1). The corresponding spin vector of the full vacuum state reads,
\begin{eqnarray}\label{OSvector}
O_{i} = (\hat{P}Z^{x} - \hat{P}Z^{y}, \hat{P}X^{y} - \hat{P}X^{z}, \hat{P}Y^{z} - \hat{P}Y^{x}).
\end{eqnarray}
The hybrid state with two vacuum arcs in a plane and one crossing line in the perpendicular axis is denoted as,
\begin{eqnarray}\label{O2Svector}
OS_{i} = (\hat{P}Z^{x} - \hat{P}Z^{y}, 0, \hat{P}X^{z} - \hat{P}Y^{z}).
\end{eqnarray}
One general knot of entangled fluxes is represented by the combination of the Ising spin states above.

\begin{figure}
\begin{center}
\includegraphics[width=0.42\textwidth]{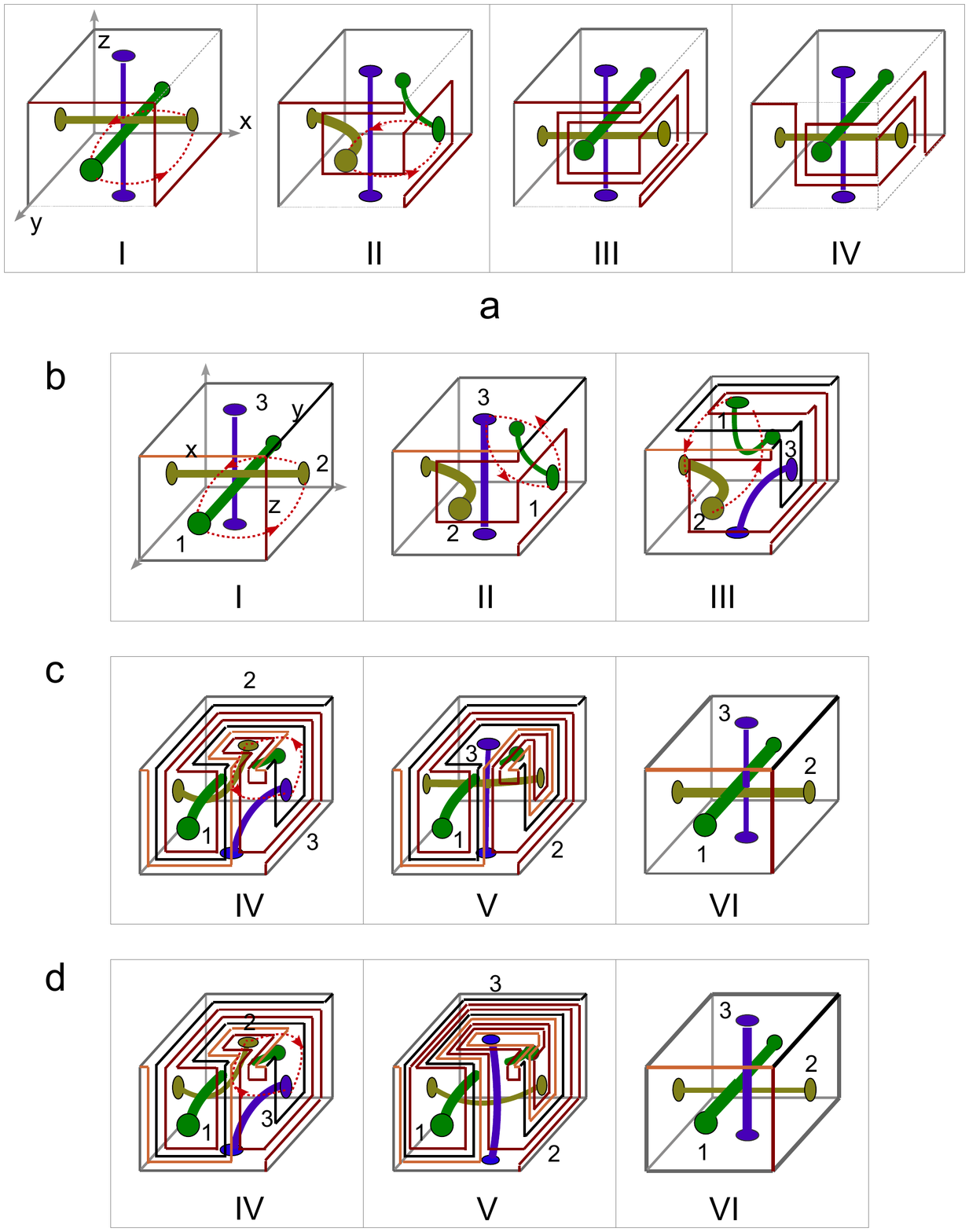}
\caption{\label{3Dfluxlattice}(a) To map the spin state $(+1,-1,+1)$ state into $(+1,-1,-1)$, i.e., to bring the green flux tube up, is realized by at least three steps of braiding operations on the two endings of X-flux and Y-flux tube. (b)I-III and (c)IV-VI, it takes a serial of braiding operations over three nearest neighboring endings to map a $(+1,-1,+1)$ state into $(-1,-1,-1)$. (b) I-III and (d) IV-VI, the braiding chain of mapping $(+1,-1,+1)$ state into $(-1,+1,-1)$ is only one step difference from that of (b)(c).}
\end{center}
\vspace{-0.3cm}
\end{figure}

Fractionally charged anyon runs along the edges of unit cubic, generating an electric current going around magnetic flux knot. Braiding the nearest neighboring ending points of two flux segments drives the electric current in the edge of cubic into winding train tracks, that bends from one face to its perpendicular neighboring face, as showed in Fig. \ref{3Dfluxlattice} (a). For the initial crossing state $S = (+1,-1,+1)$ in Fig. \ref{3Dfluxlattice} (a)-(I), two counter-clockwise braiding generate the train track in Fig. \ref{3Dfluxlattice} (a) (I-IV) that breaks one elementary charge apart into a pair of fractional charges, $Q = e/3$ and $Q = e2/3$. This braiding exchanges the relative location of the two crossing fluxes oriented in X and Y axis, with their projection to Z-axis exchanged correspondingly, flipping the initial spin vector of crossing fluxes to its mirror configuration $S = (+1,-1,-1)$ as a reflection by the X-Y plane. On the most left vertical edge in Fig. \ref{3Dfluxlattice} (a) (IV), the anyon with $Q = e2/3$ first flows downward along Z-axis, and then turns to the right hand side along X-axis. It splits into two anyons with $Q = e/3$ in the middle edge along Z-axis in Fig. \ref{3Dfluxlattice} (a) (IV), one runs along Y-edge with the other one going up along the Z-edges. $m$ period of braiding operations generates a serial of fractional charges along the cubic edges, with a fractional charge $Q = e/m$ runs in the Z-edge between two flux docking sites. The Hall conductance in three dimensional lattice is defined in the same way as the two dimensional case,
\begin{eqnarray}\label{Rxyz=VI}
R_{\alpha\beta} = V_{\alpha}/{I_{\beta}},\;\;\; \alpha,\beta = x,y,z.
\end{eqnarray}
For the train track pattern of Fig. \ref{3Dfluxlattice} (a) (IV), the Hall conductance on the edges are listed as following,
\begin{eqnarray}\label{Rzx=VI}
R_{zx} = R_{0} /(\frac{2e}{3}),\;\;R_{yz} = R_{0} /(\frac{2e}{3}),\;\;R_{zz} = R_{0} /(\frac{e}{3}).
\end{eqnarray}
Other serial of quantized Hall conductance is derived by the same braiding procedure above. Even though the braiding operation here is very similar to that in two dimensional space, three dimensional braiding operations are typical non-Abelian operations, since two sequent braiding operations are not commutable.

To study the limit charge of anyons after infinite number of braiding operations over an arbitrary initial train track, we start with a general cubic with its edges assigned by a general weight factor. Three different flux docking endings are braided, as labeled by the green disk, yellow disk and blue disc at the center of three perpendicular planes in Fig. \ref{3Dfluxlattice} (b) - (I). The border current between each pair of the three flux docking endings are represented by the red, black and yellow edges of the cubic in Fig. \ref{3Dfluxlattice} (b) - (I). A counterclockwise braiding exchange the two endings No.1 and 2, mapping an initial spin state of $S = (+1, -1, +1)$ into a hybrid vacuum state, $S = (0, 0, 0)$ in \ref{3Dfluxlattice} (b) - (II). A second counterclockwise braiding exchanges the two endings No. 1 and No. 3 and drives the flux knot to a full vacuum state (as showed in \ref{3Dfluxlattice} (b) - (III)). A third braiding brings the yellow vacuum arc and green vacuum arc into a crossing state (as showed in \ref{3Dfluxlattice} (c) (d)- (IV)). It takes one final braiding to map the hybrid vacuum state to a fully crossing state. This final braiding operation has two optional ways, one way is still a counterclockwise braiding over the endings No. 2 and No. 3 in \ref{3Dfluxlattice} (c) - (IV),  it maps into the crossing state, $S = (-1,-1,-1)$. The other way is a clockwise braiding over the two endings No. 2 and No. 3 in Fig. \ref{3Dfluxlattice} (d) - (IV), leading to a full crossing state, $S = (-1,+1,-1)$, which is exactly the spatial inversion of the initial crossing configuration.

\begin{figure}
\begin{center}
\includegraphics[width=0.35\textwidth]{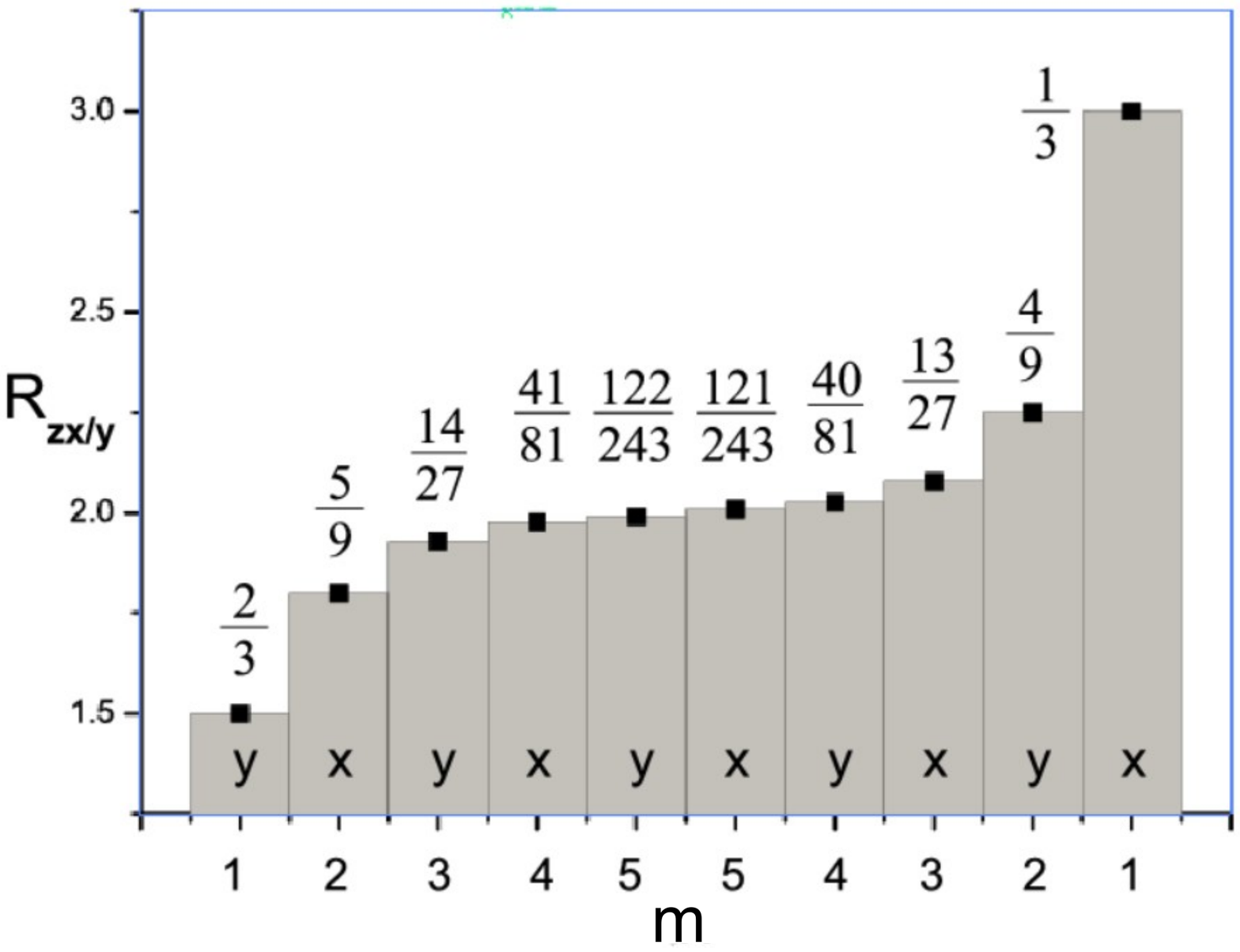}
\caption{\label{Hall3Ddata} The fractional Hall resistance in the X- and Y- edge $(R_{zx}$ and $R_{zy})$ as a descrete distribution with respect to the number of braiding periods $m$. }
\end{center}
\vspace{-0.3cm}
\end{figure}

Different braiding sequence drives the train track distribution into different dynamic path. For the first braiding chain mapping $S = (+1,-1,+1)$ to $S = (-1,-1,-1)$, the initial weight vector of the three edges, ($a_{x}({0}), a_{y}({0}), a_{z}({0})$) represented by the yellow, the red and the black edge in Fig. \ref{3Dfluxlattice} (b)-(I), is assigned by a new weight vector after track fusion (Fig. \ref{3Dfluxlattice} (c)-(VI)),($a_{x}({1}), a_{y}({1}), a_{z}({1})$). The weight vector after many rounds of repeated braiding chain obeys a difference equation,
\begin{eqnarray}\label{a3D}
a_{x} (m)&=&  a_{z}(m-1) + a_{y}(m-1),\nonumber \\
a_{y} (m)&=&  a_{x}(m-1) + a_{z}(m-1),  \nonumber \\
a_{z} (m)&=& a_{x}(m-1) + a_{y}(m-1) + 2 a_{z}(m-1), \nonumber \\
\end{eqnarray}
where $m$ is an integer indicating the total number of braiding periods. The conservation equation,  $a_{x} (m)+a_{y} (m) = a_{z} (m)$, holds at any period, indicating an integral charge on the Z-edge and two fractional charges on the X- and Y-edges,
\begin{eqnarray}\label{Qxyzn}
Q_{x} (m)= \frac{a_{x}(m)}{a_{z} (m)},\;\;Q_{y} (m)= \frac{a_{y}(m) }{a_{z}(m)},\;\;Q_{z} (m)= 1.
\end{eqnarray}
The specific fractional charge sequence is determined by the initial charges, as suggested by the exact solution of this difference equation,
\begin{eqnarray}\label{a3Dnnn}
a_{x} (m)&=& \frac{1}{4}[3^{m} - (-1)^{m}] [a_{y}(0)+a_{z}(0)]\nonumber \\
&+& \frac{1}{4}[3^{m} + 3(-1)^{m}] a_{x}(0), \nonumber \\
a_{y} (m)&=&\frac{1}{4}[3^{m} - (-1)^{n}] [a_{x}(0)+a_{z}(0)]\nonumber \\
&+& \frac{1}{4}[3^{m} - 3(-1)^{m}] a_{y}, \nonumber \\
a_{z} (m)& = & \frac{1}{4}[3^{m} - (-1)^{m}]  [a_{x}(0)+a_{y}(0)]\nonumber \\
&+&\frac{1}{4}[3*3^{m} + (-1)^{m}] a_{z}(0).
\end{eqnarray}
For instance, the special initial charge distribution, $a_{x}(0) = a_{z}(0) =1, a_{y}(0) = 0$, leads to the fractional charge sequence,
\begin{eqnarray}\label{Q3Dn}
Q_{x} (m)= \frac{3^{m} + (-1)^{m}}{2*3^m},\;\;Q_{y} (m)= \frac{3^{m} - (-1)^{m}}{2*3^m},\;\;
\end{eqnarray}
and $Q_{z} (m) = 1,$ an internal charge on Z-edge. In the limit of infinite number of braiding, the fractional charge on the X-edge and Y-edge reduces to half-charge,
\begin{eqnarray}\label{1Qxyinfty}
Q_{x} (\infty) = Q_{y} (\infty) =\frac{1}{2},
\end{eqnarray}
which is independent of initial charges. The current on the X- and Y-edge comes from the splitting current in Z-edge. As Fig. \ref{Hall3Ddata} shows, the 3D Hall resistance also shows a serial of plateaus but much less than that of two dimensional case. This fractional serial is not the only fractional charge serial, a different fractional charge serial is induced by a different initial state, for instance, take an eigenvector as initial state \begin{eqnarray}\label{axyz0}
a_{x}({0}) = 1, \;\;\; a_{y}({0}) = 1, \;\;\; a_{z}({0}) = 2.
\end{eqnarray}
The weight distribution on the X- and Y-edge evolves according to the following equation,
\begin{eqnarray}\label{axyzn}
a_{x}({m}) = 3^{m}, \;\;\; a_{y}({m}) = 3^{m}, \;\;\; a_{z}({m}) = 2\ast3^{m},
\end{eqnarray}
after $m$ rounds of braiding chain operations. This distribution constantly admits a half charge $Q = 1/2$ on both X- and Y-edge despite of the number of braiding operations. The two half charges in X-edge and Y-edge fused into the Z-edge without losing any fractions, i.e., $a_{x}(m) + a_{y}(m) = a_{z}(m)$. This brief conservation relation only holds for the first braiding chain following Fig. \ref{3Dfluxlattice} (b)(c).

For the second braiding chain that maps $S = (+1,-1,+1)$ to $S = (-1,+1,-1)$ showed in Fig. \ref{3Dfluxlattice} (b)(d), the fractional charge serial is different from that above even though the two braiding chains only differ at one step of operation. The transition equation of weight vectors after $n$ rounds of braiding reads,
\begin{eqnarray}\label{seca3D}
a_{x} (m)&=& 2 a_{x}(m-1) + a_{y}(m-1) + 3 a_{z}(m-1),\nonumber \\
a_{y} (m)&=&  a_{x}(m-1) + a_{z}(m-1),  \nonumber \\
a_{z} (m)&=& 2 a_{x}(m-1) + 2 a_{y}(m-1) + 3 a_{z}(m-1).\nonumber \\
\end{eqnarray}
This difference equation is directly read out from the track distribution in Fig. \ref{3Dfluxlattice} (d) - (V). The brief conservation equation for the first braiding chain, $a_{x} (m) + a_{y} (m) = a_{z} (m)$, failed to meet the this recurrence process. The same initial charge distribution as before, $a_{x}(0) = a_{z}(0) =1, a_{y}(0) = 0$, results in a different distribution, $a_{x} (1) = 5, a_{y} (1) = 2, a_{z} (1)= 5, $ which obviously falls out of the conservation equation. While another initial state, $(a_{x}(0) = 0, a_{y}(0) = a_{z}(0) =1)$, leads to weight distribution after one period of operation, $a_{x} (1) = 4, a_{y} (1) = 1, a_{z} (1)= 5,$ which obeys $a_{x} (1) + a_{y} (1) = a_{z} (1)$. But one more period of braiding chain drives the solution out of conservation law, $a_{x} (2) = 24, a_{y} (2) = 9, a_{z} (2)= 26.$ Thus the conservation equation of currents that meet at one node, similar to Kirchhoff law in electric circuit theory, is not a solid physical equation in this train track theory. Despite of different initial states and its corresponding fractional serial, the fractional charge on the X-edge and Y-edge finally converges to the same stable value,
\begin{eqnarray}\label{Qx1y13}
\lim_{n\rightarrow\infty}Q_{x} (m) &= &\lim_{m\rightarrow\infty}\frac{a_{x}(m)}{a_{z} (m)} = 1,\nonumber \\
\lim_{n\rightarrow\infty}Q_{y} (m)& = & \lim_{m\rightarrow\infty}\frac{a_{y}(m)}{a_{z} (m)} = \frac{1}{3}.
\end{eqnarray}
The Z-edge carries an integral charge, $Q_{z} (m) = 1$. This limit charge is completely different from that of the first braiding chain. Since this limit charge is independent of initial state, it can be used to characterize the special action of different braiding chain.

The oscillation behavior of the weight vector can be understood from the exact solution of the recurrence equation, which is derived by the Putzer algorithm. A matrix representation of this equation is expressed as $\vec{a} (m) = A \vec{a} (m-1)$, where $A$ is the coefficient matrix,
\begin{eqnarray}
A=\begin{pmatrix}
2 & 1 & 3 \\
1 & 0 & 1 \\
2 & 2 & 3 \\
\end{pmatrix}.
\end{eqnarray}
The eigenvalue of this matrix is derived by the characteristic equation $det(A-\lambda'{I}) = 0$, which has an explicit form, $1+3\lambda'+5\lambda'^2-\lambda'^3 = 0$. It yields one real eigenvalue and two complex eigenvalue, approximated by their numerical values for simplicity,
\begin{eqnarray}\label{lambda123}
\lambda_{1} = {\lambda}e^{-i\theta}, \lambda_{2} = {\lambda} e^{i\theta}, \lambda_{3} = 5.57,  {\lambda}=0.42,\theta = 2.32.\nonumber\\
\end{eqnarray}
The exact weight distribution on the three edges after n periods of braiding chain operations is $\vec{a} (m) = A^{m} \vec{a} (0)$, where $A^{m}$ is determined by the three eigenvalues above,
\begin{eqnarray}\label{An}
&&A^{m}={\lambda}^{m}e^{-in\theta}\mathbb{I}+{\lambda}^{m}\frac{\sin(m\theta)}{\sin(\theta)}(A-{\lambda}e^{-i\theta}\mathbb{I})  \nonumber\\
&+&\frac{{\lambda_{3}^{m-1}}}{\lambda}\sum_{j = 0}^{m-1}(\frac{\lambda}{\lambda_{3}})^{j}\frac{\sin(j\theta)}{\sin(\theta)}(A-{\lambda}e^{i\theta}\mathbb{I})(A-{\lambda}e^{-i\theta}\mathbb{I})\mathbb{I},\nonumber\\
\end{eqnarray}
where $\mathbb{I}$ is the identity matrix. The first and second terms decay very fast as the number of braiding chain grows, because their amplitude is proportional to ${\lambda}^{n}$ with $({\lambda} = 0.42 < 1)$, which approaches to zero. The dominant terms is the third term in the right hand side of Eq. (\ref{An}), revealing an oscillating weight distribution on the three edges as well as oscillating fractional charges that converges to the limit value.

The two exemplar cases above revealed the widespread existence of fractional charges in three dimensional network of magnetic fluxes. To observe the fractional charge in three dimensions in experiment, the temperature has to be low enough to make sure the mean free path of electron is long enough to complete the braiding. A homogeneous magnetic field is only suitable for fractional charge in two dimensions, a strong field of  crossing magnetic fluxes in three dimensions is a key technology challenge for experimental observation. Such a strong anisotropic magnetic field in nature may be exist on the surface of the Sun.

\subsection{The fractional charges generated from the vacuum states of electromagnetic current knot lattice in three dimensions}

\begin{figure}
\begin{center}
\includegraphics[width=0.4\textwidth]{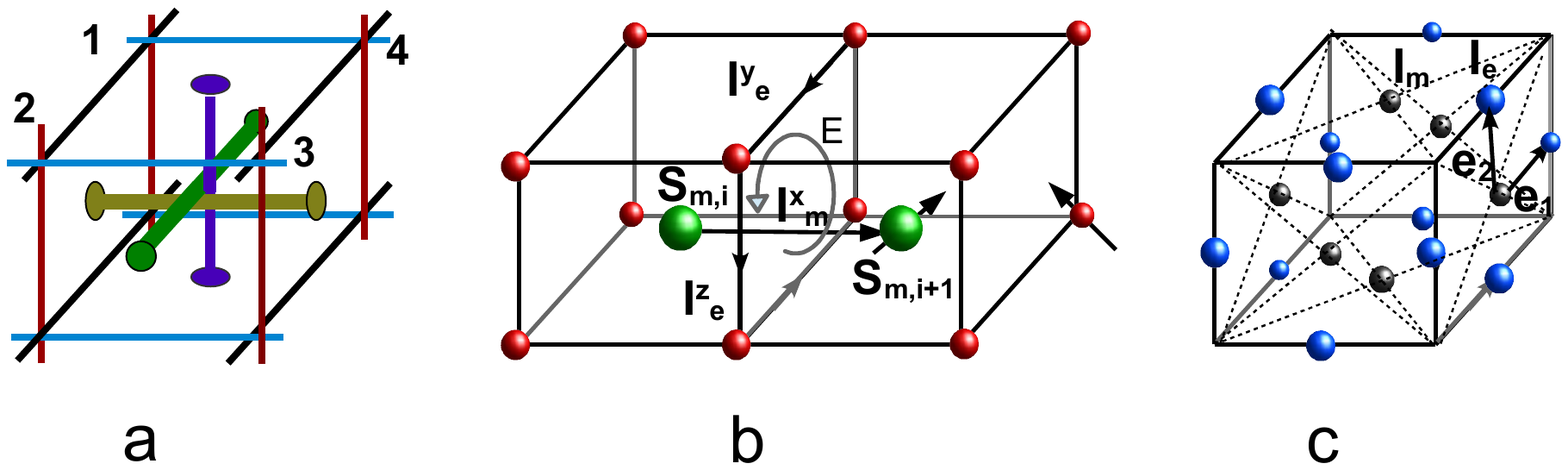}
\caption{\label{3DnetEM} (a)The three dimensional knot lattice of magnetic flux tubes interlocks with the knot lattice of electric currents. (b) The crossing nodes of electric currents forms a cubic lattice, with the crossing node of magnetic fluxes placed at the center of the square plaquette. (c) The dual face-centered cubic lattice of current segments, with each current segment represented by its middle point (the small blue and red balls).}
\end{center}
\vspace{-0.5cm}
\end{figure}

Similar to the winding train tracks emerged as vacuum state of two dimensional knot lattice of electric current, the winding train track of fractional charges also emerge in the vacuum state of three dimensional knot lattice. The three dimensional network of non-intersecting magnetic flux tubes in Fig. \ref{3DdualEM} interlocks with a dual network of electric current, as showed by the intersecting edges of the cubic. If we replace the intersecting electric edges by non-intersecting electric current, it constructs exactly the dual three dimensional knot lattice interlocking with that of magnetic fluxes as Fig. \ref{3DnetEM} shows. The crossing states of three electric currents in Fig. \ref{3DnetEM} is classified by the same Ising spins of crossing magnetic fluxes in Fig. \ref{3DdualEM}. These electric crossing nodes (represented by the red dots in Fig. \ref{3DnetEM}) are located on the vertex of cubic lattice, while the magnetic crossing nodes locates at the center of every unit cubic (represented by the green dots in Fig. \ref{3DnetEM}). Every magnetic flux can be equivalently viewed as a magnetic current of a running magnetic monopole. According to electromagnetic field theory, a magnetic (electric) current generate a circling electric (magnetic) field around it, this mutual induction effect induced strong coupling interaction between the electric current segment and magnetic current segment in the three dimensional knot lattice. Specifically speaking, the magnetic current $I_{m}^{x}$ induces a circular electric field $E$ (Fig. \ref{3DnetEM}) in the plaquette of electric current, which either accelerate the electric current segment in the direction of circular field or decelerate it in the opposite direction of circular field. This coupling interaction is governed by the Hamiltonian,
\begin{eqnarray}\label{Hami3DIII}
H_{1}=U\sum_{\textbf{r}}\vec{I}_{m,\textbf{r}}\cdot(\vec{I}_{e,\textbf{r}+\textbf{e}_{1}}\times\vec{I}_{e,\textbf{r}+\textbf{e}_{2}}),
\end{eqnarray}
where the current segments are located on the dual face-centered cubic lattice in Fig. \ref{3DnetEM}. This coupling terms is derived naturally from electromagnetic field theory but shares a similar form as Abelian Chern-Simons term, revealing the topological character of the three dimensional knot network of electromagnetic currents.

The above coupling term of electromagnetic current segments does not capture the internal state of the crossing node. As Fig. \ref{3DnetEM} (a) showed, there exist a topological repulsion between the same crossing states of the nearest neighboring nodes No. 1 and No. 2, because the vertical currents $I_{e,1}^{z}$ and $I_{e,2}^{z}$ (represented by the vertical red line) can not fuse into each other without cutting the horizontal current $I_{e,1}^{x}$. However, if the two nearest neighboring nodes have opposite crossing state, such as $I_{e,3}^{z}$ and $I_{e,4}^{z}$ in Fig. \ref{3DnetEM} (a), the two vertical currents $I_{e,3}^{z}$ and $I_{e,4}^{z}$ can gets closer enough to fuse into one current, indicating a topological attraction. Thus the interaction between the nearest neighboring nodes is summarized into the following Hamiltonian,
\begin{eqnarray}\label{Hami3DSS}
H_{2}=h\sum_{\langle\textbf{ij}\rangle}\vec{S}_{\alpha,\textbf{i}}\cdot\vec{S}_{\alpha,\textbf{j}},
\end{eqnarray}
here $\vec{S}_{\alpha,\textbf{i}},(\alpha=e,m)$ is the three dimensional Ising spin. The two hamiltonian parts together give a complete description on the three dimensional knot lattice of electromagnetic currents, $H=H_{1}+H_{2}$. When the crossing current at a node is mapped into non-intersecting vacuum arc pairs (as Fig. (\ref{3DdualEM}) showed) and assembly into a collective current pattern in three dimensional space, a vortex-path pattern around a curved magnetic flux tube can always be constructed self-consistently to visualize the FQHE in a tilted magnetic field \cite{yang2002}. If there exist no other magnetic flux tubes blocked in between neighboring current segment, these electromagnetic current segment can fuse into one bundle under topological transformation, leading to fractional charges in three dimensional space. This construction protocol based on lattice structure has a straightforward relation with quantum many body models on three dimensional lattice.

\section{Conclusion}

A topological path fusion theory is developed to generate serial of fractional charges that is similar to those in FQHE and beyond. This physical theory is rooted in the path integral theory of quantum mechanics, topology theory of train tracks and knot lattice model of anyon, providing a mathematical insight on the physical origin of fractional charges in quantum system in magnetic flux lattice. The existed serials of fractional charge in FQHE are well explained by topological fusion of winding paths around flux pairs, revealing the topological root of Jain's composite fermion theory. Similar winding path also exists as energy flow in momentum space of boson-fermion pairing model \cite{Si2019}. The winding path can be mapped into knot lattice. The regular knot path induced the plateau of Hall resistance, while the chaotic paths result in the fractal structure of energy spectrum of Hofstadter model. The corresponding knot lattice model in this theory also provides a systematic way of constructing collective wave function of anyon, such as the Laughlin wavefunction.

This topological path fusion theory predicted irrational charges and fractionally charged anyon in three dimensions, despite of the long-term belief that anyon can not exist in three dimensions. Fractional charges can also be implemented in a multi-connected domain without magnetic fluxes, such as a porous material with forbidden zones. The topological path splitting and fusion not only generate fractional charges, but also generate factional masses of elementary particles. The rapid development of experimental technology for detecting anyon \cite{Bartolomei2020}\cite{Nakamura2020} provides a promising future to verify the predictions above, even though it is quite a challenge to construct such a strong magnetic field in laboratory. The topological mixing of two quantum fluids or fluid of light in optical cavity maybe is also promising for implementing this theory. The topological fractional charge is a fundamental character of quantum system with multi-connected spatial domain, therefore there exist many physical systems for topological quantum computation.

\bibliographystyle{apsrev}

\bibliography{references}

\end{document}